\documentclass[final,3p,times,authoryear]{elsarticle}
\makeatletter\let\ps@pprintTitle\ps@plain\makeatother

\usepackage{amssymb}
\usepackage{amsmath}
\usepackage{amsthm}
\usepackage{aliascnt}
\usepackage{mathtools}
\usepackage{bbm}
\DeclareFontFamily{U}{bbm}{}
\DeclareFontShape{U}{bbm}{m}{n}{<-> bbm10}{}
\usepackage{mathabx}

\usepackage{xcolor}
\usepackage{tikz}
\usetikzlibrary{arrows.meta,positioning,fit,backgrounds,calc}
\usepackage{multicol}
\usepackage{float}
\usepackage{array}
\usepackage{enumerate}
\usepackage[final]{microtype}
\usepackage{cmap}          
\usepackage[T1]{fontenc}

\usepackage[hidelinks]{hyperref}
\usepackage[capitalize,noabbrev,nameinlink,sort]{cleveref}
\newcommand{\Crefpair}[3]{\crefname{#1}{#2}{#3}\Crefname{#1}{#2}{#3}}
\Crefpair{theorem}{Theorem}{Theorems}
\Crefpair{lemma}{Lemma}{Lemmas}
\Crefpair{corollary}{Corollary}{Corollaries}
\Crefpair{definition}{Definition}{Definitions}
\Crefpair{condition}{Condition}{Conditions}
\Crefpair{remark}{Remark}{Remarks}
\Crefpair{example}{Example}{Examples}
\Crefpair{proposition}{Proposition}{Propositions}
\Crefpair{section}{Section}{Sections}
\Crefpair{subsection}{Section}{Sections}
\Crefpair{subsubsection}{Section}{Sections}
\Crefpair{appendix}{Appendix}{Appendices}
\crefformat{appendix}{#2#1#3}
\Crefformat{appendix}{#2#1#3}
\Crefpair{figure}{Figure}{Figures}
\Crefpair{table}{Table}{Tables}

\Crefpair{equation}{Equation}{Equations}
\creflabelformat{equation}{#2(#1)#3}
\crefrangeformat{equation}{Equations~#3(#1)#4--#5(#2)#6}
\Crefrangeformat{equation}{Equations~#3(#1)#4--#5(#2)#6}
\crefmultiformat{equation}
  {Equations~#2(#1)#3}
  { and~#2(#1)#3}
  {, #2(#1)#3}
  {, and~#2(#1)#3}
\Crefmultiformat{equation}
  {Equations~#2(#1)#3}
  { and~#2(#1)#3}
  {, #2(#1)#3}
  {, and~#2(#1)#3}

\crefformat{enumi}{#2(#1)#3}
\Crefformat{enumi}{#2(#1)#3}
\crefrangeformat{enumi}{#3(#1)#4--#5(#2)#6}
\Crefrangeformat{enumi}{#3(#1)#4--#5(#2)#6}
\crefmultiformat{enumi}{#2(#1)#3}{ and~#2(#1)#3}{, #2(#1)#3}{, and~#2(#1)#3}
\Crefmultiformat{enumi}{#2(#1)#3}{ and~#2(#1)#3}{, #2(#1)#3}{, and~#2(#1)#3}
\crefformat{enumii}{#2(#1)#3}
\Crefformat{enumii}{#2(#1)#3}
\crefrangeformat{enumii}{#3(#1)#4--#5(#2)#6}
\Crefrangeformat{enumii}{#3(#1)#4--#5(#2)#6}
\crefmultiformat{enumii}{#2(#1)#3}{ and~#2(#1)#3}{, #2(#1)#3}{, and~#2(#1)#3}
\Crefmultiformat{enumii}{#2(#1)#3}{ and~#2(#1)#3}{, #2(#1)#3}{, and~#2(#1)#3}
\crefformat{enumiii}{#2(#1)#3}
\Crefformat{enumiii}{#2(#1)#3}
\crefrangeformat{enumiii}{#3(#1)#4--#5(#2)#6}
\Crefrangeformat{enumiii}{#3(#1)#4--#5(#2)#6}
\crefmultiformat{enumiii}{#2(#1)#3}{ and~#2(#1)#3}{, #2(#1)#3}{, and~#2(#1)#3}
\Crefmultiformat{enumiii}{#2(#1)#3}{ and~#2(#1)#3}{, #2(#1)#3}{, and~#2(#1)#3}
\crefformat{enumiv}{#2(#1)#3}
\Crefformat{enumiv}{#2(#1)#3}
\crefrangeformat{enumiv}{#3(#1)#4--#5(#2)#6}
\Crefrangeformat{enumiv}{#3(#1)#4--#5(#2)#6}
\crefmultiformat{enumiv}{#2(#1)#3}{ and~#2(#1)#3}{, #2(#1)#3}{, and~#2(#1)#3}
\Crefmultiformat{enumiv}{#2(#1)#3}{ and~#2(#1)#3}{, #2(#1)#3}{, and~#2(#1)#3}

\newcommand{\ProofLink}[1]{%
  \leavevmode\hfill\textnormal{\footnotesize
    \mbox{\hyperref[pf:#1]{\texorpdfstring{($\to$ Proof, p.~\pageref*{pf:#1})}{-> Proof, p.~\pageref*{pf:#1}}}}%
  }%
}

\numberwithin{equation}{section}

\theoremstyle{plain}
\newtheorem{theorem}{Theorem}[section]
\newaliascnt{corollary}{theorem}
\newtheorem{corollary}[corollary]{Corollary}
\aliascntresetthe{corollary}
\newaliascnt{lemma}{theorem}
\newtheorem{lemma}[lemma]{Lemma}
\aliascntresetthe{lemma}
\newaliascnt{proposition}{theorem}
\newtheorem{proposition}[proposition]{Proposition}
\aliascntresetthe{proposition}

\theoremstyle{definition}

\newtheorem{condition}{Condition}[section]

\theoremstyle{plain}
\newtheorem{remark}{Remark}[section]

\AddToHook{cmd/appendix/after}{%
}

\renewcommand{\hat}{\widehat}
\renewcommand{\bar}{\overline}

\renewcommand{\P}{\mathbb{P}}
\DeclareMathOperator{\Var}{Var}
\DeclareMathOperator{\Cov}{Cov}
\newcommand{\given}{\;\middle|\;}

\newcommand{\pto}{\xrightarrow{\P}}
\newcommand{\dto}{\xrightarrow{d}}

\newcommand{\R}{\mathbb{R}}

\newcommand{\dd}{\;\mathrm{d}}

\newcommand{\bx}{\mathbf{x}}

\newcommand{\calB}{\mathcal{B}}

\newcommand{\calF}{\mathcal{F}}

\newcommand{\calH}{\mathcal{H}}

\newcommand{\calW}{\mathcal{W}}

\newcommand{\vecalpha}{\boldsymbol{\alpha}}
\newcommand{\veceta}{\boldsymbol{\eta}}
\newcommand{\vecpsi}{\boldsymbol{\psi}}

\newcommand{\veca}{\boldsymbol{a}}
\newcommand{\vecm}{\boldsymbol{m}}

\DeclarePairedDelimiterX{\abs}[1]{\lvert}{\rvert}{\mathopen{}\thinspace #1 \thinspace\mathclose{}}
\DeclarePairedDelimiterX{\norm}[1]{\lVert}{\rVert}{\mathopen{}\thinspace #1 \thinspace\mathclose{}}
\DeclarePairedDelimiterX{\ip}[1]{\langle}{\rangle}{\mathopen{}\thinspace #1 \thinspace\mathclose{}}
\DeclarePairedDelimiterX{\opnorm}[1]{\lVert}{\rVert_{\mathrm{op}}}{\mathopen{}\thinspace #1 \thinspace\mathclose{}}
\DeclarePairedDelimiterX{\floor}[1]{\lfloor}{\rfloor}{\mathopen{}\thinspace #1 \thinspace\mathclose{}}
\DeclarePairedDelimiterX{\ceil}[1]{\lceil}{\rceil}{\mathopen{}\thinspace #1 \thinspace\mathclose{}}
\DeclarePairedDelimiterX{\set}[1]{\lbrace}{\rbrace}{\mathopen{}\thinspace #1 \thinspace\mathclose{}}

\DeclarePairedDelimiterXPP{\Ex}[1]{\mathbb{E}}{\lbrack}{\rbrack}{}{%
  \renewcommand\given{\nonscript\:\delimsize\vert\nonscript\:\mathopen{}}%
  \mathopen{}#1\mathclose{}%
}
\DeclarePairedDelimiterXPP{\Exi}[2]{\mathbb{E}_{#1}}{\lbrack}{\rbrack}{}{%
  \renewcommand\given{\nonscript\:\delimsize\vert\nonscript\:\mathopen{}}%
  \mathopen{}#2\mathclose{}%
}

\DeclarePairedDelimiterXPP{\Pb}[1]{\P}{(}{)}{}{%
  \renewcommand\given{\nonscript\:\delimsize\vert\nonscript\:\mathopen{}}%
  \mathopen{}#1\mathclose{}%
}
\DeclarePairedDelimiterXPP{\Pbi}[2]{\P_{#1}}{(}{)}{}{%
  \renewcommand\given{\nonscript\:\delimsize\vert\nonscript\:\mathopen{}}%
  \mathopen{}#2\mathclose{}%
}

\DeclarePairedDelimiterXPP{\Varb}[1]{\Var}{(}{)}{}{%
  \mathopen{}#1\mathclose{}%
}
\DeclarePairedDelimiterXPP{\Varbi}[2]{\Var_{#1}}{(}{)}{}{%
  \mathopen{}#2\mathclose{}%
}

\DeclarePairedDelimiterXPP{\Covp}[1]{\Cov}{(}{)}{}{%
  \mathopen{}#1\mathclose{}%
}
\DeclarePairedDelimiterXPP{\Covpi}[2]{\Cov_{#1}}{(}{)}{}{%
  \mathopen{}#2\mathclose{}%
}

\DeclarePairedDelimiterXPP{\1}[1]{\mathbbm{1}}{\lbrace}{\rbrace}{}{%
  \mathopen{}\thinspace #1 \thinspace\mathclose{}%
}

\DeclarePairedDelimiterXPP{\OP}[1]{O_{\P}}{(}{)}{}{%
  \mathopen{}\thinspace #1 \thinspace\mathclose{}%
}
\DeclarePairedDelimiterXPP{\oP}[1]{o_{\P}}{(}{)}{}{%
  \mathopen{}\thinspace #1 \thinspace\mathclose{}%
}

\newcommand{\neffx}{n_{\mathrm{eff}}}

\newcommand{\pkg}[1]{\textsc{#1}}
\newcommand{\routetag}[2]{\pkg{#1}\ensuremath{[#2]}}
\newcommand{\routelink}[3]{\hyperref[#3]{\routetag{#1}{#2}}}
\newcommand{\AffRoute}{\mathrm{Aff}}
\newcommand{\SmRoute}{\mathrm{Sm}}
\newcommand{\NSRoute}{\mathrm{NS}}
\newcommand{\ZeroRoute}{0}
\newcommand{\GapRoute}{\mathrm{Gap}}
\newcommand{\AnalyticRoute}{\mathrm{Analytic}}
\newcommand{\DerivativeRoute}{\mathrm{Derivative}}
\newcommand{\FDRoute}{\mathrm{FD}}
\newcommand{\DeclareRouteCref}[3]{%
  \crefformat{#1}{#2 route in #3~##2##1##3}%
  \Crefformat{#1}{#2 route in #3~##2##1##3}%
}
\DeclareRouteCref{smtAff}{\routetag{SM--T}{\AffRoute}}{Condition}
\DeclareRouteCref{smtSm}{\routetag{SM--T}{\SmRoute}}{Condition}
\DeclareRouteCref{smtNS}{\routetag{SM--T}{\NSRoute}}{Condition}
\DeclareRouteCref{psfsAff}{\routetag{PS--FS}{\AffRoute}}{Condition}
\DeclareRouteCref{psfsSm}{\routetag{PS--FS}{\SmRoute}}{Condition}
\DeclareRouteCref{psfsNS}{\routetag{PS--FS}{\NSRoute}}{Condition}
\DeclareRouteCref{mmidZero}{\routetag{MM--ID}{\ZeroRoute}}{Condition}
\DeclareRouteCref{mmidGap}{\routetag{MM--ID}{\GapRoute}}{Condition}
\DeclareRouteCref{brAnalytic}{\routetag{BR}{\AnalyticRoute}}{Definition}
\DeclareRouteCref{brDerivative}{\routetag{BR}{\DerivativeRoute}}{Definition}
\DeclareRouteCref{brFD}{\routetag{BR}{\FDRoute}}{Condition}

\newcounter{assumproute}[condition]

\makeatletter
\@ifundefined{theHcondition}
  {}
  {}
\@ifundefined{theHassumproute}
  {}
  {}

\newcommand{\routecondlabel}[3][]{%
  \refstepcounter{assumproute}%
  \def\@currentlabelname{#1}%
  \label[#2]{#3}%
}

\newcounter{defroute}[definition]

\@ifundefined{theHdefinition}
  {}
  {}
\@ifundefined{theHdefroute}
  {}
  {}

\newcommand{\routedeflabel}[3][]{%
  \refstepcounter{defroute}%
  \def\@currentlabelname{#1}%
  \label[#2]{#3}%
}
\makeatother

\DeclareMathOperator*{\argmin}{arg\,min} 
\newcommand{\indep}{\mathrel{\perp\mspace{-10mu}\perp}}

\usepackage{placeins}

\newcommand{\CoupledPaperAuthor}{Jakob R. Juergens}
\newcommand{\CoupledPaperAffiliation}{%
  organization={University of Wisconsin-Madison},
  addressline={1180 Observatory Drive},
  city={Madison},
  postcode={53706},
  state={Wisconsin},
  country={United States}%
}
\newcommand{\CoupledPaperEmail}{jrjuergens@wisc.edu}
\newcommand{\CoupledPaperWebsite}{https://jakobjuergens.com}

\newcommand{\AffinePaperTitleLead}{%
  Orthogonal Inference for Conditional Z-Estimation%
}
\newcommand{\AffinePaperTitleTail}{%
  A Distributional Nearest-Neighbor Approach%
}
\newcommand{\AffinePaperTitle}{%
  \AffinePaperTitleLead
  \ifpreprint\\\else\space\fi
  without Cross-Fitting:\\
  \AffinePaperTitleTail%
}

\input{affine_external_references.tex}

\journal{Journal of Econometrics}
\begin{document}
\pagestyle{plain}
\setcounter{page}{1}
\begin{frontmatter}
  \title{\AffinePaperTitle}
  \author{\CoupledPaperAuthor\texorpdfstring{\corref{affine-corresponding}}{}}
  \cortext[affine-corresponding]{Corresponding author.}
  \expandafter\affiliation\expandafter{\CoupledPaperAffiliation}
  \ead{\CoupledPaperEmail}
  \ead[url]{\CoupledPaperWebsite}
  \begin{abstract}
    Estimating conditional parameters, such as heterogeneous treatment effects, often requires rich covariate adjustment, even when the parameters are reported along only a few dimensions.
    Fitting nuisance functions on the full sample makes all observations available for adjustment, but creates dependence between fitted scores and the observations at which they are evaluated, which is typically avoided by cross-fitting.
    We develop pointwise inference with full-sample nuisance predictions for exactly identified conditional moment equations, recovering conditional average treatment effects as a special case.
    We localize the estimating equations around the reporting point using distributional nearest neighbors (DNN), a U-statistic-based averaging procedure.
    Under sufficient conditions on the local design, orthogonality, prediction accuracy and quantitative deletion stability, the estimator is asymptotically normal, and a frozen-score jackknife consistently estimates its limiting covariance with nuisance predictions held fixed.
    In simulations, full-sample and cross-fitted inference perform similarly, with coverage shortfalls from nuisance approximation error, spatial bias and heavy studentized tails.
    An application to 401(k) eligibility using Survey of Income and Program Participation data finds similar estimates across nuisance prediction methods, with wide normal-reference intervals.
\end{abstract}

\end{frontmatter}

\section{Introduction}\label{sec:affine_introduction}

Many parameters of economic interest are reported conditional on a small number of observable characteristics, even when estimation requires adjustment for a much richer set of covariates.
Conditional average treatment effects (CATE), for example, can describe heterogeneity of treatment effects by income and age while propensity scores and outcome regressions account for additional characteristics such as education and family size.
Estimating these effects requires households with characteristics similar to the profile of interest, and the number of such households can vary widely across profiles, meaning that local comparisons can remain imprecise even when the full sample is large.

Neyman-orthogonal scores make the estimating equation first-order insensitive to nuisance estimation error.
Cross-fitting separates nuisance training from score evaluation by predicting each observation with a fit that excludes its fold \citep{chernozhukov_doubledebiased_2018}.
However, each fold fit uses fewer training observations, even when all observations are retained in the final local equation.
We study whether the nuisance functions can instead be fitted once, on the full sample, throughout conditional estimation and inference.
An observation can then influence the nuisance predictions used to evaluate its own score.
In this paper, we give sufficient conditions for valid inference with full-sample nuisance predictions and distributional nearest-neighbor (DNN) weights in exactly identified conditional moment equations that are affine in a fixed-dimensional parameter of interest.
Orthogonality removes first-order nuisance bias, and quantitative deletion stability of the learner bounds the dependence created by using the same observations to fit the nuisance functions and to solve the local equation.

Our affine score class includes CATE, where nuisance functions enter only the adjusted outcome signal, and conditional local average treatment effects (CLATE) and residualized slopes, where they also enter the Jacobian coefficient.
DNN averages nearest-neighbor scores over all subsamples of a fixed size, while only depending on distance ranks, so the estimator and its jackknife covariance can be computed directly, without enumerating subsamples.
A frozen-score jackknife changes the local weights while retaining the parameter estimate and fitted nuisance predictions, so estimating uncertainty requires no nuisance refits.

We verify the learner conditions in a model covering CATE and a scalar conditional slope, using a growing clipped ridge-series learner.
In simulations and an application to 401(k) eligibility, we compare estimates based on full-sample and cross-fitted nuisance predictions while holding the observations and weights in the final estimating equation fixed.
In the reported simulations, full-sample reuse gives inference similar to cross-fitting; the coverage shortfalls reflect nuisance approximation error, spatial bias and heavy studentized tails affecting both methods.
The application estimates are also close across prediction methods, with wide intervals that point to limited local information.

\paragraph{Related literature}
Our paper contributes to the literature on estimating heterogeneous treatment effects with flexible nuisance models.
\citet{belloni_uniform_2015} develop uniformly valid inference for Z-estimation with estimated nuisance functions using orthogonal scores and restrictions on nuisance complexity.
\citet{fan_estimation_2022_affine} study CATE reported along a small number of covariates while allowing adjustment for a much richer set, using local-linear estimation with either full-sample or cross-fitted nuisance predictions.
Their full-sample analysis restricts nuisance-class complexity through entropy bounds which, while theoretically elegant, can be difficult to verify in practice.
Our pointwise DNN analysis instead requires quantitative deletion stability and allows nuisance-dependent affine coefficients.
\citet{semenova_debiased_2021} develop series estimation and simultaneous inference for orthogonal signals, and \citet{kennedy_towards_2023} gives oracle error bounds for regression with estimated pseudo-outcomes.
\citet{nie_quasi-oracle_2021} use outcome and treatment residuals to learn treatment-effect heterogeneity.
Closest to our question, \citet{chen_debiased_2022-2} show that valid inference without sample splitting is possible when the nuisance learner is sufficiently stable, so that omitting one observation has only a small effect on its predictions.
We develop this stability argument for local affine estimating equations with nuisance-dependent coefficients.
We relate the required prediction and deletion bounds to the local standard error and establish inference using a frozen-score jackknife.
\citet{velez_debiased_2026_affine} studies the role of the number of cross-fitting folds, including leave-one-out fitting, where almost the entire sample is available for each nuisance fit.
Each prediction still comes from a fit that excludes the scored observation; our predictions use a fit that includes it.

Forest methods provide another approach to estimating conditional effects, using the data to construct neighborhoods of comparable observations \citep{wager_estimation_2018,athey_generalized_2019,Oprescu2019}.
Our use of DNN draws on work on averaging nearest-neighbor estimators, including the exact averaging formulas of \citet{steele_exact_2009} and the convergence rates studied by \citet{biau_rate_2010}.
\citet{demirkaya_optimal_2024} develop estimation and inference for two-scale DNN regression.
We study DNN estimating equations with fitted nuisance functions and apply the generalized-\(U\)-statistic jackknife results of \citet{juergens_jackknife_2025} to their score covariance.

Sections~\ref{sec:affine_setup}--\ref{sec:affine_results} develop estimation and inference, Section~\ref{sec:affine_applications} verifies the series learner, and Section~\ref{sec:affine_numerical} reports the simulations and 401(k) application.
Proofs appear in \ref{app:affine_technical} through \ref{app:affine_series_construction}, and \ref{app:affine_production_numerics} gives the simulation design.

\section{Conditional Affine Scores}\label{sec:affine_setup}

\subsection{The general coefficient equation}

Let a generic observation \(W\) contain reporting covariates \(X\in\R^{k_x}\), and fix a point \(\bx \in\R^{k_x}\) that specifies the covariate profile of interest.
At a reporting profile \(u\), the parameter of interest \(\vecalpha_0(u)\in\R^p\) has fixed dimension \(p\) and is identified by a conditional moment equation, with nuisance functions \(\veceta_0\) entering an exactly identified score of the form
\begin{equation}\label{eq:affine_score_structure}
    \vecpsi(W;\veca,\veceta)
    = B(W;\veceta)\veca+b(W;\veceta),
    \qquad B(W;\veceta)\in\R^{p\times p},\quad b(W;\veceta)\in\R^p.
\end{equation}
The score is affine in the parameter, and \(B\) and \(b\) may vary across observations and depend nonlinearly on \(\veceta\).
With the reporting covariates fixed at \(u\), the conditional moment equation averages these coefficients over the remaining components of \(W\).
For fixed nuisance functions, define
\begin{equation}\label{eq:affine_conditional_coefficients}
    \bar B(u;\veceta)=\Ex*{B(W;\veceta)\given X=u},
    \qquad
    \bar b(u;\veceta)=\Ex*{b(W;\veceta)\given X=u}.
\end{equation}
We fix versions of these conditional means near \(\bx\), giving the conditional moment map \(\vecm(u;\veca,\veceta)=\bar B(u;\veceta)\veca+\bar b(u;\veceta)\), and define the parameter at each profile by
\begin{equation}\label{eq:affine_target}
    \bar B(u;\veceta_0)\vecalpha_0(u)+\bar b(u;\veceta_0)=0.
\end{equation}
We rely on nonsingularity of \(\bar B(u;\veceta_0)\) to identify the parameter locally, with a Jacobian \(\Gamma_{\bx}=\bar B(\bx;\veceta_0)\).

For nonnegative local weights \(w_i(\bx)\) summing to one and a full-sample nuisance fit \(\hat{\veceta}\), we estimate the conditional coefficient means by
\begin{equation}\label{eq:affine_fitted_coefficients}
    \hat B_{\bx}=\sum_{i=1}^n w_i(\bx)B(W_i;\hat{\veceta}),
    \qquad
    \hat b_{\bx}=\sum_{i=1}^n w_i(\bx)b(W_i;\hat{\veceta}).
\end{equation}
The resulting weighted moment equation remains affine in \(\veca\), so solving it gives an explicit estimator and a Jacobian equal to the fitted coefficient matrix,
\begin{equation}\label{eq:affine_exact_inverse}
    \hat{\vecalpha}(\bx)=-\hat B_{\bx}^{-1}\hat b_{\bx},
    \qquad \hat\Gamma_{\bx}=\hat B_{\bx}.
\end{equation}
A fixed measurable convention defines the estimator on the event that \(\hat B_{\bx}\) is singular; this event has probability tending to zero under our assumptions.
With \(\mathbb P_{n,w}f=\sum_iw_i(\bx)f(W_i)\) and \(\vecalpha_0=\vecalpha_0(\bx)\), the same affine structure yields, on the nonsingular event, the exact error representation
\begin{equation}\label{eq:affine_exact_error}
    \hat{\vecalpha}(\bx)-\vecalpha_0
    =-\hat B_{\bx}^{-1}
    \mathbb P_{n,w}\vecpsi(\vecalpha_0,\hat{\veceta}).
\end{equation}
The inverse Jacobian translates score variation into uncertainty about the parameter of interest.
To establish inference, we control the weighted difference between fitted and oracle scores relative to the local standard error and show that \(\hat B_{\bx}\) consistently estimates \(\Gamma_{\bx}\).

\subsection{A treatment-effect question}

For treatment-effect estimation, take \(W=(X,Z,T,Y)\), with binary treatment \(T\), outcome \(Y\), reporting covariates \(X\in\R^{k_x}\) and remaining adjustment variables \(Z\), and write \(V=(X,Z)\) for the full adjustment vector.
We use the potential-outcome framework \citep[Sections~1.1 and~1.3]{rosenbaum_central_1983}, and let \(Y(0)\) and \(Y(1)\) denote the outcomes a unit would realize without and with treatment, respectively.
We assume that one unit's treatment does not affect another's outcome, with consistency linking the observed outcome to the potential outcome under the treatment actually received:
\[
    Y=TY(1)+(1-T)Y(0).
\]
At a given reporting profile, our parameter of interest is the mean difference between these two potential outcomes, giving the conditional average treatment effect (CATE)
\[
    \tau_0(\bx)=\Ex*{Y(1)-Y(0)\given X=\bx}.
\]
We assume independent and identically distributed observations and unconfounded treatment given the full adjustment vector \(V\), meaning that treatment assignment is conditionally independent of the pair of potential outcomes:
\[
    (Y(0),Y(1))\indep T\mid V.
\]
The nuisance functions are the treatment propensity and the two outcome regressions given the full adjustment vector \(V\):
\[
    \pi_0(v)=\Pb*{T=1\given V=v},\qquad
    \mu_{0,t}(v)=\Ex*{Y\given T=t,V=v},\quad t=0,1.
\]
We impose overlap, \(\underline\pi\le\pi_0(v)\le1-\underline\pi\), for some \(\underline\pi>0\), on the adjustment support.
Together, these assumptions identify the treatment comparison after conditioning on all of \(V\).
The CATE averages that comparison conditional only on \(X\).

We estimate this conditional average using an augmented inverse-probability-weighted signal that combines the outcome regressions' adjusted treatment comparison with weighted residual corrections, defined for \(\veceta=(\mu_0,\mu_1,\pi)\) by
\begin{equation}\label{eq:affine_cate_signal}
    \phi(W;\veceta)
        ={}  \mu_1(V)-\mu_0(V)
        +\frac{T}{\pi(V)}\{Y-\mu_1(V)\} -\frac{1-T}{1-\pi(V)}\{Y-\mu_0(V)\}.
\end{equation}
At the true nuisance functions, this construction gives \(\Ex*{\phi(W;\veceta_0)\given X=u}=\tau_0(u)\), so averaging the signal locally around \(\bx\) estimates the effect at the reporting profile.
Writing the score as \(\psi(W;a,\veceta)=a-\phi(W;\veceta)\) places CATE in our affine framework with \(B=1\), \(b=-\phi\) and Jacobian \(\Gamma_{\bx}=1\), so the estimator reduces to the weighted signal average
\begin{equation}\label{eq:affine_cate_estimator}
    \hat\tau(\bx)=\sum_{i=1}^n w_i(\bx)\phi(W_i;\hat{\veceta}).
\end{equation}

\section{DNN Localization and Tuning}\label{sec:affine_dnn}

\subsection{Exact nearest-neighbor weights}

We use distributional nearest neighbors (DNN) to supply the local weights in the estimating equation.
We measure proximity to \(\bx\) by Euclidean distance in the reporting coordinates, which can be standardized to put them on comparable scales.
For subsample size \(s\), which we call the DNN order, DNN averages the nearest observation's score over all such subsamples.
An observation's weight is therefore the fraction of subsamples in which it is closest to \(\bx\).
The resulting bagged nearest-neighbor construction is of the kind studied by \citet{steele_exact_2009,biau_rate_2010}, with regression distribution theory developed by \citet{demirkaya_optimal_2024}.
Assume that \(\norm*{X-\bx}_2\) has a continuous distribution and order observations by this distance.
An observation of rank \(r\) is nearest within a subsample exactly when it is included and the remaining \(s-1\) members are selected from the \(n-r\) more distant observations.
Counting these subsamples gives the exact DNN rank weights:
\begin{equation}\label{eq:affine_dnn_weights}
    w_{s,(r)}(\bx)=
    \begin{cases}
        \displaystyle\frac{\binom{n-r}{s-1}}{\binom ns},
           & 1\le r\le n-s+1, \\[0.4em]
        0, & r>n-s+1.
    \end{cases}
\end{equation}
These weights sum to one and can be evaluated by a distance sort, which allows us to compute a complete-subsample average without enumerating its constituent subsamples.
In implementation, we handle exact distance ties by averaging the assigned rank weights in a tied block, including zero rank weights.
This, however, is an implementation convention motivated by practical considerations and simulation evidence; our theory assumes continuous distances and does not cover ties occurring with positive probability.

The role of \(s\) differs from that of the neighbor count in ordinary \(k\)-nearest-neighbor averaging: \(s\) counts the observations competing to be nearest to \(\bx\), so increasing it concentrates weight on closer observations.
At the extremes, \(s=1\) gives the global mean and \(s=n\) gives the full-sample nearest neighbor.
We summarize the resulting weight concentration by the local effective sample size,
\begin{equation}\label{eq:affine_effective_size}
    \neffx=\left\{\sum_{i=1}^n w_i(\bx)^2\right\}^{-1}.
\end{equation}
With equal weights, \eqref{eq:affine_effective_size} recovers the number of observations averaged; concentrating the weights reduces the effective size.
For a deterministic sequence \(s=s_n\to\infty\) satisfying \(s_n/n\to0\), the leading rank weights become approximately geometric:
\[
    w_{s,(r)}(\bx)
    =\frac{s}{n}\left(1-\frac{s}{n}\right)^{r-1}\{1+o(1)\},
    \qquad r\le M n/s,
\]
uniformly for each fixed \(M\).
This approximation gives
\begin{equation}\label{eq:affine_dnn_scales}
    \sum_iw_i(\bx)^2\sim\frac{s}{2n},
    \qquad \neffx\sim\frac{2n}{s}.
\end{equation}
With distinct distances, the rank-weight values depend only on \(n\) and \(s\), so effective size is constant across reporting points even though the observations receiving those weights and their distances can change.
DNN therefore behaves like a smoother with a location-dependent bandwidth: the same rank weights cover shorter distances in dense regions and longer distances in sparse regions.
Under local small-ball bounds \(c r^{k_x}\le\Pb*{\norm*{X-\bx}_2\le r}\le C r^{k_x}\), this rank concentration corresponds to a localization scale \(h_s=s^{-1/k_x}\), which is also the probabilistic order of a radius containing a fixed weight fraction.
Effective sample size therefore sets the scale of the local limit, but equal effective sizes need not imply equal precision because the covariance also depends on treatment overlap, signal dispersion and the conditioning of the identifying coefficient.

\subsection{Localization bias and pointwise inference}

To express the spatial bias induced by averaging over nearby profiles, write \(F_{\bx}(r)=\Pb*{\norm*{X-\bx}_2\le r}\), which gives the population weight corresponding to a subsample winner as
\begin{equation}\label{eq:affine_population_localizer}
    \tilde\tau_{s,\bx}(u)
    =s\{1-F_{\bx}(\norm*{u-\bx}_2)\}^{s-1},
    \qquad \Ex*{\tilde\tau_{s,\bx}(X)}=1.
\end{equation}
For a fixed response function \(f\), exchangeability then relates the weighted sample average to its population counterpart through
\[
    \Ex*{\mathbb P_{n,w}f}=\Ex*{\tilde\tau_{s,\bx}(X)f(W)}.
\]
Writing \(P_{s,\bx}f\) for this population average, we compare it with the conditional mean at \(\bx\) to obtain the localization bias, which for CATE takes the form
\[
    P_{s,\bx}\phi(\veceta_0)-\tau_0(\bx)
    =\Ex*{\tilde\tau_{s,\bx}(X)
        \{\tau_0(X)-\tau_0(\bx)\}}.
\]
When the relevant conditional moment varies by at most \(C\norm*{u-\bx}_2^\beta\) near \(\bx\), for \(\beta\in(0,1]\), and has integrable tails, localization gives a bias bound of order \(s^{-\beta/k_x}\).
For inference on the parameter at \(\bx\), we require this bound to vanish on the local standard-error scale, giving the undersmoothing condition
\begin{equation}\label{eq:affine_undersmoothing}
    \sqrt{\neffx}\,s^{-\beta/k_x}\longrightarrow0,
    \quad\text{equivalently}\quad
    \sqrt{n/s}\,s^{-\beta/k_x}\longrightarrow0.
\end{equation}
Since \(h_s=s^{-1/k_x}\) and \(\neffx\asymp nh_s^{k_x}\), this condition takes the familiar kernel-smoothing form \(\sqrt{nh_s^{k_x}}\,h_s^\beta\to0\), requiring smoothing bias to be negligible relative to the local standard error.
Increasing \(s\) reduces the bias bound by concentrating weight closer to \(\bx\), but also lowers \(\neffx\).
A choice aimed at estimation accuracy therefore need not make spatial bias negligible for inference.

\paragraph{Tuning and reporting}

Our inference results use a prespecified deterministic sequence \(s_n\), keeping the same order in every jackknife deletion so that deletion changes the observations entering the local equation without retuning the neighborhood.
For Lipschitz conditional moments, a practical choice is \(s=\floor*{\lambda_s n^{(k_x+1)/(k_x+2)}}\).
The bounded reference simulations use \(\lambda_s=1\), with a curvature comparison at 0.4; the SIPP application uses 0.4 as its default.
A fixed nearby grid of \(\lambda_s\) can describe sensitivity to the balance between localization bias and variance, but choosing \(s\) from estimated effects or interval lengths requires a separate inference argument.
For each reporting profile, one can summarize weight concentration by \(\neffx\) and the largest weight, describe the extent of the neighborhood by a physical radius, and assess local identification through the conditioning of \(\hat B_{\bx}\).

\section{Conditions for Local Inference}\label{sec:affine_conditions}

After accounting for localization bias through undersmoothing, inference requires the contribution of nuisance estimation to be negligible on the local standard-error scale determined by the design and oracle score.
The conditions below bound conditional prediction errors, the nuisance mean remainder and the dependence created by full-sample fitting.
Let \(P_X\) be the law of \(X\), \(\calW\) the observation space, and \(\calH\) a class of nuisance functions containing \(\veceta_0\) and the fitted functions encountered below.
We define the conditional coefficient means for each fixed \(\veceta\in\calH\) and equip \(\calH\) with a fixed sigma-field for which the coefficient and conditional-moment maps are jointly measurable in their arguments.
Vector norms are Euclidean, matrix prediction norms are Frobenius, and matrix consistency is in operator norm.
Throughout, constants \(C\) are finite and independent of \(n\), though their values may change between bounds.

\subsection{Local design and oracle moments}

\begin{condition}[Local design]\label{cond:affine_design}
  Observations are i.i.d.; the dimensions \(p,k_x\) are fixed; and \(\bx\) is a fixed interior point of the support of \(X\).
  The distribution of \(\norm{X-\bx}_2\) is continuous.
  There are \(r_0>0\), \(0<c_{\bx}\le C_{\bx}<\infty\), and \(\beta\in(0,1]\) such that \(\calB_0=\{u:\norm{u-\bx}_2\le r_0\}\) lies in that support and
  \[
    c_{\bx}r^{k_x}\le\Pb*{\norm{X-\bx}_2\le r}\le C_{\bx}r^{k_x},
    \qquad 0<r\le r_0.
  \]
  The oracle conditional means have versions satisfying, for every \(u,v\in\calB_0\),
  \[
    \norm*{\bar B(u;\veceta_0)-\bar B(v;\veceta_0)}_F
    +\norm*{\bar b(u;\veceta_0)-\bar b(v;\veceta_0)}
    \le C_{\bx}\norm{u-v}_2^\beta.
  \]
  The matrix \(\Gamma_{\bx}=\bar B(\bx;\veceta_0)\) is nonsingular.
  The DNN subsample size \(s=s_n\) is deterministic, with \(s_n\to\infty\), \(s_n/n\to0\), and \eqref{eq:affine_undersmoothing} holding for this \(\beta\).
\end{condition}

The local identification and regularity statements below follow from Condition~\ref{cond:affine_design}.
Continuity of \(\bar B\) and its nonsingularity at \(\bx\) give a uniformly bounded inverse on \(\calB_0\), after reducing \(r_0\) if necessary.
The conditional moment equation uniquely determines \(\vecalpha_0(u)\).
The uniform inverse bound and H\"older regularity of the conditional coefficient means then imply H\"older regularity of the identified parameter.
Consequently, for some \(C_\alpha<\infty\),
\[
  \begin{gathered}
    \sup_{u\in\calB_0}\opnorm*{\bar B(u;\veceta_0)^{-1}}<\infty,\\
    \vecalpha_0(u)=-\bar B(u;\veceta_0)^{-1}\bar b(u;\veceta_0),
    \quad u\in\calB_0,\\
    \norm*{\vecalpha_0(u)-\vecalpha_0(v)}
    \le C_\alpha\norm{u-v}_2^\beta,\quad u,v\in\calB_0.
  \end{gathered}
\]
These versions define the parameter even at conditioning events of probability zero, and we write \(\vecpsi_0(W)=\vecpsi(W;\vecalpha_0(\bx),\veceta_0)\) for the oracle score evaluated at the reporting point.

\begin{condition}[Oracle moments and covariance]\label{cond:affine_oracle}
  For some \(q\in(2,4]\) and \(M_q<\infty\), the oracle coefficients have finite moments and conditional-moment versions satisfying
  \[
    \Ex*{\norm{B(W;\veceta_0)}_F^q+\norm{b(W;\veceta_0)}^q}  <\infty,\qquad \sup_{u\in\calB_0}\Ex*{\norm{B(W;\veceta_0)}_F^q
                          +\norm{b(W;\veceta_0)}^q\given X=u}         \le M_q.
  \]
  The oracle score covariance \(C_0(u)=\Varb*{\vecpsi_0(W)\given X=u}\) has a version with
  \[
    \lim_{u\to\bx}\opnorm*{C_0(u)-C_0(\bx)}=0,
    \qquad \Sigma_{\bx}=C_0(\bx)\text{ positive definite}.
  \]
\end{condition}

The moment bounds hold, for example, when the oracle coefficients are bounded.
For CATE, overlap and outcome moments give corresponding moment bounds for the signal, while the covariance restriction also depends on the first-stage functions and the conditional law of \(Z\mid X\).
Together with the local design, these restrictions give oracle normality and jackknife covariance consistency, as established in Proposition~\ref{prop:affine_oracle}.

\subsection{Neyman orthogonality and spatial drift}

Estimating the nuisance functions can shift the score's conditional mean and hence the local solution, so we use orthogonality to remove the first-order mean effect at the true conditional parameter while allowing individual fitted scores to change.
A nuisance direction \(h\) is admissible if \(\veceta_0+t h\in\calH\) for every \(t\in[0,\epsilon_h)\), for some \(\epsilon_h>0\).
In the applications, \(\veceta\) and \(h\) are functions of \(V=(X,Z)\), so conditioning on \(X=u\) evaluates the path as \(\veceta_0(u,Z)+t h(u,Z)\).
Admissibility is checked in the full nuisance class \(\calH\); the conditional derivative varies \(t\) while holding \(u\), \(\vecalpha_0(u)\) and the conditional law of the data fixed.
Conditional Neyman orthogonality then requires that, for each such \(h\) and \(P_X\)-almost every \(u\in\calB_0\),
\begin{equation}\label{eq:affine_conditional_orthogonality}
  D_{\veceta}\vecm(u;\vecalpha_0(u),\veceta_0)[h] :=\lim_{t\downarrow0}
    \frac{\vecm(u;\vecalpha_0(u),\veceta_0+t h)
      -\vecm(u;\vecalpha_0(u),\veceta_0)}{t}=0.
\end{equation}
Because orthogonality applies to the complete score, first-order changes in the two coefficient means can cancel, leaving the remainder
\[
  R(u;\veceta)=\{\bar B(u;\veceta)-\bar B(u;\veceta_0)\}
  \vecalpha_0(u)+\bar b(u;\veceta)-\bar b(u;\veceta_0).
\]
A quantitative bound on \(R(u;\veceta)\) controls mean bias at the parameter value \(\vecalpha_0(u)\) for a nearby profile, whereas the estimating equation evaluates scores at \(\vecalpha_0(\bx)\).
Keeping this parameter fixed at the reporting point leaves an additional spatial term,
\begin{equation}\label{eq:affine_mixed_drift}
  \vecm(u;\vecalpha_0(\bx),\veceta)-\vecm(u;\vecalpha_0(\bx),\veceta_0) =R(u;\veceta)+\{\bar B(u;\veceta)-\bar B(u;\veceta_0)\}
    \{\vecalpha_0(\bx)-\vecalpha_0(u)\}.
\end{equation}
The product in \eqref{eq:affine_mixed_drift} links coefficient error to spatial variation in the parameter, so its contribution is controlled jointly by smoothness near \(\bx\) and localization of the weights.

For CATE, \(B=1\) removes the additional spatial term, and the remaining mean bias is a product of propensity and outcome-regression errors.
Writing \(\delta_\pi=\pi-\pi_0\), \(\delta_{\mu,t}=\mu_t-\mu_{0,t}\), \(t\in\{0,1\}\), and conditioning first on \(V\) gives
\begin{equation}\label{eq:affine_cate_bias_identity}
  \Ex*{\phi(W;\veceta)-\phi(W;\veceta_0)\given X=u} =\Ex*{\delta_\pi(V)
               \left\{\frac{\delta_{\mu,1}(V)}{\pi(V)}
               +\frac{\delta_{\mu,0}(V)}{1-\pi(V)}\right\}\given X=u}.
\end{equation}
We use conditional-mean versions satisfying this identity.
If \(\epsilon\le\pi(v)\le1-\epsilon\) for all \(v\), with fixed \(\epsilon\in(0,1/2)\), the absolute mean error is at most
\(C\norm{\delta_\pi}_{u,2}(\norm{\delta_{\mu,1}}_{u,2}+\norm{\delta_{\mu,0}}_{u,2})\),
where \(\norm{f}_{u,2}^2=\Ex*{f(V)^2\given X=u}\) is the conditional \(L^2\) norm of a fixed function under the distribution of \(V\mid X=u\).

\subsection{Quantitative deletion conditions}

Full-sample nuisance fits make score errors dependent on the observations being averaged, so we use deletion stability to control the effect of reusing each observation.
For fixed nuisance functions, we measure changes in the two coefficients by the conditional prediction norms
\begin{align}\label{eq:affine_component_prediction_norms}
  d_{B,u}(\veceta,\veceta')^2
   & =\Ex*{\norm*{B(W;\veceta)-B(W;\veceta')}_F^2\given X=u}, \\
  d_{b,u}(\veceta,\veceta')^2
   & =\Ex*{\norm*{b(W;\veceta)-b(W;\veceta')}^2\given X=u}.
\end{align}
These norms integrate the coefficient differences with respect to the conditional distribution of \(W\) given \(X=u\), holding the fitted function fixed.
They control the score error from replacing \(\veceta_0\) by \(\veceta\), with the parameter held at its true value \(\vecalpha_0(\bx)\).
The affine representation \eqref{eq:affine_score_structure} gives this increment as
\begin{equation}\label{eq:affine_component_score_increment}
  d_{\veceta}(W)=\{B(W;\veceta)-B(W;\veceta_0)\}\vecalpha_0(\bx)
  +b(W;\veceta)-b(W;\veceta_0).
\end{equation}
Its conditional \(L^2\) norm is at most \(\norm{\vecalpha_0(\bx)}d_{B,u}(\veceta,\veceta_0)+d_{b,u}(\veceta,\veceta_0)\), by the triangle inequality.
To center it, fix a regular conditional distribution \(\mathsf P_u\) of \(W\) given \(X=u\) and define the jointly measurable mean \(\mu_{\veceta}(u)=\int d_{\veceta}(w)\,\mathsf P_u(\mathrm{d}w)\).
The required comparison with oracle scores is then
\begin{equation}\label{eq:affine_centered_comparison}
  \sqrt{\neffx}\sum_iw_i\{d_{\hat{\veceta}}(W_i)
  -\mu_{\hat{\veceta}}(X_i)\}=\oP*{1}.
\end{equation}
We establish this centered comparison by replacing the full fit with fits omitting one or two observations.
Write \([n]=\{1,\ldots,n\}\) and \(\hat{\veceta}^{[-I]}\) for the fit after deleting \(I\subset[n]\), \(|I|\le2\).
We use \(\hat{\veceta}^{[-\varnothing]}=\hat{\veceta}\) and \(\hat{\veceta}^{[-I,-j]}=\hat{\veceta}^{[-(I\cup\{j\})]}\), with common learner randomness \(U\) independent of \(W_1,\ldots,W_n\).
Conditioning on \(\calF_{-I}=\sigma\{X_1,\ldots,X_n,U,W_j:j\notin I\}\) fixes the weights and retained fit while preserving the deleted observations' original conditional laws given their own \(X_i\) and conditional independence of their remaining components.
Since each deletion fit uses only retained observations and \(U\), score increments at the omitted observations can be centered at their conditional means, with deletion stability controlling the difference from the full fit used by the estimator.

\begin{condition}[Bounded prediction and deletion]\label{cond:affine_bounded_learner}
  For every \(I\subset[n]\), \(|I|\le2\), the fit \(\hat{\veceta}^{[-I]}\) is measurable with respect to \(\sigma\{U,W_j:j\notin I\}\).
  At each \(n\), all fits use the same deterministic tuning parameters and the same \(U\).
  For boundedness, assume there is a fixed \(M<\infty\) such that, almost surely, for every \(n\) and \(\veceta\in\{\veceta_0\}\cup\{\hat{\veceta}^{[-I]}:|I|\le2\}\),
  \[
    \sup_{w\in\calW}\{\norm{B(w;\veceta)}_F+\norm{b(w;\veceta)}\}\le M.
  \]
  For deletion stability, assume there are events \(E_n\in\sigma\{U,W_1,\ldots,W_n\}\) and deterministic \(\delta_n\ge0\) such that, on \(E_n\),
  \[
    \sup_{w\in\calW}\big\{  \norm*{B(w;\hat{\veceta}^{[-I]})
                                 -B(w;\hat{\veceta}^{[-I,-j]})}_F +\norm*{b(w;\hat{\veceta}^{[-I]})
                                  -b(w;\hat{\veceta}^{[-I,-j]})}\big\}
      \le\delta_n,
  \]
  simultaneously for every \(I\subset[n]\), \(|I|\le1\), and \(j\in[n]\setminus I\).

  To measure local prediction error and mean bias, fix \(L>1\) and \(A_L>0\) and define
  \[
    r_{n,L}             =(A_L\log n/s)^{1/k_x},\qquad
      \calB_{n,L}=\{u:\norm{u-\bx}_2\le r_{n,L}\},\qquad W_{n,\mathrm{out}}  =\sum_iw_i\1*{X_i\notin\calB_{n,L}}.
  \]
  Choose \(A_L\) large enough that \(\Ex{W_{n,\mathrm{out}}}=O(n^{-L})\), as permitted by Condition~\ref{cond:affine_design} and Lemma~\ref{lem:affine_dnn_geometry}; \(\calB_{n,L}\subset\calB_0\) eventually.
  The maps \((u,\veceta)\mapsto d_{B,u}(\veceta,\veceta_0),d_{b,u}(\veceta,\veceta_0)\) have jointly measurable versions continuous in \(u\) on this window for each fixed \(\veceta\in\calH\).
  For all sufficiently large \(n\), define \(\rho_n\) to measure coefficient prediction error over the full and retained fits, and \(t_n\) to measure the full fit's mean remainder:
  \[
    \rho_n  =\max_{|I|\le2}\sup_{u\in\calB_{n,L}}
      \{d_{B,u}(\hat{\veceta}^{[-I]},\veceta_0)
      +d_{b,u}(\hat{\veceta}^{[-I]},\veceta_0)\},\qquad t_n     =\sup_{u\in\calB_{n,L}}\norm*{R(u;\hat{\veceta})}.
  \]
  These suprema are assumed to be measurable random variables, and we impose the following rate restrictions:
  \begin{equation}\label{eq:affine_bounded_learner_rates}
    \Ex{\rho_n^2}            \to0,\qquad \sqrt{\neffx}\,t_n=\oP*{1},\qquad \sqrt{\neffx}\,\delta_n  \to0,\qquad \neffx\Pb{E_n^c}\to0.
  \end{equation}
  Expectations and probabilities here include the training data and \(U\); conditional prediction norms hold the fitted function fixed.
\end{condition}

The nuisance learner assumptions in \eqref{eq:affine_bounded_learner_rates} control both the centered sum in \eqref{eq:affine_centered_comparison} and the weighted mean error \(\sum_iw_i\mu_{\hat{\veceta}}(X_i)\).
The centered part affects sampling variation, while the mean part captures the systematic change in the estimating equation.
Both must be negligible relative to the local standard error.

Prediction accuracy, measured by \(\rho_n\), controls the conditional \(L^2\) score errors through the bounds for \(B\) and \(b\).
To apply these bounds at observation \(i\), the proof uses the fit \(\hat{\veceta}^{[-i]}\).
Conditional on \(X_1,\ldots,X_n,U\), \(W_i\) is independent of the retained data and of \(\hat{\veceta}^{[-i]}\), so the corresponding centered score error has conditional mean zero given \(\calF_{-i}\).
Since \(\neffx\sum_iw_i^2=1\), the sum of the individual variance terms in the scaled centered sum is bounded by \(C\Ex{\rho_n^2}\) inside the local window.
The assumption \(\Ex{\rho_n^2}\to0\) makes this bound vanish, with no additional convergence rate involving \(\neffx\) required for these terms.

The centered score errors from once-deleted fits may still have nonzero covariance because the fit used at observation \(i\) can depend on observation \(j\); the proof controls this covariance by comparing both errors with a common fit omitting \(i\) and \(j\).
Deletion stability bounds these changes and the difference from the full fit, yielding an upper bound for the expected squared norm of the scaled centered sum proportional to
\[
  \Ex{\rho_n^2}+\neffx\delta_n^2
  +\neffx\Pb{E_n^c}+\Ex{W_{n,\mathrm{out}}}.
\]
The first two terms reflect prediction error and the effect of deletion.
Fixed envelopes bound the squared scaled error on \(E_n^c\) by \(C\neffx\), giving the failure-probability term; the last term is the negligible weight outside the prediction window.
The assumed rates make every term vanish, establishing the centered comparison in Lemma~\ref{lem:affine_centered_loo} and Proposition~\ref{prop:affine_bounded_comparisons}.

The mean error requires separate control because it can accumulate under averaging.
Orthogonality removes the first-order nuisance bias at \(\vecalpha_0(u)\), and \(t_n\) bounds the remaining local mean error.
Using the common parameter value \(\vecalpha_0(\bx)\) also introduces the spatial interaction in \eqref{eq:affine_mixed_drift}, multiplying coefficient prediction error by variation in the parameter.
H\"older regularity and DNN localization give this interaction order \(\rho_n s^{-\beta/k_x}\).
Under Conditions~\ref{cond:affine_design} and \ref{cond:affine_bounded_learner}, the mean-error controls are therefore
\begin{equation}\label{eq:affine_nuisance_bias_rates}
  \sqrt{\neffx}\,t_n=\oP*{1},\qquad
  \sqrt{\neffx}\,\rho_n s^{-\beta/k_x}=\oP*{1}.
\end{equation}
The first relation is assumed in \eqref{eq:affine_bounded_learner_rates}; the second follows from prediction consistency and undersmoothing.
Deletion stability transfers the mean comparison to the full fit, as established in Proposition~\ref{prop:affine_bounded_comparisons}.

For CATE, \(B=1\) removes the spatial interaction, so the mean-error restriction depends only on the orthogonal remainder.
To express this restriction in terms of propensity and outcome-regression accuracy, choose conditional norm versions so that the full-fit error radii
\[
  r_{\pi,n}=\sup_{u\in\calB_{n,L}}\norm{\hat\pi-\pi_0}_{u,2},
  \qquad
  r_{\mu,n}=\sup_{u\in\calB_{n,L}}
  \sum_{t=0}^1\norm{\hat\mu_t-\mu_{0,t}}_{u,2}
\]
are measurable.
The product identity \eqref{eq:affine_cate_bias_identity} gives \(t_n\le C r_{\pi,n}r_{\mu,n}\), so the mean error is negligible relative to the local standard error when \(\sqrt{\neffx}r_{\pi,n}r_{\mu,n}=\oP*{1}\).
This recovers the familiar product-rate condition from double/debiased machine learning \citep{chernozhukov_doubledebiased_2018}, with the local effective sample size \(\neffx\) replacing \(n\).
In particular, \(r_{\pi,n}=\oP*{\neffx^{-1/4}}\) and \(r_{\mu,n}=\oP*{\neffx^{-1/4}}\) suffice, although the product condition permits unequal rates.
These errors are measured in conditional \(L^2\) norm uniformly over the shrinking reporting window, so ordinary unconditional prediction rates alone do not establish this requirement.

\subsection{Coefficient consistency, covariance and rates}

The fitted coefficient matrix \(\hat B_{\bx}\) is the Jacobian used to solve the weighted estimating equation, so inference also requires it to consistently estimate the nonsingular matrix \(\Gamma_{\bx}\).
We establish this by comparing the fitted coefficient average with its oracle counterpart and then identifying the oracle average's limit:
\begin{equation}\label{eq:affine_coefficient_bridge}
  \opnorm*{\mathbb P_{n,w}\{B(\hat{\veceta})-B(\veceta_0)\}}=\oP*{1},
  \quad \mathbb P_{n,w}B(\veceta_0)\pto\Gamma_{\bx}.
\end{equation}
The first relation follows from the prediction and deletion bounds for \(B\); the second follows from the oracle moment conditions and localization.
Together, they give \(\hat B_{\bx}\pto\Gamma_{\bx}\), so the fitted inverse exists with probability tending to one and converges to \(\Gamma_{\bx}^{-1}\).
The same controls give the weighted squared error bound needed to compare fitted and oracle score covariances:
\begin{equation}\label{eq:affine_fitted_square}
  \sum_iw_i\norm*{\vecpsi(W_i;\hat{\vecalpha}(\bx),\hat{\veceta})
    -\vecpsi_0(W_i)}^2=\oP*{1}.
\end{equation}
Besides nuisance error at the true parameter, this comparison includes the contribution from parameter estimation,
\[
  B(W_i;\hat{\veceta})\{\hat{\vecalpha}(\bx)-\vecalpha_0(\bx)\}.
\]
The weighted squared norm of this term vanishes because the parameter estimate is consistent and the coefficients are bounded.
Using \eqref{eq:affine_fitted_square} and the jackknife inequality in Lemma~\ref{lem:affine_frozen_quadratic}, we show that the feasible score covariance estimator is consistent.
For CATE, this parameter-estimation term is constant across observations, so centering pseudo-values cancels the common parameter value exactly.

Table~\ref{tab:affine_requirements} expresses the sufficient conditions as polynomial-rate restrictions when \(s\asymp n^\gamma\), with \(\gamma\in(0,1)\), and \(r_{\pi,n}=\OP*{n^{-a}}\), \(r_{\mu,n}=\OP*{n^{-b}}\), \(\delta_n=O(n^{-d})\), and \(\Pb{E_n^c}=O(n^{-c})\).
The restrictions follow by multiplying each bias or stability bound by \(n^{(1-\gamma)/2}\), and the failure probability by \(n^{1-\gamma}\).
Prediction accuracy for the full fit and every retained fit must additionally satisfy \(\Ex{\rho_n^2}\to0\).
Equality at an exponent boundary generally requires an additional little-o factor.
Under the default \(\gamma=(k_x+1)/(k_x+2)\), the localization restriction requires \(\beta>k_x/\{2(k_x+1)\}\), which for \(k_x=2\) means \(\beta>1/3\).
Stronger localization lowers the effective sample size and raises the local standard error.
The parameter estimate then converges more slowly, so a given nuisance product error is smaller relative to the local standard error and the required nuisance rates can be weaker.

\begin{table}[tbp]
  \centering\small
  \begin{tabular}{@{}>{\raggedright\arraybackslash}p{0.37\linewidth}>{\raggedright\arraybackslash}p{0.53\linewidth}@{}}
    \hline
    Quantity, with \(s\asymp n^\gamma\) & Sufficient restriction          \\
    \hline
    Localization bias, H\"older exponent \(\beta\)
                                        & \(\gamma>k_x/(k_x+2\beta)\)     \\[0.4em]
    CATE product error \(n^{-a-b}\)
                                        & \(a+b>(1-\gamma)/2\), \(a,b>0\) \\[0.4em]
    Stability bound \(n^{-d}\)
                                        & \(d>(1-\gamma)/2\)              \\[0.4em]
    Failure probability \(O(n^{-c})\)
                                        & \(c>1-\gamma\)                  \\
    \hline
  \end{tabular}
  \caption{Sufficient polynomial-rate restrictions, with \(\neffx\asymp n^{1-\gamma}\).
    Prediction accuracy additionally requires vanishing expected local error.}
  \label{tab:affine_requirements}
\end{table}

\section{Estimation and Inference}\label{sec:affine_results}

\subsection{Frozen-score covariance}

The frozen-score jackknife estimates uncertainty by deleting each observation in turn and recomputing the exact DNN weights at the same order \(s\), while holding the parameter estimate and full-sample nuisance predictions fixed.
Writing \(\hat{\vecpsi}_i=\vecpsi(W_i;\hat{\vecalpha}(\bx),\hat{\veceta})\) for the frozen fitted scores and \(w_i^{[-j]}\) for the weights after deleting observation \(j\), we form the weighted score averages and their jackknife pseudo-values as
\begin{align}
      \hat S        & =\sum_iw_i\hat{\vecpsi}_i,                            &
      \hat S^{[-j]} & =\sum_{i\ne j}w_i^{[-j]}\hat{\vecpsi}_i,
      \label{eq:affine_frozen_scores}                                         \\
      Q_j           & =n\hat S-(n-1)\hat S^{[-j]},                          &
      \bar Q        & =n^{-1}\sum_jQ_j.\label{eq:affine_score_pseudovalues}
\end{align}
On the event that \(\hat B_{\bx}\) is nonsingular, the estimator in \eqref{eq:affine_exact_inverse} solves the full-sample weighted moment equation, giving
\[
      \hat S=\hat B_{\bx}\hat{\vecalpha}(\bx)+\hat b_{\bx}
      =\hat B_{\bx}\bigl(-\hat B_{\bx}^{-1}\hat b_{\bx}\bigr)+\hat b_{\bx}=0.
\]
The deletion averages \(\hat S^{[-j]}\) generally remain nonzero because they use the deletion weights while retaining the full-sample parameter and nuisance estimates.
The centered pseudo-values define the feasible covariance estimators
\begin{equation}\label{eq:affine_feasible_covariance}
      \hat\Sigma_{\bx}  =\frac{\neffx}{n(n-1)}
            \sum_j(Q_j-\bar Q)(Q_j-\bar Q)^\top,\qquad \hat\Omega_{\bx}  =\hat B_{\bx}^{-1}\hat\Sigma_{\bx}\hat B_{\bx}^{-T}.
\end{equation}
The factor \(\neffx\) puts score covariance on the scale of the local limit, and the inverse coefficient matrices translate that variation into uncertainty about the parameter of interest.
Dividing \(\hat\Omega_{\bx}\) by \(\neffx\) then gives the asymptotic covariance estimate for \(\hat{\vecalpha}(\bx)\).
Once distances are sorted, we use prefix and shifted-suffix sums to compute all frozen-score deletions in \(O(n)\) time for fixed \(p\), so including the sort gives \(O(n\log n)\) cost per reporting point, apart from nuisance fitting.

For CATE, freezing \(\hat\phi_i=\phi(W_i;\hat{\veceta})\) gives the full-sample signal average \(\hat H=\sum_iw_i\hat\phi_i\) and deletion averages \(\hat H^{[-j]}=\sum_{i\ne j}w_i^{[-j]}\hat\phi_i\), whose pseudo-values yield
\[
      P_j=n\hat H-(n-1)\hat H^{[-j]},\qquad
      \hat\Sigma_{\bx}=\frac{\neffx}{n(n-1)}\sum_j(P_j-\bar P)^2.
\]
At a common parameter value \(a\), the CATE score pseudo-value is \(a-P_j\), so centering cancels \(a\) exactly because it contributes the same constant to every pseudo-value.

\subsection{Inference under sufficient conditions}

The jackknife covariance estimate is consistent for oracle scores by Proposition~\ref{prop:affine_oracle}, which applies the generalized-\(U\)-statistic theorem of \citet{juergens_jackknife_2025}.
The weighted squared error from nuisance and parameter estimation vanishes by \eqref{eq:affine_fitted_square}, so the jackknife bound in Lemma~\ref{lem:affine_frozen_quadratic} gives consistency of the feasible covariance estimator as well.

\begin{theorem}[Pointwise affine inference]\label{thm:affine_inference}
      Let the parameter of interest at the fixed reporting point and its exact estimator be given by \eqref{eq:affine_target} and \eqref{eq:affine_exact_inverse}.
      Suppose Conditions~\ref{cond:affine_design}, \ref{cond:affine_oracle} and \ref{cond:affine_bounded_learner} hold.
      Then
      \begin{equation}\label{eq:affine_linear_representation}
            \sqrt{\neffx}\{\hat{\vecalpha}(\bx)-\vecalpha_0(\bx)\}
            =-\Gamma_{\bx}^{-1}\sqrt{\neffx}\,\mathbb P_{n,w}\vecpsi_0+\oP*{1}.
      \end{equation}
      Moreover,
      \[
            \sqrt{\neffx}\{\hat{\vecalpha}(\bx)-\vecalpha_0(\bx)\}
            \dto\mathcal N(0,\Omega_{\bx}),\qquad
            \hat\Omega_{\bx}\pto\Omega_{\bx},
      \]
      where \(\Omega_{\bx}=\Gamma_{\bx}^{-1}\Sigma_{\bx}\Gamma_{\bx}^{-T}\).
      Every fixed nonzero \(c\in\R^p\) satisfies
      \[
            \frac{\sqrt{\neffx}\,c^\top
                  \{\hat{\vecalpha}(\bx)-\vecalpha_0(\bx)\}}
            {\{c^\top\hat\Omega_{\bx}c\}^{1/2}}\dto\mathcal N(0,1).
      \]
\end{theorem}

Normality and covariance consistency give the pointwise interval for a fixed nonzero contrast \(c\),
\begin{equation}\label{eq:affine_contrast_interval}
      c^\top\hat{\vecalpha}(\bx)\ \pm\ z_{1-\xi/2}
      \sqrt{\frac{c^\top\hat\Omega_{\bx}c}{\neffx}}.
\end{equation}
For CATE, \(\Gamma_{\bx}=1\), so the interval uses the estimated signal variance directly.
The vector result also gives the usual Wald tests for fixed full-row-rank restrictions at a fixed reporting point.
When a small number of observations carry substantial local weight, asymptotic covariance consistency alone gives little guidance on finite-sample accuracy, so we assess coverage and interval length in Section~\ref{sec:affine_numerical}.

\section{CATE and a Conditional Slope with Growing Series}\label{sec:affine_applications}

We use one bounded model and a growing clipped ridge-series learner to verify the sufficient conditions for two estimators of the same conditional mean treatment effect.
The CATE estimator has score coefficient \(B=1\), while the residualized slope has a coefficient that depends on the nuisance functions and varies across observations.
The examples therefore illustrate inference with both a fixed Jacobian and a Jacobian estimated from nuisance-dependent coefficients.

\subsection{The model and series learner}

The model separates treatment-effect heterogeneity in two reporting coordinates from nonlinear adjustment for four additional covariates.
Fix \(\chi\in\{0,1\}\), let \(X\sim\mathrm{Unif}([0,1]^2)\) and \(Z\sim\mathrm{Unif}([0,1]^4)\) be independent, write \(V=(X,Z)\), and set
\begin{align}
    \pi_0(z)     & =\tfrac12+h(z_4),\qquad h(z_4)=\tfrac18\sin(2\pi z_4),
    \label{eq:affine_series_primitives}                                   \\
    g_\ell(z)    & =\tfrac18\sin(2\pi z_{\ell+1}),\quad \ell=0,1,2,\notag \\
    \mu_{0,0}(v) & =g_0(z)+x_1g_1(z)+x_2g_2(z)+\chi h(z_4),\notag         \\
    \mu_{0,1}(v) & =\mu_{0,0}(v)+\theta_0(x),\qquad
    \theta_0(x)=\tfrac18(x_1+x_2).
    \label{eq:affine_series_means}
\end{align}
Conditional on \(V\), draw \(T\) with success probability \(\pi_0(Z)\), independently of potential-outcome errors \(\varepsilon_0,\varepsilon_1\).
The two errors are mutually independent and independent of \((X,Z)\), each taking values \(\pm1/8\) with equal probability.
With \(Y(t)=\mu_{0,t}(V)+\varepsilon_t\) and \(Y=TY(1)+(1-T)Y(0)\), this construction satisfies unconfoundedness, \(3/8\le\pi_0\le5/8\) and \(\abs{Y}\le7/8\), and gives the marginal CATE \(\tau_0(x)=\theta_0(x)\).
The choices \(\chi=0\) and \(\chi=1\) give models C0 and C1, respectively.
In C1, \(Z_4\) affects both the treatment propensity and baseline outcomes.

We use series regression to estimate the nuisance functions entering the CATE and slope scores.
The treatment propensity and outcome means are approximated by linear combinations of a common dictionary \(p_n(V)\) of basis functions, with coefficients fitted separately by ridge least squares.
The dictionary grows with \(n\), allowing the approximation error to vanish asymptotically.
For this model, we choose cell indicators for the nonlinear dependence on \(Z\), with a separate intercept and two slopes in \(X\) within each cell.
Partition \([0,1]^4\) into \(J_n=M_n^4\) equal cubes \(C_j\), using half-open cells with the outer faces included, and define the dictionary
\[
    q(x)      =\{1,\sqrt{12}(x_1-1/2),\sqrt{12}(x_2-1/2)\}^{\top},\qquad p_n(x,z)  =q(x)\otimes
        \{\sqrt{J_n}\1{z\in C_j}:1\le j\le J_n\}^{\top}.
\]
With three coefficients per cell, the normalized dictionary has dimension \(K_n=3J_n\) and satisfies
\[
    \sup_v\norm{p_n(v)}_2^2\le\zeta_n^2=7J_n,\qquad
    \Ex{p_n(V)p_n(V)^\top}=I_{K_n}.
\]
The linear dependence on \(X\) preserves continuity of the conditional prediction norms after clipping, as required by Condition~\ref{cond:affine_bounded_learner}.
We fix the partition resolution, DNN order and ridge penalty before observing the data:
\begin{equation}\label{eq:affine_series_schedule}
    M_n=\max\{1,\floor{n^{1/12}}\},\qquad
    s_n=\floor{n^{3/4}},\qquad \lambda_n=n^{-1/2}.
\end{equation}
Under this schedule, \(1<s_n<n\) and \(K_n\asymp n^{1/3}\) for sufficiently large \(n\).

For the coefficient regressions, \(D_{ji}\) denotes the observation weight selecting the sample for regression \(j\), and \(R_{ji}\) the response.
For the propensity and arm-specific outcome regressions, set
\[
    (D_{\pi i},R_{\pi i})=(1,T_i),\qquad
    (D_{ti},R_{ti})=(\1{T_i=t},Y_i),\quad t=0,1.
\]
The propensity uses all observations, and each outcome regression uses its own treatment arm.
The slope also needs a pooled outcome regression, using \((D_{mi},R_{mi})=(1,Y_i)\) to fit
\[
    m_0(V)=\Ex*{Y\given V}=\mu_{0,0}(V)+\pi_0(Z)\theta_0(X).
\]
For \(j\in\{\pi,0,1,m\}\) and \(\abs{I}\le2\), the common ridge training rule is
\begin{equation}\label{eq:affine_ridge_fit}
    \hat\beta_j^{[-I]}=\argmin_{\beta\in\R^{K_n}}
    \left\{\frac1n\sum_{i\notin I}D_{ji}
    \{R_{ji}-p_n(V_i)^\top\beta\}^2
    +\lambda_n\norm{\beta}_2^2\right\}.
\end{equation}
Ridge regularization makes the fit unique even when the sample Gram matrix is singular.
To obtain bounded nuisance predictions and score coefficients, we clip the fitted propensity to \([1/4,3/4]\) and the fitted outcome means to \([-1,1]\).
The true functions \(f_{\pi0}=\pi_0\), \(f_{t0}=\mu_{0,t}\) and \(f_{m0}=m_0\) lie in these intervals, so clipping cannot increase prediction error.
We denote the clipped functions by \(\hat f_j^{[-I]}\), using the full-sample versions in the estimator and the retained versions for deletion comparisons; keeping the dictionary, penalty, clipping intervals and \(1/n\) normalizer fixed across these fits ensures that the comparisons isolate the effect of removing observations.

Conditional \(L^2\) norms under \(Z\mid X=u\), with fitted functions held fixed, measure prediction error for full-sample and retained fits.
Taking the maximum over these fits and the supremum over reporting profiles gives the common error radius
\[
    \rho_n=\max_{\abs{I}\le2}\sup_{u\in[0,1]^2}
    \sum_{j\in\{\pi,0,1,m\}}
    \norm{\hat f_j^{[-I]}-f_{j0}}_{u,2}.
\]
For each score below, a fixed multiple of this radius bounds the conditional prediction errors of \(B\) and \(b\) in Condition~\ref{cond:affine_bounded_learner}.
The conditional norm maps have jointly measurable versions continuous in \(u\), so \(\rho_n\) is measurable.

A uniform lower bound on the penalized Gram eigenvalues controls the inverse matrices in the ridge normal equations and hence the sensitivity of the coefficients to deleting observations.
The following proposition combines this deletion bound with prediction-error control.

\begin{proposition}[Retained series predictions]\label{prop:affine_series_predictions}
    For the model and training rule above, there is one event \(E_n\) such that
    \[
        \Pb{E_n^c}\le CK_n\exp(-cn/K_n),
    \]
    and, on \(E_n\), every penalized Gram matrix with at most two observations removed has smallest eigenvalue at least \(\kappa=3/32\), for all sufficiently large \(n\).
    The actual population least-squares projections, including the treatment-arm-weighted outcome projections, have uniformly bounded coefficient norms and residuals, with uniform approximation error \(\epsilon_{K_n}\le CK_n^{-1/4}\).
    Moreover,
    \[
        \Ex{\rho_n^2}
        \le C\{K_n/n+K_n^{-1/2}+\lambda_n^2+\Pb{E_n^c}\}.
    \]
    On \(E_n\), every adjacent retained pair satisfies
    \[
        \sup_v\abs*{\hat f_j^{[-I]}(v)-\hat f_j^{[-I,-k]}(v)}
        \le Ca_n,\qquad \abs{I}\le1,\ k\notin I,
    \]
    where
    \begin{equation}\label{eq:affine_ridge_deletion_scale}
        a_n=\frac{\zeta_n^2}{n\kappa}
        \left(1+\frac{\zeta_n}{\sqrt{\kappa}}\right)
        =O(K_n^{3/2}/n).
    \end{equation}
\end{proposition}

Removing at most two observations changes the Gram matrix by \(O(K_n/n)\), so one full-sample eigenvalue bound controls all the auxiliary fits.
Substituting the specified schedule into the proposition gives the rates used in both applications (\ref{app:affine_series}):
\begin{equation}\label{eq:affine_ridge_rates}
    \begin{aligned}
        \Ex{\rho_n^2}                & =O(n^{-1/6}),     &
        \sqrt{\neffx}\,\Ex{\rho_n^2} & =O(n^{-1/24}),      \\
        \sqrt{\neffx}\,a_n           & =O(n^{-3/8}),     &
        \neffx\Pb{E_n^c}             & \longrightarrow0,   \\
        \sqrt{\neffx}\,s_n^{-1/2}    & =O(n^{-1/4}).
    \end{aligned}
\end{equation}
Both scores below have orthogonal mean remainders bounded by \(C\rho_n^2\).
The bound \(\sqrt{\neffx}\,\Ex{\rho_n^2}\to0\) therefore makes these remainders negligible in probability relative to the local standard error.
The deletion and failure-probability bounds control dependence from nuisance estimation, while the final bound makes spatial bias negligible for the Lipschitz parameter in two reporting coordinates.
Approximation in the four adjustment coordinates determines the slowest scaled rate, \(n^{-1/24}\).

\subsection{CATE inference}

Substituting the true nuisance functions into \eqref{eq:affine_cate_signal} leaves the conditional mean effect and the potential-outcome errors:
\[
    \phi_0(W)=\theta_0(X)
    +\frac{T\varepsilon_1}{\pi_0(Z)}
    -\frac{(1-T)\varepsilon_0}{1-\pi_0(Z)}.
\]
The residual corrections have conditional mean zero, giving mean \(\theta_0(u)\) at \(X=u\) and the positive constant conditional variance
\begin{equation}\label{eq:affine_series_cate_variance}
    \Sigma_{\bx}
    =\frac1{64}\Ex*{\{\pi_0(Z)(1-\pi_0(Z))\}^{-1}}.
\end{equation}
These moments, the bounded signal and the uniform reporting design verify the design and oracle conditions at interior reporting points.
For estimated nuisance functions, the product identity \eqref{eq:affine_cate_bias_identity} bounds the conditional mean error by \(C\rho_n^2\).
Clipping makes the score Lipschitz in the nuisance predictions, so the learner's deletion bound gives a score change of at most \(Ca_n\).
The rates in \eqref{eq:affine_ridge_rates} therefore verify the remaining learner and undersmoothing conditions of Theorem~\ref{thm:affine_inference}, giving the following corollary (\ref{app:affine_series}).

\begin{corollary}[Growing-series CATE inference]\label{cor:affine_series_cate}
    Under \eqref{eq:affine_series_primitives}--\eqref{eq:affine_ridge_fit}, for either fixed \(\chi\in\{0,1\}\), with the specified clipping and any fixed \(\bx\in(0,1)^2\), use the full fitted CATE signal, exact DNN order \(s_n\), and frozen-signal jackknife of Section~\ref{sec:affine_results}.
    Then
    \[
        \sqrt{\neffx}\{\hat\tau(\bx)-\tfrac18(x_1+x_2)\}
        \dto\mathcal N(0,\Sigma_{\bx}),\qquad
        \hat\Sigma_{\bx}\pto\Sigma_{\bx},
    \]
    with \(\Sigma_{\bx}\) given by \eqref{eq:affine_series_cate_variance}.
    The studentized statistic is asymptotically standard normal by Theorem~\ref{thm:affine_inference}, so the pointwise interval \eqref{eq:affine_contrast_interval} has limiting coverage \(1-\xi\).
\end{corollary}

\subsection{A residualized conditional slope}

An alternative estimator of the same conditional mean treatment effect uses a local regression of the outcome residual on the treatment residual, after adjusting both variables for \(V\).
In our model,
\[
    Y=\theta_0(X)T+g(V)+\varepsilon,\qquad g(V)=\mu_{0,0}(V),\qquad \Ex*{\varepsilon\given V,T}=0,\qquad
        \Ex*{\varepsilon^2\given V,T}=1/64.
\]
The nuisance functions are the propensity \(e_0(V)=\pi_0(Z)\) and the pooled outcome mean \(m_0(V)=g(V)+\pi_0(Z)\theta_0(X)\).
Subtracting these conditional means removes \(g(V)\) and gives \(Y-m_0(V)=\theta_0(X)\{T-e_0(V)\}+\varepsilon\), so the residual regression has slope \(\theta_0(X)\).
Following \citet{robinson_root-n-consistent_1988,nie_quasi-oracle_2021}, we replace the conditional means by fitted functions and use the corresponding least-squares normal equation to obtain the affine score
\begin{equation}\label{eq:affine_residual_score}
    \psi(W;a,(e,m))=\{T-e(V)\}^2a
    -\{T-e(V)\}\{Y-m(V)\}.
\end{equation}
Here \(B=(T-e)^2\) and \(b=-(T-e)(Y-m)\), so both coefficients vary across observations and depend on the nuisance predictions.
Solving the DNN-weighted equation with full-sample nuisance predictions gives the local least-squares slope through the origin,
\[
    \hat\theta(\bx)
    =\frac{\sum_iw_i(\bx)\{T_i-\hat e(V_i)\}
        \{Y_i-\hat m(V_i)\}}
    {\sum_iw_i(\bx)\{T_i-\hat e(V_i)\}^2},
    \qquad \hat e=\hat f_\pi,\quad \hat m=\hat f_m.
\]
The denominator is the local average squared treatment residual and also the empirical Jacobian.
It measures the treatment variation available after adjusting for \(V\), which determines how precisely the slope can be estimated.
In this construction, clipping the propensity predictions keeps every squared treatment residual at least \(1/16\), so the denominator is bounded away from zero.

Because the propensity estimate enters the Jacobian as well as the score mean, we need to control its effect on both.
With \(\delta_e=e-e_0\) and \(\delta_m=m-m_0\), a direct expansion gives the conditional mean score increment at a fixed \(a\) and the change in the mean coefficient:
\[
    \Ex*{a\delta_e(V)^2-\delta_e(V)\delta_m(V)\given X=u},\qquad \bar B(u;(e,m))-\bar B(u;(e_0,m_0))
        =\Ex*{\delta_e(V)^2\given X=u}.
\]
The first-order nuisance terms cancel in conditional expectation, leaving only products of estimation errors.
At \(a=\theta_0(u)\), the mean remainder is therefore bounded by \(C\rho_n^2\).
Holding the parameter at \(\theta_0(\bx)\) across nearby profiles \(u\) introduces the additional spatial term \(\{\theta_0(\bx)-\theta_0(u)\}\Ex*{\delta_e^2\given X=u}\).
The prediction and undersmoothing rates in \eqref{eq:affine_ridge_rates} make the remainder and spatial interaction negligible relative to the local standard error.
Clipping also keeps the coefficient maps bounded and Lipschitz in \((e,m)\), so the prediction and deletion bounds in Proposition~\ref{prop:affine_series_predictions} apply to \(B\) and \(b\).
The oracle variance follows directly from the residual regression.
Conditional on \(X=\bx\), the oracle score is \(-(T-e_0(V))\varepsilon\).
The identifying coefficient is the expected squared treatment residual, and the conditional error variance \(1/64\) makes the score variance equal to \(\Gamma_{\bx}/64\):
\[
    \Gamma_{\bx}=\Ex*{\pi_0(Z)(1-\pi_0(Z))}=31/128,
    \qquad \Sigma_{\bx}=\Gamma_{\bx}/64=31/8192.
\]
Dividing by the squared Jacobian gives the slope's asymptotic variance, \(\Sigma_{\bx}/\Gamma_{\bx}^2=2/31\), and the preceding bounds allow us to estimate it with the frozen-score sandwich.

\begin{corollary}[Growing-series varying-slope inference]\label{cor:affine_series_slope}
    Under the same primitives, dictionary, schedule, and clipping as Corollary~\ref{cor:affine_series_cate}, fit \((e,m)\) using regressions \((\pi,m)\) in \eqref{eq:affine_ridge_fit}.
    At any fixed \(\bx\in(0,1)^2\), the residualized estimator and general frozen-score sandwich satisfy
    \[
        \sqrt{\neffx}\{\hat\theta(\bx)-\tfrac18(x_1+x_2)\}
        \dto\mathcal N(0,2/31),\qquad
        \hat\Omega_{\bx}\pto2/31.
    \]
    The corresponding studentized statistic is asymptotically standard normal by Theorem~\ref{thm:affine_inference}, so the pointwise interval \eqref{eq:affine_contrast_interval} has limiting coverage \(1-\xi\).
\end{corollary}

The coefficient's variation across observations also explains why slope estimation enters the frozen-score covariance argument.
The additional score term \(B(W_i;\hat e) \{\hat\theta(\bx)-\theta_0(\bx)\}\) can vary with \(i\), but its weighted squared norm tends to zero because the coefficients are bounded and the slope estimate is consistent.

Both estimators recover the same conditional mean treatment effect here because \(\theta_0(X)\) is constant across adjustment profiles at a given \(X\).
If the effect also varies with \(Z\), the residualized score weights profiles by residual treatment variance, \(e_0(V)\{1-e_0(V)\}\), giving more weight to propensities nearer one-half.
The resulting weighted effect can then differ from marginal CATE.

\begin{remark}[Conditional local average treatment effects]\label{rem:affine_clate}
    Conditional local average treatment effects provide a further example of an affine equation with a varying coefficient.
    Let \(A\) be a binary instrument for binary treatment \(D\), with adjustment covariates \(V=(X,Z)\), and write \(D(a)\) for potential treatment under instrument value \(a\).
    Under consistency, exclusion, conditional instrument independence given \(V\), monotonicity and overlap, the effect among compliers at \(X=\bx\) is
    \[
        \theta_C(\bx)
        =\Ex*{Y(1)-Y(0)\given D(1)>D(0),X=\bx}
        =\frac{\Delta_Y(\bx)}{\Delta_D(\bx)},
    \]
    with \(\Delta_D(\bx)>0\) \citep{imbens_identification_1994,ogburn_doubly_2015_affine}.
    We use \(e(v)=\Pb*{A=1\given V=v}\) for the instrument propensity, with \(\mu_a(v)=\Ex*{Y\given A=a,V=v}\) and \(\nu_a(v)=\Ex*{D\given A=a,V=v}\) denoting the outcome and treatment regressions in arm \(a\), respectively.
    Averaging the regression contrasts over the adjustment distribution at reporting profile \(u\), with suppressed arguments evaluated at \(V\), gives
    \[
        \begin{aligned}
            \Delta_Y(u) & =\Ex*{\mu_1-\mu_0\given X=u}, \\
            \Delta_D(u) & =\Ex*{\nu_1-\nu_0\given X=u}.
        \end{aligned}
    \]

    For \(\veceta=(e,\mu_0,\mu_1,\nu_0,\nu_1)\), augmenting these contrasts with inverse-probability-weighted residual corrections gives the orthogonal score
    \[
        \begin{aligned}
            \psi(W;\theta,\veceta)
            ={} & \theta\set[\big]{\nu_1-\nu_0
                            +\frac{A}{e}(D-\nu_1)
                            -\frac{1-A}{1-e}(D-\nu_0)} \\
                & -\set[\big]{\mu_1-\mu_0
                       +\frac{A}{e}(Y-\mu_1)
                       -\frac{1-A}{1-e}(Y-\mu_0)}.
        \end{aligned}
    \]
    Writing \(\phi_D\) and \(\phi_Y\) for the first and second bracketed signals, respectively, we have \(B=\phi_D\), \(b=-\phi_Y\) and \(\Gamma_{\bx}=\Delta_D(\bx)\), so solving the DNN-weighted equation gives
    \[
        \hat\theta_C(\bx)
        =\frac{\sum_iw_i(\bx)\hat\phi_{Y,i}}
        {\sum_iw_i(\bx)\hat\phi_{D,i}},
    \]
    where the fitted signals use the full-sample nuisance estimates.
    Both contrasts average over the same adjustment distribution given \(X=\bx\), so the ratio weights stratum effects by their complier probabilities.
    The augmented contrasts have product-form nuisance bias, and under the conditions of Theorem~\ref{thm:affine_inference}, full-sample nuisance estimates give pointwise inference with a frozen-score covariance that retains the joint variation of the reduced form and first stage.
\end{remark}

\section{Numerical Evidence}\label{sec:affine_numerical}

\subsection{A bounded experiment with limited local information}

We use the bounded construction in Section~\ref{sec:affine_applications} to examine how nuisance approximation, spatial bias and limited local information affect pointwise inference.
We report effects at \(\bx=(1/2,1/2)\) in two reporting coordinates \(X\), using four additional coordinates \(Z\) for nuisance adjustment.
Since the treatment effect depends only on \(X\), CATE and the residualized slope both equal \(1/8\) at this point.
Starting from the bounded reference setups C0 and C1 introduce shared confounding to examine bias from nuisance approximation, while C2 adds reporting curvature that the smooth dictionary factors can represent.
Two further designs vary the noise and reporting density: C0H has heteroskedastic noise with the same asymptotic variance at the reporting point, and C2S lowers reporting density while preserving positive support and treatment overlap.
The formulas and schedules for these designs appear in \ref{app:affine_production_numerics}.

\begin{figure}[!htb]
  \centering
  \begingroup%
\makeatletter%
\let\Gread@transgrouptrue\Gread@transgroupfalse%
\makeatother%
\includegraphics[width=\linewidth]{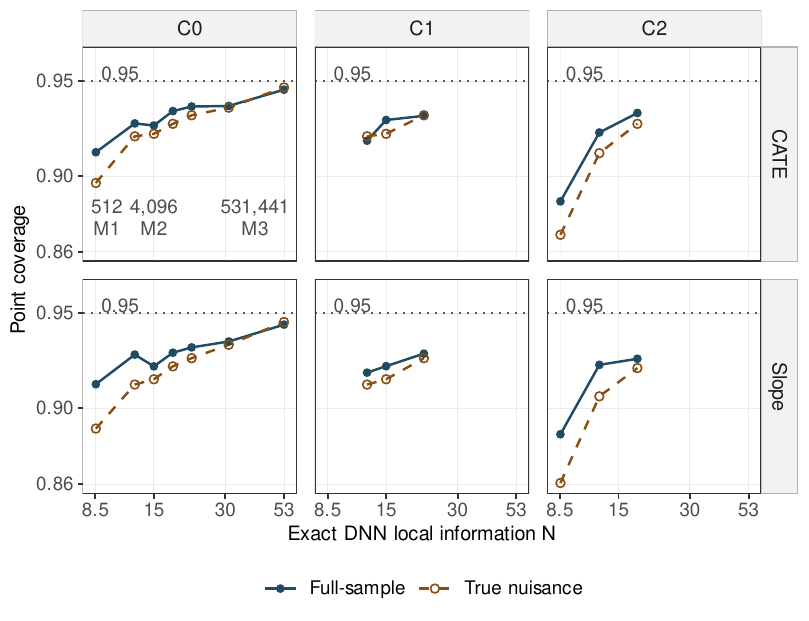}%
\endgroup
  \caption{Point coverage and local information, with 10,000 replications
    per cell. Full-sample and true-nuisance estimates use the same final DNN equation and
    frozen-score jackknife. C0 is the bounded reference, C1 adds shared
    confounding, and C2 adds represented curvature. The order multiplier is
    one. Labels show selected sample sizes and partition resolution \(M\). The
    nominal coverage line is 0.95.}
  \label{fig:affine_simulation}
\end{figure}

To assess the role of nuisance fitting, we compare estimates based on full-sample, pooled five-fold and true-nuisance predictions while keeping the same observations and weights in the final equation.
The fitted versions use the clipped ridge series in \eqref{eq:affine_ridge_fit}, with smooth \(X\) factors, histogram growth only in \(Z\) and penalty \(\lambda_n=n^{-1/2}\).
The five-fold version pools the out-of-fold scores in the local equation described in \ref{app:affine_fold_aggregation} and uses smaller nuisance training samples, so its comparison with full-sample fitting also reflects a change in effective regularization.
The reference simulation order is \(s=\floor{n^{3/4}}\); in C2, we also use \(s=\floor{0.4n^{3/4}}\) to examine broader localization and its effects on bias due to the added curvature.
Under the reference schedule, the local effective sample size \(N=(\sum_iw_i^2)^{-1}=\neffx\) grows only as \(N\asymp n^{1/4}\), so pointwise noise decreases as \(n^{-1/8}\).
This slow growth leaves much less local information than the total sample size might suggest: the effective sample size rises from about 8.5 at \(n=512\) to 31 at \(n=65{,}536\), reaching only 53 at \(n=531{,}441\).
We use this information scale in Figure~\ref{fig:affine_simulation} to report coverage of the parameter at the reporting point (point coverage) for CATE and the slope.

\subsection{Agreement between estimates, spatial bias and interval coverage}

\begin{figure}[!htb]
  \centering
  \begingroup%
\makeatletter%
\let\Gread@transgrouptrue\Gread@transgroupfalse%
\makeatother%
\includegraphics[width=\linewidth]{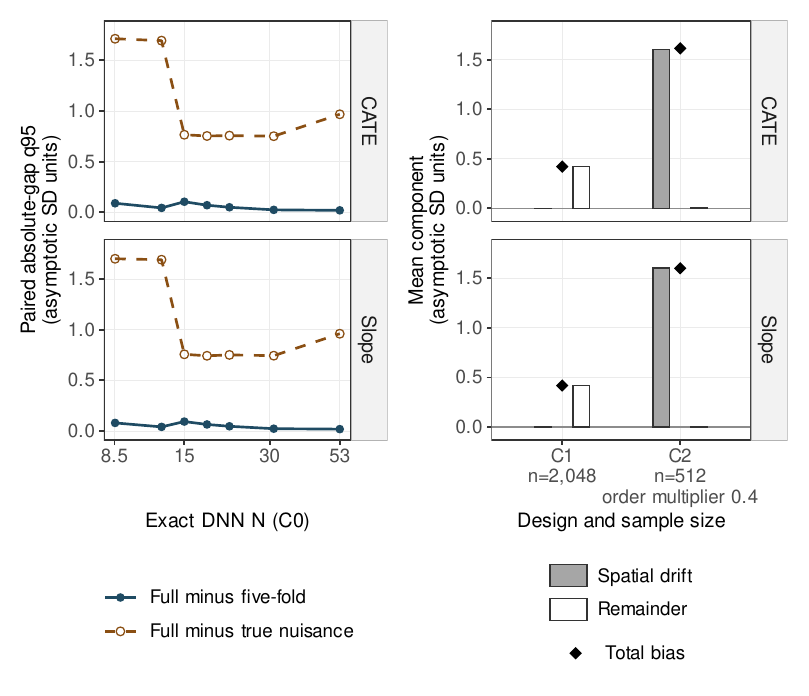}%
\endgroup
  \caption{Differences between estimates and point centering for CATE and the slope.
    Left: the 95th percentile of absolute paired C0 gaps, multiplied by
    \(\sqrt N/\sqrt{\Omega_{\bx}}\), with 10,000 pairs at each size.
    Right: mean realized spatial drift and mean remainder, with total bias
    marked, all scaled by \(\sqrt N/\sqrt{\Omega_{\bx}}\), in C1 at
    \(n=2048\) and C2 at \(n=512\), order multiplier 0.4.
    The remainder includes oracle noise and nuisance effects. CATE and the slope
    use their respective weighted targets.}
  \label{fig:affine_simulation_mechanisms}
\end{figure}

Full-sample and five-fold estimates can agree while sharing nuisance approximation error, so Figure~\ref{fig:affine_simulation_mechanisms} compares their paired differences with those between full-sample and true-nuisance estimates, measured in asymptotic standard-deviation units \(\sqrt N(\hat a_L-\hat a_R)/\sqrt{\Omega_{\bx}}\).
Along this schedule, full-sample estimates are closer to pooled five-fold estimates than to their true-nuisance counterparts.
The C1 results show why this agreement need not imply small bias: with \(M=1\), the dictionary retains the reporting-coordinate factors but leaves the confounding variation in \(Z\) unresolved.
At \(n=2048\), the fitted CATE bias of about 0.030 is close to the population projection's signal bias of 0.031, with negligible mean spatial drift, linking the bias to nuisance approximation.
Dictionary resolution also helps interpret the C0 results, where the early drop in the gap between full-sample and true-nuisance estimates coincides with the first histogram refinement from \(M=1\) to \(M=2\) along the joint schedule.

C2 illustrates a different source of bias that remains even with known nuisance functions: curvature shifts the neighborhood average away from the point target.
To measure this spatial contribution, we form a weighted target in each draw by averaging treatment effects as defined in \eqref{eq:affine_realized_weighted_truth}.
CATE uses the DNN weights, while the slope uses these weights multiplied by squared treatment residuals evaluated at the true propensity and normalized to sum to one.
The decomposition
\[
  \hat a-a_0(\bx)
  =\{a_w-a_0(\bx)\}+\{\hat a-a_w\}
\]
separates realized spatial drift from a remainder containing oracle sampling noise and nuisance effects.

With \(n=512\) and order multiplier \(\lambda_s=0.4\), this spatial component accounts for nearly all the CATE bias: mean drift is about 0.0857, compared with total bias of 0.0863.
The difference between the weighted target and the point parameter has a large effect on coverage: the same intervals cover the point parameter in about 75.4 percent of draws and the weighted target in about 93.9 percent (\AffineExternalRef{supplement}{tab:affine_production_calibration}).
Increasing the multiplier to one reduces spatial drift by concentrating weights nearer the point, at the cost of a smaller effective sample size and an increase in Monte Carlo SD from about 0.072 to 0.115.

The extent of this spatial averaging also depends on reporting density, so a common effective sample size need not imply comparable spatial bias.
We measure the extent by the DNN weighted radius, \(\sum_iw_i(\bx)\norm{X_i-\bx}_2\), the weighted mean distance from the reporting point.
At \(n=2048\), moving from C2 to C2S doubles the mean DNN weighted radius at unchanged effective sample size, while full-sample CATE coverage falls to 87.1 percent, a paired decline of 5.2 percentage points (Monte Carlo SE 0.28 points).

Coverage also depends on how well the normal approximation describes the studentized errors, even when average standard errors appear well calibrated.
For CATE in C0 at \(n=512\), mean estimated SE is close to Monte Carlo SD, yet point coverage is only 91.3 percent.
This agreement on average does not extend to tail behavior: the studentized errors have heavier tails than errors standardized by Monte Carlo SD (\AffineExternalRef{supplement}{app:affine_production_calibration}).
Coverage improves as local information grows, but even at \(n=65{,}536\), full-sample CATE coverage is still 93.7 percent and the local effective sample size is only about 31.

To examine how these findings depend on the choice of local weights, we compare DNN with a Gaussian local-constant kernel using the same nuisance predictions and a deterministic bandwidth matched to DNN's population effective sample size.
Matching this population quantity still allows the realized effective sample sizes, radii and smoothing bias to differ, so differences in coverage can also reflect changes in spatial averaging.
Table~\ref{tab:affine_kernel_comparison} reports coverage and mean interval length.
This change in localization also matters for point-estimation accuracy: relative MSE varies across designs, with lower CATE MSE for the kernel in C2 at \(n=512\), multiplier 0.4.

Holding the local estimate fixed lets us isolate the contribution of covariance estimation to coverage.
For the C0 kernel estimate at \(n=2048\), replacing conventional squared-weight score covariance by the jackknife raises coverage from 89.32 to 92.35 percent, a gain of 3.03 percentage points (Monte Carlo SE 0.17 points), with longer intervals.
The corresponding DNN comparison in C0 at \(n=512\) gives the same direction of change: replacing direct-moment covariance by the jackknife raises CATE coverage and increases mean interval length without changing the estimate.
The DNN coverage difference largely disappears at \(n=65{,}536\), where direct-moment covariance gives nearly identical coverage and slightly shorter intervals (\AffineExternalRef{supplement}{tab:affine_dnn_covariance_routes}).

Coverage continues to improve at the largest C0 sample size, reaching 94.6 and 94.4 percent for CATE and the slope.
This improvement coexists with a persistent nuisance-estimation gap: on the original CATE scale, the 95th percentile of the absolute gap between full-sample and true-nuisance CATE estimates changes little as local uncertainty shrinks.
Comparing the two largest sample sizes combines increasing local information with changes in nuisance approximation because \(n,N,M,K,s\) change together; in particular, moving from two to three bins per adjustment coordinate changes cell boundaries and increases the population projections' mean squared approximation error.
\AffineExternalRef{supplement}{tab:affine_production_endpoint} reports the coverage intervals, risk figures and gap quantiles, together with ratios of estimated and Monte Carlo variance to the asymptotic point variance.


\begin{table}[!htb]
  \centering\small
\begin{tabular}{@{}llrrr@{}}
\hline
Case $(n)$ & App. & \shortstack{DNN\\jackknife} & \shortstack{Kernel\\jackknife} & \shortstack{Kernel\\conventional} \\
\hline
C0 $(2\,048)$ & CATE & 0.9277 (0.372) & 0.9235 (0.372) & 0.8932 (0.331) \\
C0 $(2\,048)$ & Slope & 0.9283 (0.371) & 0.9225 (0.371) & 0.8923 (0.331) \\
C2 $(512)$ & CATE & 0.8866 (0.453) & 0.8809 (0.450) & 0.8297 (0.379) \\
C2 $(512)$ & Slope & 0.8863 (0.451) & 0.8806 (0.447) & 0.8279 (0.377) \\
C0H $(8\,192)$ & CATE & 0.9335 (0.255) & 0.9269 (0.255) & 0.9091 (0.236) \\
C0H $(8\,192)$ & Slope & 0.9282 (0.254) & 0.9226 (0.254) & 0.9052 (0.235) \\
C2S $(2\,048)$ & CATE & 0.8705 (0.376) & 0.8637 (0.374) & 0.8295 (0.334) \\
C2S $(2\,048)$ & Slope & 0.8691 (0.376) & 0.8634 (0.374) & 0.8285 (0.333) \\
\hline
\end{tabular}

  \caption{Coverage (mean interval length) using full-sample nuisance predictions, with 10,000
    replications per cell. DNN and Gaussian kernel jackknives freeze the fitted
    scores. The conventional kernel covariance uses squared normalized weights.
    All intervals use the normal 95 percent reference. The order multiplier is
    one. C0H changes noise variance and C2S lowers reporting density.
    Paired uncertainty and loss contrasts appear in
    \AffineExternalRef{supplement}{tab:affine_production_pairs}.}
  \label{tab:affine_kernel_comparison}
\end{table}

\ifpreprint\else\FloatBarrier\fi
\subsection{401(k) eligibility: nuisance predictions and local information}
\label{sec:affine_sipp}

We illustrate the CATE signal in \eqref{eq:affine_cate_signal} using 401(k) eligibility and net financial assets for 9,915 households from wave 4 of the 1990 SIPP panel, interviewed in 1991 \citep{chernozhukov_hdm_2016,benjamin_401k_2003}.
Eligibility records employer access, including eligible nonparticipants, and the outcome is an accumulated financial stock measured in 1991 dollars; interpreting this contrast causally requires unconfounded eligibility given the adjustment covariates and overlap.
To describe variation across income profiles, we report contrasts at age 40 and the full-sample 25th, 50th and 75th income percentiles, using standardized age and inverse-hyperbolic-sine income as reporting coordinates.
The nuisance regressions use richer income terms and household controls; \ref{app:affine_sipp} gives the adjustment, standardization and computational details.

To examine the role of nuisance prediction, we compare all-tree, single-fit out-of-bag (OOB) and five-fold held-out predictions from regression forests, with propensity estimates clipped away from zero and one.
All-tree predictions use every tree in the fitted forests, whereas OOB predictions evaluate the propensity regression and the outcome regression for an observation's treatment group at its covariates using only trees trained without that observation.
All three methods share the final weights at the default order \(s=397\), with \(N\approx49\), so their comparison holds localization fixed.
The OOB estimates are about \$2,800 and \$5,900 at the lower and median incomes and about \(-\$3,400\) at the upper income, where the SE is \$18,400.
Switching to all-tree predictions moves these estimates modestly and gives SEs about 11--15 percent smaller (\AffineExternalRef{supplement}{tab:affine_sipp} and \AffineExternalRef{supplement}{tab:affine_sipp_pairs}).
Even with the smaller all-tree SEs, all nine default intervals from the frozen-score covariance include zero.

\begin{figure}[!htb]
  \centering
  \begingroup%
\makeatletter%
\let\Gread@transgrouptrue\Gread@transgroupfalse%
\makeatother%
\includegraphics[width=.94\textwidth]{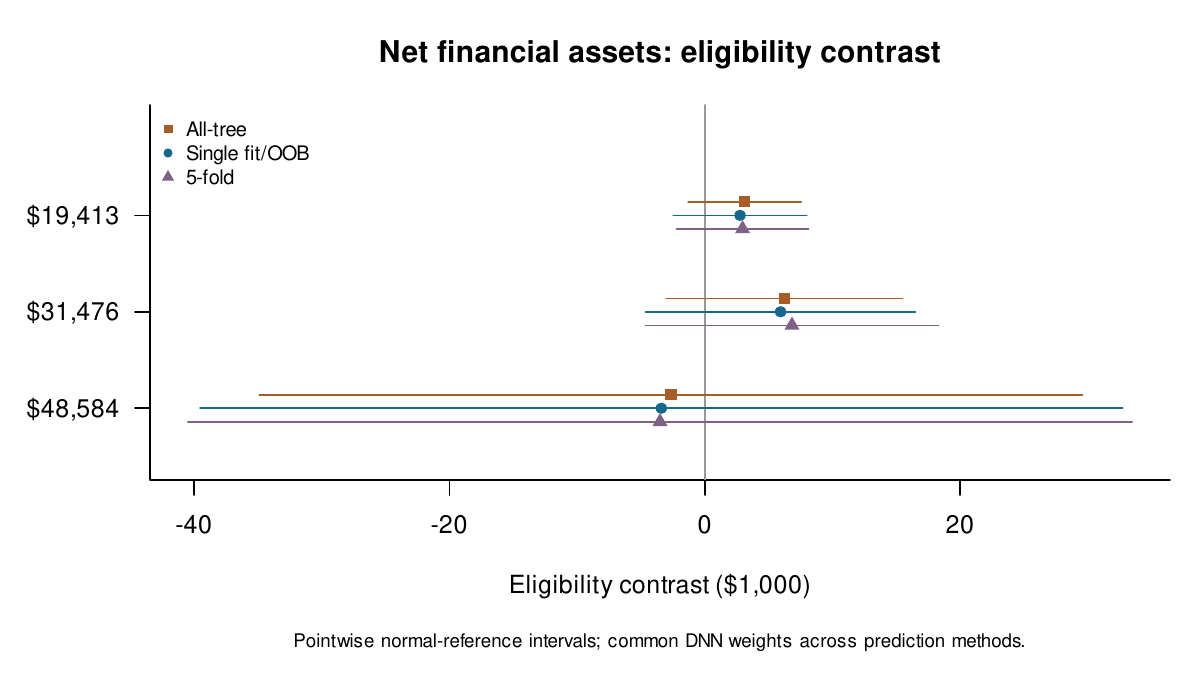}%
\endgroup
  \caption{Eligibility contrasts in net financial assets at three age-40
    income profiles. All-tree, single-fit/OOB and five-fold nuisance predictions
    share the final DNN weights, with \(s=397\). Bars are pointwise
    normal-reference 95-percent intervals from frozen-score covariance, in
    thousands of 1991 dollars.}
  \label{fig:affine_sipp_intervals}
\end{figure}

\begin{figure}[!htb]
  \centering
  \begingroup%
\makeatletter%
\let\Gread@transgrouptrue\Gread@transgroupfalse%
\makeatother%
\includegraphics[width=.94\textwidth]{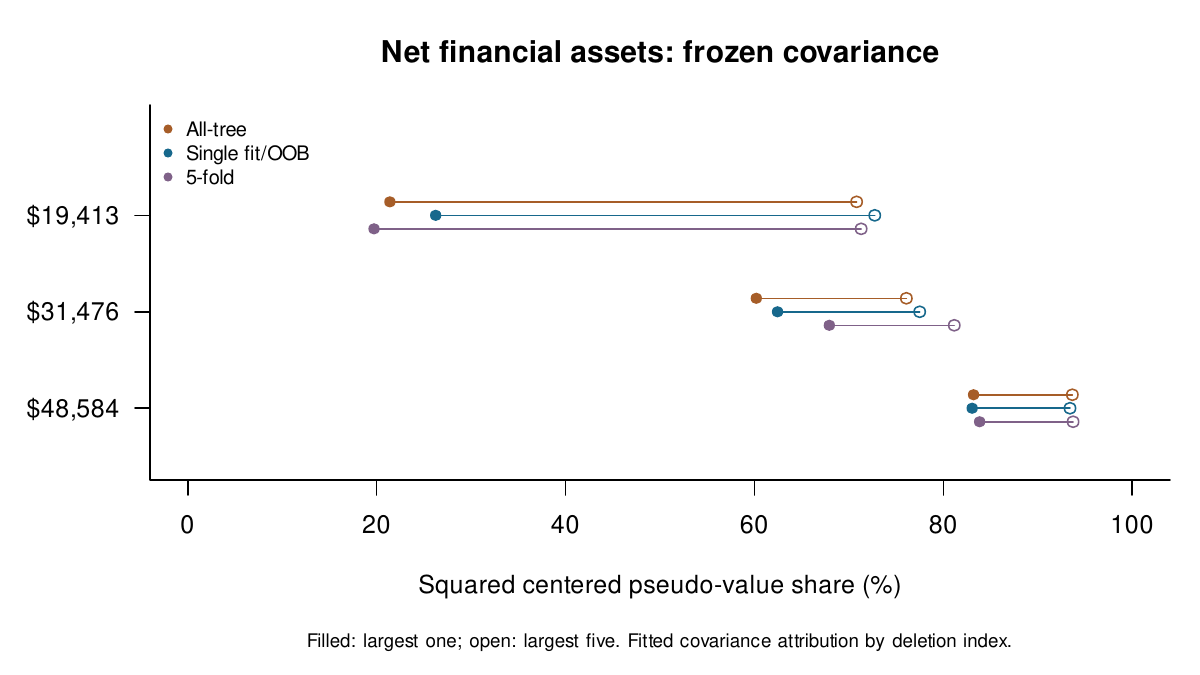}%
\endgroup
  \caption{Concentration of the frozen-score jackknife covariance at the
    same profiles and nuisance prediction methods. Filled markers show the largest share
    of the sum of squared centered pseudo-values. Open markers show the combined
    share of the largest five.
    Shares attribute estimated variance to deletion indices, including
    the rank changes caused by deleting a household.}
  \label{fig:affine_sipp_concentration}
\end{figure}

The upper profile's large uncertainty is accompanied by substantial variance concentration, which the local effective sample size alone does not describe.
At nearly identical \(N\), the largest OOB jackknife contribution accounts for 26, 62 and 83 percent of fitted variance at the lower, median and upper incomes (Figure~\ref{fig:affine_sipp_concentration}).
Each contribution is indexed by the deleted household and includes the rank changes among retained observations.
At the upper profile, one ineligible household combines the largest local weight with a large control residual and dominates fitted variance under all three nuisance prediction methods.
Deleting this household while holding fitted signals fixed and recomputing retained ranks moves the OOB estimate from about \(-\$3,400\) to \$13,400, illustrating the sensitivity to a single observation and warranting caution with normal-reference intervals.
This concentration motivates examining broader neighborhoods, which affect uncertainty at the upper profile more than switching between the tested nuisance prediction methods.
With fitted signals fixed, reducing \(s\) from 397 to 99 lowers the nearest household's weight \(s/n\) from 4.0 to 1.0 percent.
The upper OOB SE falls from about \$18,400 to \$5,700 and its largest covariance share from 83 to 52 percent, while the radius containing 90 percent of the DNN weight roughly doubles.
The reduction in uncertainty therefore comes with averaging over a wider range of age and income.

Broader pooling affects variance concentration differently across profiles: at the lower income, SE falls while the largest share rises from 26 to 42 percent (Table~\ref{tab:affine_sipp_localization_core}).
Thus lower standard errors can coexist with a greater relative contribution from an individual household deletion.
Some intervals are positive at exploratory smaller orders (\AffineExternalRef{supplement}{supp:affine_sipp}), but these results also involve broader pooling and potentially greater spatial bias.

We interpret the intervals as illustrative normal-reference comparisons; \ref{app:affine_sipp} explains the departures from the conditions of Theorem~\ref{thm:affine_inference}.
\FloatBarrier

\section{Conclusion}\label{sec:affine_conclusion}

When a parameter is reported along only a few characteristics, the nuisance functions may still have to account for a much richer set of covariates.
The observations that help estimate these functions need not be confined to the neighborhood used for the conditional comparison.
Cross-fitting separates nuisance training from score evaluation, but does so by fitting each nuisance model on fewer observations, even when all observations remain in the final local equation.
Keeping the full training sample is especially worth considering when the adjustment is rich and the local comparison already relies on relatively few observations.

Whether this reuse supports inference depends on prediction accuracy near the reporting point and on how much each observation influences the prediction used in its own score.
This is a local requirement on a learner trained with the full sample: its prediction errors must be controlled in the region relevant for the conditional parameter.
A good average fit across the covariate distribution need not establish this if predictions remain poor near the reporting point.

Under our conditions for affine DNN equations, orthogonality and quantitative deletion stability ensure that these nuisance effects are negligible relative to the local standard error.
The full-sample estimator then has the same first-order distribution as if the nuisance functions were known, even though those functions are fitted on the observations used in the local equation.
This remains possible when nuisance predictions enter the identifying coefficient, because the conditions control the estimated Jacobian along with the score.
The same fitted functions can then be used for the local estimate and its frozen-score covariance calculation, with no nuisance refits.

Keeping the full training sample leaves the locality of the final comparison unchanged.
The neighborhood still has to balance spatial bias against the variation available for estimating the parameter, so a large sample for nuisance adjustment can coexist with limited precision at a reporting point.

The simulations reflect this distinction: full-sample and cross-fitted inference can perform similarly while both remain affected by nuisance approximation and spatial bias.
Comparisons using the true nuisance functions help distinguish errors in the learned adjustment from spatial bias that persists even with the adjustment known.
In the curvature design, the CATE intervals cover the weighted target much more often than the point parameter, illustrating the cost of using nearby profiles to learn about the reporting point.
In the same design, reducing spatial bias by concentrating weight closer to the reporting point also raises sampling uncertainty.
The resulting trade-off concerns how much information can be borrowed from nearby profiles while still learning about the point parameter, and persists even when the nuisance adjustment is known.
Agreement between training methods therefore tells us less about coverage than about how much that particular change in nuisance fitting matters.
Avoiding cross-fitting keeps more observations available to learn the adjustment; the precision of the conditional estimate is still determined by the local information in the data.

\clearpage
\appendix
\section{Technical Lemmas}\label{app:affine_technical}

\subsection{Notation and conditional evaluation}

Fix the reporting point \(\bx\) throughout and write \(a_0=\vecalpha_0(\bx)\), \(N_n=\neffx\), \(B_0(W)=B(W;\veceta_0)\), \(b_0(W)=b(W;\veceta_0)\), and \(\vecpsi_0=B_0a_0+b_0\).
The notation \(a_0\) denotes a vector even when the score is scalar in an example.
For an array \(z=(z_1,\ldots,z_n)\), put \(A(z)=\sum_iw_i z_i\), while for a deterministic function \(f\), write \(A(f)=\mathbb P_{n,w}f\) and \(P_sf=P_{s,\bx}f\).
All vector norms are Euclidean, matrix prediction norms are Frobenius, and matrix consistency is in operator norm.
Constants denoted by \(C\) may change between displays but do not depend on \(n\).

Let \(U\) be any algorithmic random seed, independent of the observations and shared across retained fits.
The corresponding conditioning field is
\begin{equation}\label{eq:affine_appendix_fields}
    \calF_{-I}=\sigma\{X_1,\ldots,X_n,U,(W_k:k\notin I)\}.
\end{equation}
The learner \(\hat{\veceta}^{[-I]}\) uses only the retained observations and \(U\), whereas the field in \eqref{eq:affine_appendix_fields} also reveals the removed observations' reporting covariates to make the DNN weights measurable.
Conditional on \(X_1,\ldots,X_n,U\), the removed observations are mutually independent and independent of the retained observations.
Their conditional distributions given \(\calF_{-I}\) therefore equal their respective conditional distributions given \(X_i\).

Fix a regular conditional distribution \(\mathsf P_u\) of \(W\) given \(X=u\).
For a fixed nuisance parameter \(\veceta\), define
\[
d_{\veceta}(W)=\vecpsi(W;a_0,\veceta)-\vecpsi_0(W),\qquad \mu_{\veceta}(u)=\int d_{\veceta}(w)\,\mathsf P_u(\mathrm{d}w),\qquad \xi_{\veceta}(W)=d_{\veceta}(W)-\mu_{\veceta}(X).
\]
These maps are jointly measurable.

\subsection{Weight geometry and localization}

\begin{lemma}[Exact weights and local averages]\label{lem:affine_dnn_geometry}
Suppose the radial distribution is continuous, the local small-ball bounds hold, and \(s\to\infty\), \(s/n\to0\).
Then
\[
    \max_iw_i=s/n,\qquad
    N_n\sim2n/s,\qquad N_n\max_iw_i\to2.
\]
For every integrable deterministic response for which the following expectations exist,
\begin{equation}\label{eq:affine_appendix_population_identity}
    \Ex*{A(f)}=P_sf
       =s\Ex*{(1-F_{\bx}(\norm*{X-\bx}_2))^{s-1}f(W)}.
\end{equation}
Choose a sufficiently small fixed ball \(A_{\bx}=\{u:\norm*{u-\bx}_2\le r_0\}\) with \(a=\Pb*{X\in A_{\bx}}\in(0,1)\).
For each fixed \(t>0\),
\begin{align}
    P_s[\norm*{X-\bx}_2^t\1*{X\in A_{\bx}}]
       &=O(s^{-t/k_x}),\label{eq:affine_appendix_radius_moment}\\
    P_s[|f|\1*{X\notin A_{\bx}}]
       &\le s(1-a)^{s-1}\Ex*{|f|\1*{X\notin A_{\bx}}}.
       \label{eq:affine_appendix_fixed_tail}
\end{align}
For \(r_{n,L}=(A_L\log n/s)^{1/k_x}\to0\), put \(W_{n,\mathrm{out}}=\sum_iw_i\1*{\norm*{X_i-\bx}_2>r_{n,L}}\).
Then
\begin{equation}\label{eq:affine_appendix_window_tail}
    \Ex*{W_{n,\mathrm{out}}}\le n^{-cA_L}.
\end{equation}
The smallest radius containing any fixed fraction in \((0,1)\) of the realized weight is of order \(s^{-1/k_x}\) in probability.
\end{lemma}

\begin{proof}
Let \(v_r=\binom{n-r}{s-1}/\binom ns\) be the rank weights.
Their ratios satisfy
\[
    v_1=s/n,\qquad v_{r+1}/v_r=1-(s-1)/(n-r).
\]
Writing \(q_n=s/n\), the product of these ratios gives \(v_r=q_ne^{-q_n(r-1)}\{1+o(1)\}\), uniformly for \(r\le M/q_n\), for each fixed \(M\).
Indeed, the sum of squared logarithmic increments is \(O(q_n)\), and replacing \(n-r\) by \(n\) has total error \(O(1/s)\) on this range.
The bound \(v_r\le q_n e^{-(s-1)(r-1)/n}\) controls its complement.
Thus
\[
    q_n^{-1}\sum_r v_r^2
       \to\int_0^\infty e^{-2t}\dd t=1/2.
\]
This proves the effective-size and maximum-weight statements, as well as the geometric approximation in the main text.

The weighted average is the complete average of the response selected by the nearest observation in a size-\(s\) subsample.
Conditional on its covariates, observation 1 wins with probability \((1-F_{\bx}(\norm*{X_1-\bx}_2))^{s-1}\).
Symmetry proves \eqref{eq:affine_appendix_population_identity}, and taking \(f=1\) gives normalization.
The radial probability transform is uniform, so the population mass outside radius \(r\) equals \((1-F_{\bx}(r))^s\) and is at most \(e^{-csr^{k_x}}\) by the small-ball lower bound.
Integration of this survival bound on \([0,r_0]\) proves \eqref{eq:affine_appendix_radius_moment}.
The possible boundary contribution is exponentially small.
On the complement of \(A_{\bx}\), the selector is bounded by \((1-a)^{s-1}\), which proves \eqref{eq:affine_appendix_fixed_tail}.
Applying the survival bound at \(r_{n,L}\) proves \eqref{eq:affine_appendix_window_tail}.

For the realized radius, the exact rank-tail identity is
\[
    \sum_{r>m}v_r=\binom{n-m}{s}/\binom ns.
\]
It places the cutoff for any fixed weight fraction at \(m\asymp n/s\).
Counts in balls of radii \(c_1s^{-1/k_x}\) and \(c_2s^{-1/k_x}\) have means of order \(n/s\), which diverges, and variances no larger than their means.
With \(c_1\) chosen small and \(c_2\) large using the two small-ball bounds, Chebyshev's inequality places the corresponding order statistic between these radii with probability tending to one.
\end{proof}

\subsection{Removing training dependence}

\begin{lemma}[Centered score comparison]\label{lem:affine_centered_loo}
Write \(d_i=d_{\hat{\veceta}^{[-i]}}\), \(\xi_i=\xi_{\hat{\veceta}^{[-i]}}(W_i)\), and, for \(i\ne j\), \(\xi_i^{(j)}=\xi_{\hat{\veceta}^{[-i,-j]}}(W_i)\).
Let \(r_i^2=\Ex*{\norm*{d_i(W_i)}^2\given\calF_{-i}}\).
Suppose nonnegative \(\calF_{-\{i,j\}}\)-measurable radii \(b_{ij}\) satisfy
\begin{equation}\label{eq:affine_appendix_pair_radius}
    \Ex*{\norm*{d_{\hat{\veceta}^{[-i]}}(W_i)
                 -d_{\hat{\veceta}^{[-i,-j]}}(W_i)}^2
                    \given\calF_{-\{i,j\}}}\le b_{ij}^2.
\end{equation}
Require
\begin{align}
    D_n&=\Ex*{N_n\sum_iw_i^2r_i^2}=o(1),\label{eq:affine_appendix_diagonal}\\
    H_n&=\sum_{i\ne j}\Ex*{N_nw_iw_jb_{ij}b_{ji}}=o(1).
       \label{eq:affine_appendix_pairs}
\end{align}
Also suppose
\begin{equation}\label{eq:affine_appendix_training_transfer}
\sqrt{N_n}\sum_iw_i\big[\norm*{d_{\hat{\veceta}}(W_i)-d_i(W_i)} +\norm*{\mu_{\hat{\veceta}}(X_i)
                    -\mu_{\hat{\veceta}^{[-i]}}(X_i)}\big]=\oP*{1}.
\end{equation}
Then \eqref{eq:affine_centered_comparison} holds.
\end{lemma}

\begin{proof}
The \(\calF_{-i}\)-measurability of the retained fit and the conditional distribution \(\mathsf P_{X_i}\) of \(W_i\) given \(\calF_{-i}\) imply \(\Ex*{\xi_i\given\calF_{-i}}=0\) and \(\Ex*{\norm*{\xi_i}^2\given\calF_{-i}}\le r_i^2\).
For \(i\ne j\), the twice-deleted fit provides a common comparison for the two once-deleted scores, so subtracting these comparisons isolates a product of deletion differences while conditional centering removes the mixed terms.
We take the following conditional expectations before any restriction to a sample-dependent good event, preserving the conditional mean zero used in the cancellations.
For these cancellations, condition successively on \(\calF_{-i}\), \(\calF_{-j}\), and \(\calF_{-\{i,j\}}\).
The weights are measurable in every field, while the twice-deleted score evaluated at \(W_j\) is \(\calF_{-i}\)-measurable, and conversely.
Conditional independence of the two removed observations gives the third cancellation.
Therefore
\begin{equation}\label{eq:affine_appendix_pair_cancellation}
    \Ex*{w_iw_j\ip{\xi_i,\xi_j}}
       =\Ex*{w_iw_j\ip{\xi_i-\xi_i^{(j)},\xi_j-\xi_j^{(i)}}}.
\end{equation}
Centering the score difference conditional on \(\calF_{-i}\) contracts its conditional second moment.
Iterating to \(\calF_{-\{i,j\}}\) and using \eqref{eq:affine_appendix_pair_radius} bounds that moment by \(b_{ij}^2\).
Conditional Cauchy--Schwarz in \eqref{eq:affine_appendix_pair_cancellation} consequently gives
\[
    \Ex*{N_n\norm*{\sum_iw_i\xi_i}^2}\le D_n+H_n=o(1).
\]
Equation~\eqref{eq:affine_appendix_training_transfer} bounds the difference between this once-deleted centered sum and the full-fit centered sum, so Chebyshev's inequality proves the result.
\end{proof}

The score radii in this lemma can be checked component by component:
\[
    r_i\le\norm{a_0}
       d_{B,X_i}(\hat{\veceta}^{[-i]},\veceta_0)
       +d_{b,X_i}(\hat{\veceta}^{[-i]},\veceta_0).
\]
The same triangle inequality bounds deletion radii using differences between retained fits; the proof of Proposition~\ref{prop:affine_bounded_comparisons} controls their conditional second moments under the pair field through the expectation bound
\begin{equation}\label{eq:affine_pairfree_payment}
 \sum_{i\ne j}\Ex*{N_nw_iw_jb_{ij}b_{ji}}=o(1).
\end{equation}

\begin{lemma}[Weighted squared error from one deletion]\label{lem:affine_fitted_square_loo}
Let \(f_{\veceta}(W)\) be any vector or matrix increment relative to its oracle value, with the corresponding Euclidean or Frobenius norm.
Set \(r_{f,i}^2=\Ex*{\norm*{f_{\hat{\veceta}^{[-i]}}(W_i)}^2 \given\calF_{-i}}\).
If
\[
    \Ex*{\sum_iw_i r_{f,i}^2}\to0,\qquad
    \sum_iw_i\norm*{f_{\hat{\veceta}}(W_i)
                      -f_{\hat{\veceta}^{[-i]}}(W_i)}^2=\oP*{1},
\]
then \(\sum_iw_i\norm*{f_{\hat{\veceta}}(W_i)}^2=\oP*{1}\).
\end{lemma}

\begin{proof}
The squared triangle inequality bounds the full-sample squared error by twice the squared change after deletion plus twice the once-deleted squared error.
Conditional on \(\calF_{-i}\), \(w_i\) is measurable and the expectation of the latter summand is \(w_i r_{f,i}^2\), so Markov's inequality completes the proof without a root-scale deletion condition or twice-deleted fit.
\end{proof}

\subsection{A deterministic jackknife inequality}

For any vector array \(z\), define
\[
    R_j(z)=\sum_{i\ne j}w_i^{[-j]}z_i,\quad
    Q_j(z)=nA(z)-(n-1)R_j(z),\quad
    \bar Q(z)=n^{-1}\sum_jQ_j(z),
\]
and let \(J_n(z)=N_n\sum_j\{Q_j(z)-\bar Q(z)\} \{Q_j(z)-\bar Q(z)\}^{\top}/\{n(n-1)\}\).

\begin{lemma}[Jackknife bound]\label{lem:affine_frozen_quadratic}
For exact DNN weights with one common order \(1\le s<n\), \(n^{-1}\sum_jR_j(z)=A(z)\), and, pathwise,
\begin{equation}\label{eq:affine_appendix_quadratic}
    \opnorm*{J_n(z)}
       \le C_n\opnorm*{\sum_iw_i z_i z_i^\top},\qquad
    C_n=\frac{3N_n(n-1)}n\left\{\frac sn+\frac{s}{n-s}\right\}\to12
\end{equation}
under the asymptotic regime of Lemma~\ref{lem:affine_dnn_geometry}.
For two arrays \(z\) and \(z^0\), put \(e=z-z^0\).
Then
\begin{equation}\label{eq:affine_appendix_covariance_perturbation}
    \opnorm*{J_n(z)-J_n(z^0)}
       \le2\opnorm*{J_n(z^0)}^{1/2}\opnorm*{J_n(e)}^{1/2}
                 +\opnorm*{J_n(e)}.
\end{equation}
These statements hold simultaneously for every array, including random arrays obtained by fitting on the full sample.
\end{lemma}

\begin{proof}
Every size-\(s\) subsample is retained in exactly \(n-s\) deletion arrays, which proves the averaging identity.
It suffices to prove the quadratic inequality for a scalar array after ordering by distance.
Put \(m=n-s+1\), \(v_r=\binom{n-r}{s-1}/\binom ns\), \(v_{m+1}=0\), and \(\alpha=n/(n-s)\).
The deletion formula is
\begin{equation}\label{eq:affine_appendix_delete_formula}
    R_j(z)=\alpha\left\{\sum_{r<j}v_{r+1}z_r
                              +\sum_{r>j}v_rz_r\right\}.
\end{equation}
We bound the prefix, diagonal and suffix terms separately.
Set \(E_z=\sum_rv_rz_r^2\), \(\beta_s=s/(n-s)\), and \(c_r=(n-sr)/\{(n-s)(n-r)\}\).
Equation \eqref{eq:affine_appendix_delete_formula} gives
\[
    R_j(z)-A(z)=\sum_{r<j}v_rc_rz_r-v_jz_j
                                      +\beta_s\sum_{r>j}v_rz_r.
\]
The prefix term has squared \(\ell_2\) norm at most \(E_z\sum_r(n-r)v_rc_r^2\), by row-wise Cauchy--Schwarz.
The exact rank identity is
\begin{equation}\label{eq:affine_appendix_prefix_identity}
    \sum_r(n-r)v_rc_r^2=\frac{s}{(s+1)(n-s)}.
\end{equation}
To verify it for \(s\ge2\), let \(b_r\) be the distribution of the minimum rank in an \((s-1)\)-subset of \(\{1,\ldots,n-1\}\).
Then \(v_r/(n-r)=sb_r/\{n(s-1)\}\), and summing its binomial tails gives \(\Exi{b}{R}=n/s\) and \(\Exi{b}{R^2}=n(2n-s+1)/\{s(s+1)\}\).
Thus \(\Exi{b}{(n-sR)^2}=(s-1)n(n-s)/(s+1)\), which proves \eqref{eq:affine_appendix_prefix_identity}.
For \(s=1\), set \(c_n=0\), since it never enters a prefix.
Direct summation gives the same identity, \(1/\{2(n-1)\}\).

The diagonal term has squared norm at most \((s/n)E_z\).
For the suffix, put \(\tau_j=\sum_{r>j}v_r\).
Cauchy--Schwarz bounds its squared norm by \(\beta_s^2 E_z\sum_j\tau_j\), where the rank-tail identity gives \(\sum_j\tau_j=(n-s)/(s+1)\).
Consequently
\[
\sum_j\{R_j(z)-A(z)\}^2
       \le3\left\{\frac sn+\frac{s}{(s+1)(n-s)}
                         +\frac{s^2}{(s+1)(n-s)}\right\}E_z =3\left\{\frac sn+\frac{s}{n-s}\right\}E_z.
\]
The averaging identity gives \(\bar Q(z)=A(z)\) and \(Q_j(z)-\bar Q(z)=-(n-1)\{R_j(z)-A(z)\}\), proving the scalar bound.
Apply it to \(\gamma^\top z_i\) for each unit \(\gamma\) to obtain the vector bound.
For the perturbation, expand the centered pseudo-values of \(z=z^0+e\) and apply matrix Cauchy--Schwarz to bound each cross term by \(\opnorm{J_n(z^0)}^{1/2}\opnorm{J_n(e)}^{1/2}\), proving \eqref{eq:affine_appendix_covariance_perturbation}.
\end{proof}

\section{Proofs of the Results}\label{app:affine_proofs}

Localization and oracle moments establish oracle normality and covariance calibration.
Deletion comparisons and the affine equation then give the fitted estimator's expansion, with the jackknife inequality transferring covariance consistency.

\subsection{Oracle normality and variance}

\begin{proposition}[Oracle normality and covariance]\label{prop:affine_oracle}
Suppose Conditions~\ref{cond:affine_design} and \ref{cond:affine_oracle} hold.
Then
\begin{align}
    \sqrt{N_n}A(\vecpsi_0)&\dto\mathcal N(0,\Sigma_{\bx}),
        \label{eq:affine_oracle_clt}\\
    N_n\Varb*{A(\vecpsi_0)}&\to\Sigma_{\bx},\qquad
    \opnorm*{J_n(\vecpsi_0)-\Sigma_{\bx}}=\oP*{1}.
        \label{eq:affine_appendix_oracle_variance}
\end{align}
Also \(A(B_0)\pto\Gamma_{\bx}\) and \(A(\norm{B_0}_F^2)=\OP*{1}\).
\end{proposition}

\begin{proof}
\textit{Conditional mean and residual.}
Put \(\mu_0(u)=\Ex*{\vecpsi_0(W)\given X=u}\) and \(\varepsilon(W)=\vecpsi_0(W)-\mu_0(X)\).
The point identifying equation gives \(\mu_0(\bx)=0\), so the local H\"older bound and Lemma~\ref{lem:affine_dnn_geometry} imply
\begin{equation}\label{eq:affine_appendix_oracle_bias}
    \norm*{P_s\vecpsi_0}
       \le Cs^{-\beta/k_x}+Cs(1-a)^{s-1}.
\end{equation}
The tail constant follows from the integrated score moment.
Undersmoothing implies \(n=o(s^{1+2\beta/k_x})\), so the exponentially small tail remains negligible after multiplication by \(\sqrt{N_n}\).

The randomness of the conditional mean is smaller than that of the score residual.
For any deterministic scalar response \(g\), the ordinary Hoeffding decomposition of its complete DNN statistic gives
\begin{equation}\label{eq:affine_appendix_mean_variance_bound}
    \Varb*{A(g)}\le(s/n)\Varb*{H_{s,g}},
\end{equation}
where \(H_{s,g}\) is the response of the winner in a size-\(s\) subsample.
Indeed, in the variance decomposition, each component's coefficient relative to its full-kernel coefficient is \(\binom sj/\binom nj\le s/n\), \(1\le j\le s\).
For a unit direction \(\gamma\), take \(g=\gamma^\top\mu_0(X)\), whose selected mean satisfies
\[
    \Ex*{|H_{s,g}|^2}=P_s|g|^2
       =O(s^{-2\beta/k_x})+O(s(1-a)^{s-1}).
\]
Since \(N_ns/n=O(1)\), it follows that \(\sqrt{N_n}\{A(\mu_0)-P_s\mu_0\}=\oP*{1}\), in fixed dimension.
Localization suffices for this variance calculation.
Undersmoothing is needed for the deterministic mean in \eqref{eq:affine_appendix_oracle_bias}.

\textit{Conditional normal approximation.}
Given \(X_1,\ldots,X_n\), the residuals are independent and centered, with normalized covariance \(N_n\sum_iw_i^2 C_0(X_i)\).
By continuity at \(\bx\), the population-weight identity, and the fixed-ball tail bound,
\[
    \Ex*{\opnorm*{N_n\sum_iw_i^2C_0(X_i)-\Sigma_{\bx}}}
       \le N_n\max_iw_i\,P_s\opnorm{C_0(X)-\Sigma_{\bx}}\to0.
\]
Conditional Jensen transfers the assumed score moments to \(\varepsilon\), and Lemma~\ref{lem:affine_dnn_geometry} gives
\begin{equation}\label{eq:affine_appendix_lyapunov}
\Ex*{N_n^{q/2}\sum_iw_i^q
                \Ex*{\norm{\varepsilon_i}^q\given X_i}}
       \le N_n^{q/2}(\max_iw_i)^{q-1}P_s\norm{\varepsilon}^q =O((s/n)^{q/2-1})\to0.
\end{equation}
Thus the conditional Lyapunov quantity tends to zero in probability and implies the conditional Lindeberg condition in every fixed direction.
Every subsequence admits a further subsequence along which it and the conditional covariance converge almost surely.
The triangular-array CLT on that subsequence gives Gaussian conditional characteristic functions.
Their boundedness gives unconditional convergence.
Cram\'er--Wold, the negligible conditional-mean fluctuation, and \eqref{eq:affine_appendix_oracle_bias} prove \eqref{eq:affine_oracle_clt}.

The law of total covariance gives
\[
    N_n\Varb*{A(\vecpsi_0)}
       =\Ex*{N_n\sum_iw_i^2C_0(X_i)}
            +N_n\Varb*{A(\mu_0)}\to\Sigma_{\bx}.
\]
This establishes the calibration of the variance of the complete statistic to the pointwise score covariance.

\textit{Oracle jackknife consistency.}
Use the fixed polarization family \(\mathcal D_p=\{e_j:1\le j\le p\}\cup \{(e_j+e_k)/\sqrt2:1\le j<k\le p\}\).
For \(d\in\mathcal D_p\), let \(R=d^\top\vecpsi_0\), and let \(H_s\) select this response from a size-\(s\) subsample.
Write \(h_s^{(1)}(W)=\Ex*{H_s\given W}-\Ex{H_s}\), \(\zeta_s^1=\Varb{h_s^{(1)}}\), and \(\zeta_s^s=\Varb{H_s}\).
The fixed ball can be shrunk so that \(\Varb*{R\given X=u}\) is bounded below there, simultaneously for this finite family, by covariance continuity and positive definiteness at \(\bx\).

Put \(U_X=F_{\bx}(\norm*{X-\bx}_2)\) and \(q_s(X)=(1-U_X)^{s-1}\).
Continuity gives \(U_X\sim\mathrm{Unif}(0,1)\) and \(\Ex{q_s(X)^c}=1/\{c(s-1)+1\}\).
Selecting exactly one response and using the local and ambient moments gives \(\Ex{|H_s|^q}=O(1)\), hence \(\zeta_s^s=O(1)\).
The first projection decomposes exactly as
\[
    h_s^{(1)}(W)=q_s(X)\{R-\Ex*{R\given X}\}+b_s(X),\qquad
    b_s(X)=\Ex*{H_s\given X}-\Ex{H_s}.
\]
The two terms are orthogonal, so the residual part and the local variance lower bound give \(\zeta_s^1\gtrsim1/s\).
Conversely, Hoeffding orthogonality for the full kernel gives \(s\zeta_s^1\le\zeta_s^s=O(1)\).

The selector identity bounds the residual projection's \(q\)-moment by \(C/s\).
To bound \(b_s\), split the conditional mean into its part inside the fixed ball and its complement.
The inside mean kernel is bounded, so its centered first projection has bounded supremum norm and, by Hoeffding orthogonality, second moment \(O(1/s)\), giving \(q\)-moment \(O(1/s)\).
The outside mean kernel has \(q\)-moment at most \(Cs(1-a)^{s-1}\).
Conditional Jensen bounds its centered projection's moment by the same order, which is also \(O(1/s)\).
Consequently
\[
    \zeta_s^1\asymp s^{-1},\qquad
    \Ex{|h_s^{(1)}|^q}=O(s^{-1}).
\]
With \(r=q/2\in(1,2]\), these bounds verify the conditions of the complete-statistic jackknife theorem of \citet{juergens_jackknife_2025}:
\[
\frac{s}{n}\left(\frac{\zeta_s^s}{s\zeta_s^1}-1\right)
       =O(s/n)\to0,\qquad n^{1-r}\Ex*{(h_s^{(1)}(W)^2/\zeta_s^1)^r}
       =O((s/n)^{r-1})\to0.
\]
Its H\'ajek-dominance and jackknife conclusions now apply, with the delete-one order requirements \(s+1\le n\), \(s=o(n)\), and positive projection and statistic variances holding eventually.
With the response and order fixed across deletions, Lemma~\ref{lem:affine_frozen_quadratic} gives
\[
    \frac{1}{n(n-1)}\sum_j\{Q_j(R)-\bar Q(R)\}^2
       =\frac{n-1}{n}\sum_j\{R_j(R)-A(R)\}^2,
\]
which gives the required normalization.
Its ratio law, multiplied by the variance calibration already proved, gives \(d^\top J_n(\vecpsi_0)d\pto d^\top\Sigma_{\bx}d\).
Polarization recovers every matrix entry: for \(d=(e_j+e_k)/\sqrt2\), the off-diagonal entry of a symmetric matrix equals \(d^\top Md-(M_{jj}+M_{kk})/2\).
Fixed dimension gives the operator convergence in \eqref{eq:affine_appendix_oracle_variance}.

\textit{Oracle coefficient.}
Applying \eqref{eq:affine_appendix_mean_variance_bound} entry by entry to \(B_0\), whose selected second moments are bounded, gives \(A(B_0)-P_sB_0=\OP*{\sqrt{s/n}}\), while localization of its continuous conditional mean gives \(P_sB_0\to\Gamma_{\bx}\).
Finally \(\Ex*{A(\norm{B_0}_F^2)}=P_s\norm{B_0}_F^2=O(1)\), so Markov's inequality gives the weighted second-moment bound.
\end{proof}

\subsection{From learner comparisons to the affine estimator}

\begin{proposition}[Sufficient weighted comparisons]\label{prop:affine_comparisons}
Assume the hypotheses of Proposition~\ref{prop:affine_oracle}.
Suppose Lemma~\ref{lem:affine_centered_loo} applies to the score increment evaluated at the true parameter, and
\begin{equation}\label{eq:affine_appendix_mean_bias_payment}
    \sqrt{N_n}\norm*{\sum_iw_i\mu_{\hat{\veceta}}(X_i)}=\oP*{1}.
\end{equation}
Suppose also that Lemma~\ref{lem:affine_fitted_square_loo} applies to \(d_{\veceta}\) and to \(B(\veceta)-B_0\).
Then coefficient consistency \eqref{eq:affine_coefficient_bridge} holds, together with
\begin{equation}\label{eq:affine_score_bridge}
 \sqrt{N_n}A\{\vecpsi(a_0,\hat{\veceta})-\vecpsi_0\}=\oP*{1}.
\end{equation}
Writing \(J_n(\vecpsi_0)\) for the oracle jackknife covariance, the fitted array satisfies
\begin{equation}\label{eq:affine_covariance_bridge}
\opnorm*{J_n(\hat{\vecpsi})-J_n(\vecpsi_0)}=\oP*{1},\qquad \opnorm*{J_n(\vecpsi_0)-\Sigma_{\bx}}=\oP*{1}.
\end{equation}

\end{proposition}

\begin{proof}
The decomposition into its conditional mean and centered part gives \eqref{eq:affine_score_bridge} from \eqref{eq:affine_appendix_mean_bias_payment} and Lemma~\ref{lem:affine_centered_loo}.
The weighted coefficient error bound and Cauchy--Schwarz give
\[
    \opnorm*{A(B(\hat{\veceta})-B_0)}
       \le\{A(\norm{B(\hat{\veceta})-B_0}_F^2)\}^{1/2}=\oP*{1}.
\]
Together with Proposition~\ref{prop:affine_oracle}, this proves \eqref{eq:affine_coefficient_bridge}: \(\hat B_{\bx}\pto\Gamma_{\bx}\), so the inverse exists with probability tending to one.
The exact identity \eqref{eq:affine_exact_error} and oracle tightness first give \(\hat{\vecalpha}-a_0=\OP*{N_n^{-1/2}}\), hence consistency.

The weighted second moment of the fitted coefficients is bounded in probability because
\[
    A(\norm{B(\hat{\veceta})}_F^2)
       \le2A(\norm{B_0}_F^2)
          +2A(\norm{B(\hat{\veceta})-B_0}_F^2)=\OP*{1}.
\]
To compare the fitted and oracle scores, use the exact varying-coefficient decomposition
\[
    \hat{\vecpsi}_i-\vecpsi_0(W_i)
       =d_{\hat{\veceta}}(W_i)
             +B(W_i;\hat{\veceta})(\hat{\vecalpha}-a_0).
\]
Its weighted square satisfies
\begin{equation}\label{eq:affine_general_square_decomposition}
A(\norm{\hat{\vecpsi}-\vecpsi_0}^2)
  \le{}2A(\norm{d_{\hat{\veceta}}}^2) +2\norm{\hat{\vecalpha}-a_0}^2
                         A(\norm{B(\hat{\veceta})}_F^2)=\oP*{1}.
\end{equation}
Here the score error bound controls the first term, while parameter consistency and the bounded coefficient moment control the second.
Applying Lemma~\ref{lem:affine_frozen_quadratic} to this difference array gives \(J_n(\hat{\vecpsi}-\vecpsi_0)=\oP*{1}\) in operator norm, so the perturbation inequality and oracle jackknife consistency prove \eqref{eq:affine_covariance_bridge}.
\end{proof}

Identity~\eqref{eq:affine_mixed_drift} follows by adding and subtracting \(\{\bar B(u;\veceta)-\bar B(u;\veceta_0)\}\vecalpha_0(u)\).
Under the bounded learner conditions, Proposition~\ref{prop:affine_bounded_comparisons} applies this identity at once-deleted fits and uses the uniform deletion bound to establish \eqref{eq:affine_appendix_mean_bias_payment} for the full fit.
The integrated radius moment gives the spatial rate \(s^{-\beta/k_x}\), avoiding the larger window supremum \(r_{n,L}^{\beta}\).

\subsection{Proof of pointwise inference}

\begin{proof}[Proof of Theorem~\ref{thm:affine_inference}]
Proposition~\ref{prop:affine_bounded_comparisons} derives the weighted comparisons from the bounded learner conditions.
Proposition \ref{prop:affine_oracle} verifies oracle normality and covariance calibration.
Coefficient convergence and nonsingularity give
\[
    \hat B_{\bx}^{-1}\pto\Gamma_{\bx}^{-1}.
\]
Use the exact equation and the score comparison at the true parameter to obtain
\[
    \sqrt{N_n}(\hat{\vecalpha}-a_0)
       =-\hat B_{\bx}^{-1}\sqrt{N_n}A(\vecpsi_0)+\oP*{1}
       =-\Gamma_{\bx}^{-1}\sqrt{N_n}A(\vecpsi_0)+\oP*{1}.
\]
Oracle normality and Slutsky's theorem give the asserted normal limit.
The covariance comparison gives \(\hat\Sigma_{\bx}\pto\Sigma_{\bx}\), and continuity of the inverse and matrix multiplication gives \(\hat\Omega_{\bx}\pto\Omega_{\bx}\).
For nonzero fixed \(c\), \(c^\top\Omega_{\bx}c>0\).
A final application of Slutsky proves the studentized contrast and the interval coverage in \eqref{eq:affine_contrast_interval}.
All statements on the event where the fitted inverse is undefined can be assigned the stated measurable convention, because that event has probability tending to zero.
\end{proof}

For completeness, let \(L\in\R^{r\times p}\) be fixed with full row rank.
Under \(H_0:L a_0=\ell\), the same argument gives
\[
    N_n(L\hat{\vecalpha}-\ell)^\top
       (L\hat\Omega_{\bx}L^\top)^{-1}
       (L\hat{\vecalpha}-\ell)\dto\chi_r^2.
\]
This proves the fixed-dimensional Wald assertion in the main text.
An approximate linear-system solution with residual \(\hat r=\hat B_{\bx}\hat{\vecalpha}+\hat b_{\bx}\) has the extra term \(\hat B_{\bx}^{-1}\hat r\).
The same conclusion holds if \(\sqrt{N_n}\norm{\hat r}=\oP*{1}\).

\subsection{Computation of the frozen-score deletions}

For the complexity claim, use \eqref{eq:affine_appendix_delete_formula}: after a distance sort, form prefix sums of \(v_{r+1}z_r\) and suffix sums of \(v_rz_r\), so that each deleted weighted score requires one prefix, one suffix, and the factor \(n/(n-s)\).
There are \(n\) such scores, so the deletion calculation is \(O(n)\) for fixed dimension, and the sort gives the overall \(O(n\log n)\) cost.
Binomial-ratio recursion computes the weights without enumerating subsamples, and the fitted array remains fixed throughout.

\subsection{Fold-specific solutions and pooled local equations}\label{app:affine_fold_aggregation}

Fix full-sample weights \((w_i)\) and evaluation folds \(I_k\), using the nuisance fit trained outside \(I_k\).
Put \(\pi_k=\sum_{i\in I_k}w_i\) and omit zero-mass folds, which contribute nothing to the pooled equation.
For each remaining fold, define \(\Gamma_k=\pi_k^{-1}\sum_{i\in I_k}w_iB(W_i;\hat{\veceta}_{-k})\) and \(g_k=\pi_k^{-1}\sum_{i\in I_k}w_ib(W_i;\hat{\veceta}_{-k})\).
Where the fold inverses exist, \(\hat{\vecalpha}_k=-\Gamma_k^{-1}g_k\).
The pooled equation has solution
\begin{equation}\label{eq:affine_fold_pooling}
 \hat{\vecalpha}_{\mathrm{pool}}
 =-\left(\sum_k\pi_k\Gamma_k\right)^{-1}\sum_k\pi_k g_k
 =\left(\sum_k\pi_k\Gamma_k\right)^{-1}
                       \sum_k\pi_k\Gamma_k\hat{\vecalpha}_k.
\end{equation}
All sums run over positive-mass folds.
The first expression requires only the pooled inverse, while the second additionally requires the individual fold inverses.
Averaging fold solutions, even by local masses, generally differs from pooling their varying-coefficient equations.
This algebra describes the distinction between DML1 and DML2 in \citet[Definitions 3.1--3.2]{chernozhukov_doubledebiased_2018}.
For global estimators, \citet[Theorems 1--2]{velez_debiased_2026_affine} derives model-dependent growing-fold mean shifts, while Theorem~3 of that paper covers global DML2 with actual out-of-fold fits, including leave-one-out under its stated conditions.

For CATE, \(\Gamma_k=1\), so pooling equals \(\sum_k\pi_k\hat\tau_k\).
Thus cross-fitted nuisances can use the same full-sample DNN second stage as full-fitted nuisances, as they do in Section~\ref{sec:affine_numerical}.
The present theory additionally permits one nuisance fit trained on every scored observation.

\section{Application and Learner Verification}\label{app:affine_verification}

\subsection{Weighted comparisons for bounded learners}

\begin{proposition}[Bounded learner comparisons]\label{prop:affine_bounded_comparisons}
Conditions~\ref{cond:affine_design}, \ref{cond:affine_oracle} and \ref{cond:affine_bounded_learner} imply the hypotheses and conclusions of Proposition~\ref{prop:affine_comparisons}.
\end{proposition}

\begin{proof}
Under Condition~\ref{cond:affine_bounded_learner}, bounded coefficients give a fixed envelope for every score increment evaluated at the true parameter.
Let \(\rho_{n,i}\) be the conditional component error bound for the once-deleted fit, taking the supremum on \(\calB_{n,L}\) only for that fit, so that it is \(\calF_{-i}\)-measurable and \(\rho_{n,i}\le\rho_n\).
Conditional on \(\calF_{-i}\), the retained fit is fixed and \(W_i\) has distribution \(\mathsf P_{X_i}\), so the conditional prediction bounds give
\[
    r_i^2\le C\rho_{n,i}^2\1*{X_i\in\calB_{n,L}}
                  +C\1*{X_i\notin\calB_{n,L}}.
\]
Consequently, using \(\rho_{n,i}\le\rho_n\) and \(N_n\sum_iw_i^2=1\),
\[
    D_n\le C\Ex{\rho_n^2}
                  +CN_n\max_iw_i\Ex{W_{n,\mathrm{out}}}=o(1).
\]

For the pair contribution, put \(p_{ij}=\Pb*{E_n^c\given\calF_{-\{i,j\}}}\).
On \(E_n\), every score deletion difference is bounded by \(C\delta_n\) at every argument.
Off \(E_n\), boundedness gives a constant bound.
Thus the symmetric, pair-measurable choice
\[
    b_{ij}=b_{ji}=C(\delta_n^2+p_{ij})^{1/2}
\]
satisfies \eqref{eq:affine_appendix_pair_radius}.
Since the weights are measurable in the pair field, the tower property gives
\[
    H_n\le CN_n\delta_n^2+CN_n\Pb{E_n^c}=o(1).
\]
We use the conditional failure probability because the good-event indicator need not be measurable in the pair field.
Integrating score increments against the same conditional distribution gives, on \(E_n\),
\[
\sup_u\norm*{\mu_{\hat{\veceta}}(u)
                  -\mu_{\hat{\veceta}^{[-i]}}(u)}
       \le\sup_W\norm*{d_{\hat{\veceta}}(W)
                         -d_{\hat{\veceta}^{[-i]}}(W)} \le C\delta_n.
\]
Together with the training score deletion bound and \(\sum_iw_i=1\), this bounds the left side of \eqref{eq:affine_appendix_training_transfer} by \(C\sqrt{N_n}\delta_n\) on \(E_n\).
The deletion rate and \(\Pb{E_n^c}\to0\) make it negligible in probability, verifying the remaining hypothesis of Lemma~\ref{lem:affine_centered_loo}.

The same one-deletion argument gives \(\Ex*{\sum_iw_i r_i^2}\le C\Ex{\rho_n^2}+C\Ex{W_{n,\mathrm{out}}}\to0\).
Since \(\delta_n\to0\) and \(\Pb{E_n^c}\to0\), the weighted squared deletion error also vanishes, so Lemma~\ref{lem:affine_fitted_square_loo} gives the weighted squared error bounds for the score and coefficient increments.

To control nuisance mean bias, first note that local nonsingularity and the H\"older coefficient means imply \(\norm{\vecalpha_0(u)-a_0}\le C\norm{u-\bx}^{\beta}\), by \(\vecalpha_0(u)=-\bar B(u;\veceta_0)^{-1} \bar b(u;\veceta_0)\) and the inverse perturbation identity.
For each fixed nuisance parameter, \eqref{eq:affine_mixed_drift} and conditional Jensen give, \(P_X\)-almost everywhere on \(\calB_{n,L}\),
\[
    \norm{\mu_{\veceta}(u)}
       \le\norm{R(u;\veceta)}
           +C d_{B,u}(\veceta,\veceta_0)\norm{u-\bx}^{\beta}.
\]
To apply this almost-everywhere drift bound at the observed \(X_i\), we use the fit with observation \(i\) removed, which is independent of \(X_i\).
Its measurable essential bound therefore applies at that observation, and uniform deletion stability controls the difference from the full fit.
Put
\[
F_{\veceta}(u)=\bigl[\norm{\mu_{\veceta}(u)}
         -C d_{B,u}(\veceta,\veceta_0)\norm{u-\bx}^{\beta}\bigr]_+,\qquad e_n(\veceta)=\operatorname*{ess\,sup}_{u\in\calB_{n,L},\,P_X}
                              F_{\veceta}(u).
\]
Joint measurability of the mean and norm maps makes \(e_n\) measurable: for rational \(r\ge0\),
\[
    \set{e_n(\veceta)>r}
       =\set*{\int_{\calB_{n,L}}\1*{F_{\veceta}(u)>r}
                                     \dd P_X(u)>0}.
\]
For each full and once-deleted pair on \(E_n\), uniform component deletion and local boundedness of the parameter function give
\[
    \norm*{R(u;\hat{\veceta}^{[-i]})-R(u;\hat{\veceta})}
       \le C\delta_n
       \quad\text{for }P_X\text{-almost every }u\in\calB_{n,L}.
\]
It follows that \(e_n(\hat{\veceta}^{[-i]})\le t_n+C\delta_n\) on \(E_n\).
The once-deleted fit depends only on \(U,W_{-i}\) and is independent of \(X_i\).
Hence
\[
    F_{\hat{\veceta}^{[-i]}}(X_i)
       \le e_n(\hat{\veceta}^{[-i]})
       \quad\text{on }\set{X_i\in\calB_{n,L}},
       \quad\text{almost surely}.
\]
Bounded score increments control the outside-window means, so using \(\rho_{n,i}\le\rho_n\) and summing gives, on \(E_n\),
\[
\norm*{\sum_iw_i\mu_{\hat{\veceta}^{[-i]}}(X_i)}
       \le{}t_n+C\delta_n +C\rho_n\sum_{i:X_i\in\calB_{n,L}}
                          w_i\norm{X_i-\bx}^{\beta}
             +CW_{n,\mathrm{out}}.
\]
The uniform mean deletion bound adds at most \(C\delta_n\) when replacing this sum by \(\sum_iw_i\mu_{\hat{\veceta}}(X_i)\).
The remainder and deletion rates control the first two terms on the \(\sqrt{N_n}\) scale.
For the spatial term, the radius moment is \(\OP*{s^{-\beta/k_x}}\) and \(\rho_n=\oP*{1}\), so undersmoothing controls their product on the same scale.
Since \(N_n\le n\) and \(\Ex{W_{n,\mathrm{out}}}=O(n^{-L})\) with \(L>1\), Markov's inequality gives \(\sqrt{N_n}W_{n,\mathrm{out}}=\oP*{1}\).
This proves \eqref{eq:affine_appendix_mean_bias_payment}.
Proposition~\ref{prop:affine_comparisons} completes the proof.
\end{proof}

\subsection{CATE identification, orthogonality, and continuity}

Consistency, unconfoundedness given \(V\), and overlap imply
\[
    \mu_{0,t}(V)=\Ex*{Y(t)\given V}.
\]
At truth, iterated expectation in \eqref{eq:affine_cate_signal} gives
\[
    \Ex*{\phi(W;\veceta_0)\given V}
       =\mu_{0,1}(V)-\mu_{0,0}(V),\qquad
    \Ex*{\phi(W;\veceta_0)\given X=u}=\tau_0(u).
\]

For fixed nuisance functions, conditioning on \(V\) gives
\[
\Ex*{\phi(W;\veceta)-\phi(W;\veceta_0)\given V}
       =(1-\pi_0/\pi)\delta_{\mu,1}
          -\{1-(1-\pi_0)/(1-\pi)\}\delta_{\mu,0} =\delta_\pi\left\{\delta_{\mu,1}/\pi
                                    +\delta_{\mu,0}/(1-\pi)\right\}.
\]
Conditioning again on \(X=u\) proves \eqref{eq:affine_cate_bias_identity}, whose derivative at the true nuisance parameter is zero in every admissible direction, establishing conditional Neyman orthogonality.
Conditional Cauchy--Schwarz with clipped propensity gives the product bound, also for the negative signal increment in the score.

With bounded \(Y\), bounded projected outcome predictions, and clipped propensities, the signal has a fixed envelope and satisfies
\begin{equation}\label{eq:affine_appendix_aipw_lipschitz}
    |\phi(W;\veceta)-\phi(W;\veceta')|
       \le C\{ |\pi(V)-\pi'(V)|
                     +|\mu_1(V)-\mu'_1(V)|+|\mu_0(V)-\mu'_0(V)|\}.
\end{equation}
To see this, subtract each residual-over-propensity term, bound its prediction difference using the clipped denominator, and use \(|1/\pi-1/\pi'|\le C|\pi-\pi'|\).
The control term is identical.
This bound transfers prediction and deletion control to the score.
\ref{app:affine_series_construction} establishes conditional continuity.

\subsection{Retained series deletion and CATE covariance}

\begin{proposition}[Rank-one series deletion]\label{prop:affine_ridge_verification}
For the training rule with fixed normalizer \(1/n\) \eqref{eq:affine_ridge_fit}, suppose \(0\le D_{ji}\le1\), responses are bounded, \(\sup_v\norm{p_n(v)}\le\zeta_n\), and every penalized Gram matrix for at most two deletions has lower eigenvalue \(\kappa_n\) on \(E_n\).
Adjacent unprojected predictions differ uniformly by at most \(C\zeta_n^2(n\kappa_n)^{-1}(1+\zeta_n/\sqrt{\kappa_n})\).
Fixed interval projection preserves the bound.
\end{proposition}

\begin{proof}
For regression \(j\), write \(A_{j,I}\) for its penalized Gram, put \(\beta_I=\hat\beta_j^{[-I]}\), and let \(\bar R\) bound the responses.
The normal equations and nonnegativity of the ridge loss give
\[
 \beta_I^\top A_{j,I}\beta_I
 \le n^{-1}\sum_{i\notin I}D_{ji}R_{ji}^2\le\bar R^2,
 \qquad \norm{\beta_I}\le\bar R/\sqrt{\kappa_n}
 \quad\text{on }E_n.
\]
With tuning and normalizer unchanged, subtraction of the normal equations gives
\begin{equation}\label{eq:affine_series_rank_one}
 A_{j,I\cup\{k\}}(\hat\beta_j^{[-I]}-\hat\beta_j^{[-I,-k]})
 =\frac{D_{jk}}n p_n(V_k)
                 \{R_{jk}-p_n(V_k)^\top\hat\beta_j^{[-I]}\}.
\end{equation}
The rank-one identity and \(\abs*{R_{jk}-p_n(V_k)^\top\beta_I}\le\bar R(1+\zeta_n/\sqrt{\kappa_n})\) give the stated prediction bound, preserved by one-Lipschitz interval projection.
\end{proof}

For CATE, the common parameter value cancels as described in Section~\ref{sec:affine_results}.
Lemma~\ref{lem:affine_frozen_quadratic} and \(\sum_iw_i(\hat\phi_i-\phi_{0i})^2=\oP*{1}\) imply that the difference between the feasible and oracle jackknife covariance estimates vanishes in probability.
Proposition~\ref{prop:affine_oracle} therefore gives consistency of the feasible covariance estimator.
The centered score comparison needs the stronger deletion argument of Lemma~\ref{lem:affine_centered_loo}.

\subsection{The varying-slope residualized example}

Let \(v=T-e_0(V)\), \(\delta_e=e-e_0\), and \(\delta_m=m-m_0\).
The model in Section~\ref{sec:affine_applications} implies \(Y-m_0(V)=\theta_0(X)v+\varepsilon\), with \(\Ex*{v\given V}=0\) and \(\Ex*{\varepsilon\given V,T}=0\).
At \(a=\theta_0(u)\), its true score is \(-v\varepsilon\) given \(X=u\), so its conditional mean is zero and its coefficient mean is \(\Ex*{\Varb*{T\given V}\given X=u}\), proving the identification claim when that quantity is positive.

For any fixed \(a\), exact expansion of the fitted score increment gives
\[
\psi(W;a,(e,m))-\psi(W;a,(e_0,m_0))
       ={}-2av\delta_e+a\delta_e^2+v\delta_m +\delta_e\{Y-m_0(V)\}-\delta_e\delta_m.
\]
All terms linear in nuisance errors have conditional expectation zero given \(V\).
Consequently
\begin{align}
    \bar B(u;(e,m))-\bar B(u;(e_0,m_0))
       &=\Ex*{\delta_e^2\given X=u},\label{eq:affine_appendix_pl_coefficient}\\
    \Ex*{\psi(a,(e,m))-\psi(a,(e_0,m_0))\given X=u}
       &=\Ex*{a\delta_e^2-\delta_e\delta_m\given X=u}.
       \label{eq:affine_appendix_pl_bias}
\end{align}
Setting \(a=\theta_0(u)\) proves the orthogonal remainder in the main text, whereas evaluation at \(a=\theta_0(\bx)\) retains the mixed term \((\theta_0(\bx)-\theta_0(u))\Ex*{\delta_e^2\given X=u}\).
The mean-coefficient error is itself second order in this example.

For the specified construction, \ref{app:affine_series_construction} verifies the component learner conditions and proves consistency of the feasible covariance estimator.
The proof accounts for parameter-estimation error when the fitted coefficient varies across observations.

Allowing the slope to depend beyond \(X\), through \(\theta_0(V)\), gives the identified coefficient
\[
    \frac{\Ex*{\Varb*{T\given V}\theta_0(V)\given X=u}}
         {\Ex*{\Varb*{T\given V}\given X=u}},
\]
which generally differs from the marginal CATE.

\section{Proof of the Growing-Series Construction}\label{app:affine_series}\label{app:affine_series_construction}

The growing dictionary allows approximation error to vanish, while ridge regularization and clipping control how the fitted predictions change after deletion.
To verify the learner conditions, we relate the population projections to the true regression functions, then use a common Gram-matrix event to bound prediction error across the retained fits.
Translating these coefficient bounds into conditional prediction norms gives the local accuracy requirement; the deletion identity supplies stability on the same event.

\subsection{Population projections and approximation}

Write \(J=J_n\), \(K=K_n=3J\), \(M=M_n\), \(p=p_n\), and \(q=(q_0,q_1,q_2)^\top\).
For \(t=0,1\), let \(w_1(z)=\pi_0(z)\) and \(w_0(z)=1-\pi_0(z)\).
The four regression targets can be written \(f_{j0}(x,z)=q(x)^\top c_j(z)\), where
\begin{align*}
    c_\pi(z)&=(\pi_0(z),0,0)^\top,\\
    c_0(z)&=\{g_0(z)+\chi h(z_4)+(g_1(z)+g_2(z))/2,
                                 g_1(z)/\sqrt{12},g_2(z)/\sqrt{12}\}^\top,\\
    d&=\{1/8,1/(8\sqrt{12}),1/(8\sqrt{12})\}^\top,\\
    c_1(z)&=c_0(z)+d,\qquad c_m(z)=c_0(z)+\pi_0(z)d.
\end{align*}
The mean \(m_0\) needed by the second corollary has the same representation, and all these coefficient functions are bounded and Lipschitz uniformly in \(n\) and \(\chi\).

Let \(\beta_{jn}^{\circ}\) minimize the population unpenalized loss \(\Ex*{D_j(R_j-p(V)^\top\beta)^2}\), and put \(f_{jn}^{\circ}=p^\top\beta_{jn}^{\circ}\).
Ordering the coefficients by cell changes the tensor order above only by a fixed permutation.
Since \(\Ex{q(X)q(X)^\top}=I_3\) and \(X\) is independent of \(Z\), the population Gram matrices are
\begin{equation}\label{eq:affine_series_population_gram}
Q_\pi=Q_m=I_K,\qquad
    Q_t=\operatorname{diag}_{1\le h\le J}\{\bar w_{t,h} I_3\},\qquad \bar w_{t,h}=J\int_{C_h}w_t(z)\dd z\in[3/8,5/8].
\end{equation}
Conditional response means give
\[
\Ex*{D_tR_t\given X,Z}=w_t(Z)\mu_{0,t}(X,Z),\qquad \Ex*{R_\pi\given X,Z}=\pi_0(Z),
 \Ex*{R_m\given X,Z}=m_0(X,Z).
\]
so the population projection has cell coefficients
\begin{equation}\label{eq:affine_series_projection_coefficients}
    \beta_{j,h}^{\circ}=J^{-1/2}\bar c_{j,h},\qquad
    \bar c_{j,h}=
    \begin{cases}
       \displaystyle\frac{\int_{C_h}w_j(z)c_j(z)\dd z}
                            {\int_{C_h}w_j(z)\dd z},&j=0,1,\\[0.8em]
       \displaystyle J\int_{C_h}c_j(z)\dd z,&j=\pi,m.
    \end{cases}
\end{equation}
The treatment-arm weights therefore enter through cell averages of the coefficient functions: for \(z\in C_h\), \(f_{jn}^{\circ}(x,z)=q(x)^\top\bar c_{j,h}\).
A positive weighted average of a Lipschitz function differs from its value anywhere in the cell by at most its Lipschitz constant times the cell diameter \(2/M\).
Since \(\norm{q(x)}_2\le\sqrt7\),
\begin{equation}\label{eq:affine_series_projection_bounds}
    \sup_v\abs*{f_{jn}^{\circ}(v)-f_{j0}(v)}
           \le C/M\le CK^{-1/4},\qquad
    \norm{\beta_{jn}^{\circ}}_2^2
           =J^{-1}\sum_h\norm{\bar c_{j,h}}_2^2\le C.
\end{equation}
Bounded projections and responses give \(\abs*{R_j-f_{jn}^{\circ}(V)}\le C\) almost surely, uniformly over these four regressions, while the population normal equations give \(\Ex*{D_jp(V)\{R_j-f_{jn}^{\circ}(V)\}}=0\).

\subsection{One event for every retained Gram matrix}

Let \(G_j=n^{-1}\sum_iD_{ji}p(V_i)p(V_i)^\top\), whose summands are positive semidefinite, independent across \(i\), and bounded in operator norm by \(7J\), with
\[
 \lambda_{\min}\left(\sum_i\Ex*{D_{ji}p(V_i)p(V_i)^\top}\right)
    =n\lambda_{\min}(Q_j)\ge3n/8.
\]
The lower-tail part of \citet[Theorem~1.1]{tropp_user_friendly_2012_affine}, with deviation \(1/2\), yields
\[
    \Pb*{\lambda_{\min}(G_j)
                      <\tfrac12\lambda_{\min}(Q_j)}
       \le K\left\{\frac{e^{-1/2}}{(1/2)^{1/2}}\right\}
                          ^{n\lambda_{\min}(Q_j)/(7J)}
       \le K\exp(-cn/K).
\]
Define
\begin{equation}\label{eq:affine_series_gram_event}
    E_n=\bigcap_{j\in\{\pi,0,1,m\}}
                         \{\lambda_{\min}(G_j)\ge3/16\}.
\end{equation}
A union over four regressions gives \(\Pb{E_n^c}\le4K\exp(-cn/K)\).
For every \(\abs{I}\le2\), the retained unpenalized Gram matrix obeys
\[
    \lambda_{\min}\left(\frac1n\sum_{i\notin I}
                             D_{ji}p(V_i)p(V_i)^\top\right)
       \ge3/16-14J/n\ge3/32
\]
on \(E_n\), eventually, because \(K/n\to0\).
Adding \(\lambda_n I_K\) preserves the margin, so one full-sample event controls every retained Gram matrix through the deterministic deletion bound.

\subsection{Maximal retained prediction error in expectation}

We control the largest retained-fit error through one full-sample estimating moment for each regression: removing at most two observations changes this moment by a bounded correction, so its second-moment bound controls all retained fits on the common Gram event.
Define the full-sample, population-centered estimating moment
\[
    S_j=\frac1n\sum_iD_{ji}p(V_i)
                              \{R_{ji}-f_{jn}^{\circ}(V_i)\}.
\]
Independence, its zero mean, and the bounded projection residual imply
\[
    \Ex{\norm{S_j}_2^2}
       =n^{-1}\Ex*{D_j^2\norm{p(V)}_2^2
                         \{R_j-f_{jn}^{\circ}(V)\}^2}\le CK/n.
\]
For a retained set, write \(S_{j,I}=S_j-n^{-1}\sum_{i\in I}D_{ji}p(V_i) \{R_{ji}-f_{jn}^{\circ}(V_i)\}\).
Every deleted summand has norm at most \(C\sqrt K\), so pathwise
\[
    \max_{\abs{I}\le2}\norm{S_{j,I}}_2^2
                           \le2\norm{S_j}_2^2+CK/n^2.
\]
Put \(A_{j,I}=n^{-1}\sum_{i\notin I}D_{ji}p(V_i)p(V_i)^\top +\lambda_n I_K\).
The fitted normal equations give the exact comparison
\[
    A_{j,I}(\hat\beta_j^{[-I]}-\beta_{jn}^{\circ})
                    =S_{j,I}-\lambda_n\beta_{jn}^{\circ}.
\]
The common Gram margin and \eqref{eq:affine_series_projection_bounds} therefore prove
\begin{equation}\label{eq:affine_series_coefficient_mse}
    \Ex*{\1{E_n}\max_{j,\abs{I}\le2}
                  \norm*{\hat\beta_j^{[-I]}-\beta_{jn}^{\circ}}_2^2}
                \le C(K/n+\lambda_n^2).
\end{equation}
Taking the maximum therefore adds no logarithmic cost for the number of deletion subsets.

To translate the coefficient bound into prediction risk at a reporting profile, hold the realized retained fit fixed and average over an independent evaluation draw of \(Z\) conditional on \(X=u\).
Independence of \(X\) and \(Z\) and the equal cell probabilities give the conditional Gram matrix \(H(u)\).
Since \(\norm{q(u)}_2^2\le7\) and projection \(\Pi_j\) onto the clipping interval containing the truth cannot increase prediction error,
\[
\begin{aligned}
    H(u)&=\Ex*{p(u,Z)p(u,Z)^\top\given X=u}\\
        &=q(u)q(u)^\top\otimes I_J,\qquad \opnorm{H(u)}\le7,\\
    \norm{\hat f_j^{[-I]}-f_{j0}}_{u,2}^2
       &\le C\{\norm*{\hat\beta_j^{[-I]}-\beta_{jn}^{\circ}}_2^2
                   +\epsilon_K^2\}.
\end{aligned}
\]
Because this bound holds simultaneously for all regressions, retained sets and reporting points, we can take their finite sum, maximum and supremum before expectation.
Using \eqref{eq:affine_series_coefficient_mse} on \(E_n\) and bounded clipped errors on its complement then gives
\begin{equation}\label{eq:affine_series_prediction_mse}
    \Ex{\rho_n^2}
       \le C\{K/n+K^{-1/2}+\lambda_n^2+\Pb{E_n^c}\}.
\end{equation}

Unique ridge solutions are measurable in the retained sample, and the clipped prediction \(\Pi_j(p(u,z)^\top\beta)\) is jointly measurable in \((u,z,\beta)\) and continuous in \(u\).
Bounded integration against the uniform \(Z\) law therefore gives jointly measurable conditional norms continuous in \(u\), whose compact suprema equal countable dense suprema, making the finite retained-fit maximum measurable.
The same argument applies to score and coefficient norms after averaging over the finite conditional laws of \(T,\varepsilon_0,\varepsilon_1\) and substituting the continuous functions \(Y(t)=\mu_{0,t}(u,Z)+\varepsilon_t\).

\subsection{Deletion stability and rate requirements}

The bounded responses, dictionary envelope \(\zeta_n^2=7J\), and retained Gram margin \(\kappa=3/32\) verify the hypotheses of Proposition~\ref{prop:affine_ridge_verification}.
Its exact rank-one identity \eqref{eq:affine_series_rank_one} yields the uniform stability bound \(Ca_n=O(K^{3/2}/n)\) for adjacent fits over all prediction arguments, including the training covariates, with one-Lipschitz clipping preserving the bound.
This completes the proof of Proposition~\ref{prop:affine_series_predictions}.

Under \eqref{eq:affine_series_schedule}, \(K\asymp n^{1/3}\), \(\neffx\asymp n^{1/4}\), and \(\Pb{E_n^c}\le Cn^{1/3}\exp(-cn^{2/3})\).
Equation \eqref{eq:affine_series_prediction_mse} yields \(\Ex{\rho_n^2}=O(n^{-1/6})\).
Its three polynomial terms have orders \(n^{-2/3}\), \(n^{-1/6}\), and \(n^{-1}\).
The approximation term therefore gives the slowest rate in this bound.
These orders give all rates in \eqref{eq:affine_ridge_rates}.
In particular, Markov's inequality gives \(\sqrt{\neffx}\rho_n^2=\oP*{1}\), in addition to \(\Ex{\rho_n^2}\to0\).
At an interior point, \(F_{\bx}(r)=\pi r^2\) for sufficiently small \(r\), and the radial distribution is continuous.
Conditional coefficient means in both examples are Lipschitz, so \(k_x=2\), \(\beta=1\), and \(\sqrt{\neffx}s_n^{-1/2}=O(n^{-1/4})\) establishes undersmoothing.
A sufficiently large common window constant supplies the outside-weight bound in Lemma~\ref{lem:affine_dnn_geometry}.

\subsection{Proof of Corollary~\ref{cor:affine_series_cate}}

The construction ensures consistency of observed outcomes, unconfoundedness given \(V\), and uniform overlap.
Directly at truth,
\[
    \phi_0=\theta_0(X)+T\varepsilon_1/\pi_0
                          -(1-T)\varepsilon_0/(1-\pi_0).
\]
The two error terms have conditional mean zero and zero cross-product since \(T(1-T)=0\), so their sum has conditional variance given \(V\) equal to \(\{1/\pi_0+1/(1-\pi_0)\}/64\).
Independence of \(X\) and \(Z\) therefore proves \eqref{eq:affine_series_cate_variance}, constant in \(X\) and strictly positive.
For the oracle score at \(a_0=\theta_0(\bx)\), the conditional mean is \(\theta_0(\bx)-\theta_0(u)\), its variance is this same constant, and its coefficient is one.
Together with uniformly bounded score moments and the preceding geometry, these calculations verify the oracle hypotheses of Proposition~\ref{prop:affine_oracle}.

On the clipping range, the fixed envelope for the CATE signal and Lipschitz bound \eqref{eq:affine_appendix_aipw_lipschitz} give conditional component errors at most \(C\rho_n\) and deletion bound \(\delta_n=Ca_n\) on \(E_n\).
The product identity \eqref{eq:affine_cate_bias_identity}, evaluated with nuisance functions held fixed, gives \(t_n\le C\rho_n^2\) by conditional Cauchy--Schwarz.
Together with \eqref{eq:affine_ridge_rates}, these bounds verify Condition~\ref{cond:affine_bounded_learner}, so Theorem~\ref{thm:affine_inference} proves the corollary.
Since the fitted and oracle coefficients are both one, centering cancels the common parameter value and gives the stated frozen-signal covariance.

\subsection{Proof of Corollary~\ref{cor:affine_series_slope}}

Let \(\varepsilon=T\varepsilon_1+(1-T)\varepsilon_0\), which has conditional mean zero and conditional variance \(1/64\) given \((V,T)\), and write \(Y=\mu_{0,0}(V)+\theta_0(X)T+\varepsilon\).
Put \(v=T-\pi_0(Z)\) and \(\gamma=\Ex*{\pi_0(Z)(1-\pi_0(Z))}\).
Since \(\Ex{\sin^2(2\pi Z_4)}=1/2\), \(\gamma=1/4-1/128=31/128\).
At the parameter value for the fixed reporting point, the oracle score is
\[
    \psi_0=v^2\{\theta_0(\bx)-\theta_0(X)\}-v\varepsilon.
\]
Because \((Z,T)\) is independent of \(X\), its conditional mean and variance have the continuous versions
\[
\Ex*{\psi_0\given X=u}
       =\gamma\{\theta_0(\bx)-\theta_0(u)\},\qquad \Varb*{\psi_0\given X=u}
       =\gamma/64+\Varb{v^2}
                          \{\theta_0(\bx)-\theta_0(u)\}^2.
\]
The cross-covariance vanishes by the conditional zero mean of \(\varepsilon\) given \((V,T)\).
At \(u=\bx\), \(\Gamma_{\bx}=\gamma\), \(\Sigma_{\bx}=\gamma/64\), and \(\Omega_{\bx}=1/(64\gamma)=2/31\).
The variance is uniformly positive, coefficients and scores are bounded, and the coefficient means \(\bar B_0(u)=\gamma\), \(\bar b_0(u)=-\gamma\theta_0(u)\) are both Lipschitz.
Together with the geometry and undersmoothing, these properties establish the oracle conditions.

For deterministic predictions \((e,m)\), put \(\delta_e=e-\pi_0\), \(\delta_m=m-m_0\).
Equations~\eqref{eq:affine_appendix_pl_coefficient}--\eqref{eq:affine_appendix_pl_bias} give the exact mean-coefficient and score increments.
At \(a=\theta_0(u)\), conditional Cauchy--Schwarz bounds the remainder by \(C\rho_n^2\), so its derivative at truth is zero.
At \(a_0=\theta_0(\bx)\), retain the additional term
\[
    \{\theta_0(\bx)-\theta_0(u)\}
                  \Ex*{\delta_e^2\given X=u}.
\]
Its weighted magnitude is at most \(C\rho_n^2\sum_iw_i\norm{X_i-\bx}_2\).
The radius moment is \(\OP*{s_n^{-1/2}}\) by Lemma~\ref{lem:affine_dnn_geometry}, with bounded support controlling its fixed tail, and \(\rho_n^2=\oP*{1}\).
Undersmoothing makes this mixed term negligible on the \(\sqrt{\neffx}\) scale, while \(\sqrt{\neffx}\rho_n^2=\oP*{1}\) controls the orthogonal remainder.

For binary \(T\), bounded \(Y\), and the fixed clipping intervals,
\[
\begin{split}
    \abs*{B(W;e)-B(W;e')}&\le C\abs{e(V)-e'(V)},\\
    \abs*{b(W;e,m)-b(W;e',m')}
       &\le C\{\abs{e(V)-e'(V)}+\abs{m(V)-m'(V)}\}.
\end{split}
\]
Hence Condition~\ref{cond:affine_bounded_learner} holds with component radius \(C\rho_n\), \(t_n\le C\rho_n^2\), and \(\delta_n=Ca_n\).
Both coefficient maps have fixed envelopes.
With the oracle verification and \eqref{eq:affine_ridge_rates}, all hypotheses of Theorem~\ref{thm:affine_inference} hold, proving the corollary.
The varying-coefficient term \(B(W_i;\hat e)\{\hat\theta(\bx)-a_0\}\) has vanishing weighted squared norm by parameter consistency and bounded coefficients, as in \eqref{eq:affine_general_square_decomposition} of Proposition~\ref{prop:affine_comparisons}.

\section{Deletion Stability of Complete Subbagging}
\label{app:affine_subbagging}

\begin{proposition}[Complete subbagging]\label{prop:affine_subbagging}
For \(A\subset[n]\), \(\abs{A}\ge n-2\), average \(h_S\) over all \(m_n\)-subsets \(S\subset A\), with fixed \(1\le m_n\le n-2\).
Each \(h_S\) uses only \(W_S\) and a data-independent subset seed, reused whenever \(S\) is unchanged, and satisfies \(\sup_v\norm{h_S(v)}\le H\).
For \(j\in A\), \(\abs{A}\ge n-1\),
\[
 g_A=\binom{|A|}{m_n}^{-1}\sum_{S\subset A:\,|S|=m_n}h_S,
 \qquad
 \sup_v\norm{g_A(v)-g_{A\setminus\{j\}}(v)}
       \le\frac{2Hm_n}{|A|}.
\]
If both affine coefficient maps are uniformly Lipschitz functions of these averaged predictions, then \(\delta_n=O(m_n/n)\) for fits that differ by one deletion, with at most two observations removed.
The scaled deletion bound vanishes if \(m_n=o(\sqrt{ns_n})\).
\end{proposition}

\begin{proof}
Partition the subsets according to whether they contain \(j\): the proportion containing it is \(m_n/|A|\), giving the exact identity
\[
 g_A=(1-m_n/|A|)g_{A\setminus\{j\}}
                   +(m_n/|A|)g_{A,j\text{ included}}.
\]
Both averages have norm at most \(H\), so the triangle inequality proves the bound, while Lipschitz composition and \(\sqrt{\neffx}\,m_n/n=O(m_n/\sqrt{ns_n})\) establish the required rate.
\end{proof}

This calculation parallels \citet[Corollary~4 and Theorem~5]{chen_debiased_2022-2} in their complete \href{https://arxiv.org/abs/2206.01825v2}{arXiv:2206.01825v2} version; scaling the deletion bound for local inference gives the threshold \(m_n=o(\sqrt{ns_n})\).
The bound requires each base prediction to be bounded before averaging, so clipping the final average alone is insufficient.
Data-dependent global preprocessing or tuning needs separate control.
Complete averaging may be computationally infeasible, while finite or OOB forests follow different averaging rules.

\section{Simulation Design}\label{app:affine_production_numerics}

The nuisance comparisons use the same observations and final DNN weights, so changes in the estimates reflect changes in the fitted adjustment.
The additional designs vary curvature, noise and reporting density to examine how localization and local information affect inference at the reporting point.

\subsection{Models, learners and bandwidths}

Write \(S_j=\sin(2\pi Z_j)\).
Except in C2S, \(X\) and \(Z\) are independent uniforms on \([0,1]^2\) and \([0,1]^4\).
In every design, \(T\mid V\) is Bernoulli with propensity \(\pi_0(Z)=1/2+S_4/8\), and
\[
 Y=g(V)+T\theta_0(X)+(1-T)\varepsilon_0+T\varepsilon_1.
\]
The errors have mutually independent, equiprobable signs, independent of covariates and treatment, with amplitudes \(1/8\) except in C0H.
The baseline and treatment-effect functions are
\[
\begin{aligned}
 g(V)&=\{S_1+X_1S_2+X_2S_3\}/8+\1{\mathrm{C1}}S_4/8,\\
 c(X)&=\{2-\cos(8\pi(X_1-1/2))-\cos(8\pi(X_2-1/2))\}/16,\\
 \theta_0(X)&=(X_1+X_2)/8+\1{\mathrm{C2\ or\ C2S}}c(X).
\end{aligned}
\]
C0 and C1 correspond to \(\chi=0\) and \(\chi=1\) in Section~\ref{sec:affine_applications}.
The resulting regression functions are \(\mu_{0,0}=g\), \(\mu_{0,1}=g+\theta_0\), and \(m_0=g+\pi_0\theta_0\).
At \(\bx=(1/2,1/2)\), the parameter of interest equals \(1/8\) for both CATE and the conditional slope because the slope depends only on \(X\).

C0H replaces each error amplitude by \(\sigma(X,Z)=\sqrt3(1+X_1+Z_1)/28\), giving noise variance conditional on \(X\) of \((3X_1^2+9X_1+7)/784\), equal to \(1/64\) at the reporting point.
C2S retains C2's functions and noise, but each independent reporting coordinate has density
\[
 f(u)=\tfrac12+\1{u<1/4\ \text{or}\ u>3/4},\qquad 0\le u\le1.
\]
The joint point density is \(1/4\), with positive support everywhere, and uniform \(Z\) remains independent of \(X\).
Across the extensions, \(3/8\le\pi_0\le5/8\), outcomes stay bounded, and the true regressions remain inside the fixed prediction intervals.

The dictionary is \(p_n(X,Z)=q(X)\otimes \{\sqrt{J_n}\1{Z\in C_j}:1\le j\le J_n\}^{\top}\), with \(M_n=\max\{1,\floor{n^{1/12}}\}\), \(J_n=M_n^4\), and the equal half-open adjustment cells defined in Section~\ref{sec:affine_applications}.
C0, C1 and C0H use \(q_3\), the vector \(q\) from that section, while C2 and C2S use
\[
 q_5(X)=\{q_3(X)^{\top},\sqrt2\cos(8\pi X_1),
                         \sqrt2\cos(8\pi X_2)\}^{\top}.
\]
The reporting factors represent curvature exactly, leaving approximation in \(Z\).
CATE fits the propensity and arm means, while the slope fits the propensity and marginal outcome mean, both using \eqref{eq:affine_ridge_fit} with \(\lambda_n=n^{-1/2}\) and clipping propensity to \([1/4,3/4]\) and means to \([-1,1]\).
The dictionary, penalty and \(1/n\) normalizer stay fixed across fits, giving an effective five-fold penalty of approximately \(\lambda_n/0.8\), and predictions are pooled in the common final equation.

\begin{table}[!htb]
\centering\small
\begin{tabular}{@{}rrrrr@{}}
\hline
$n$ & $M$ & $K$ & $s$ & $N$ \\
\hline
512 & 1 & 3 & 107 & 8.54 \\
2\,048 & 1 & 3 & 304 & 12.46 \\
4\,096 & 2 & 48 & 512 & 14.99 \\
8\,192 & 2 & 48 & 861 & 18.02 \\
16\,384 & 2 & 48 & 1\,448 & 21.62 \\
65\,536 & 2 & 48 & 4\,096 & 31.00 \\
531\,441 & 3 & 243 & 19\,683 & 53.00 \\
\hline
\end{tabular}

\caption{C0 schedule. \(K=3M^4\), \(s=\floor{n^{3/4}}\),
\(N=(\sum_iw_i^2)^{-1}\), and \(\lambda_n=n^{-1/2}\).
The corresponding curved-design dimension is \(5M^4\).}
\label{tab:affine_production_schedule}
\end{table}

C1 uses \(n=2048,4096,16384\).
C2 uses \(n=512,2048,8192\) with order multipliers one and 0.4.
C0H and C2S use \(n=2048,8192\) with multiplier one.
Only C0 includes \(n=531{,}441\), where \(M=3\).
All five designs have the conditional point constants
\[
 \Omega_{\bx}^{\rm CATE}=\frac1{4\sqrt{15}},\qquad
 \Gamma_{\bx}^{\rm slope}=\frac{31}{128},\qquad
 \Sigma_{\bx}^{\rm slope}=\frac{31}{8192},\qquad
 \Omega_{\bx}^{\rm slope}=\frac2{31}.
\]
In C0H, independence of the noise-scale coordinate \(Z_1\) from the propensity coordinate \(Z_4\) preserves these point references.

To distinguish spatial averaging from estimation error, we compute sample-specific weighted targets that average the treatment effects over each realized neighborhood.
The weights follow the estimating equation: CATE uses the DNN weights directly, while the slope also weights by squared treatment residuals evaluated at the true propensity.
The resulting targets are
\begin{equation}\label{eq:affine_realized_weighted_truth}
a_w^{\rm CATE}=\sum_iw_i\theta_0(X_i),\qquad a_w^{\rm slope}=\frac{\sum_iw_i(T_i-\pi_{0i})^2\theta_0(X_i)}
                         {\sum_iw_i(T_i-\pi_{0i})^2},
 \qquad \pi_{0i}=\pi_0(Z_i).
\end{equation}

The Gaussian kernel uses \(\kappa_h(X)=\exp\{-\norm{X-\bx}_2^2/(2h^2)\}\) and \(w_i^K=\kappa_h(X_i)/\sum_j\kappa_h(X_j)\).
Its deterministic, design-specific bandwidth solves
\[
 n\frac{\{\Ex{\kappa_h(X)}\}^2}{\Ex{\kappa_h(X)^2}}
   =N^{\rm DNN}.
\]
The expectations use the reporting density and square boundary.
This matches the kernel's population effective sample size to the effective size determined by the DNN rank weights, while allowing realized effective sizes, neighborhood radii and spatial bias to differ.
Kernel jackknife deletions keep \(h\) fixed and renormalize retained weights, while DNN deletions keep \(s\) fixed.
Both freeze the parameter estimate and nuisance predictions.
The conventional covariance estimate below changes the uncertainty calculation while leaving the kernel point estimate fixed.
The conventional kernel variance estimate is
\[
 \hat V_K=\hat\Gamma^{-2}\sum_i(w_i^K)^2\hat\psi_i^2,
 \qquad \hat\Gamma=\sum_iw_i^K B(W_i;\hat\eta),
\]
evaluated at the parameter estimate.
This squared-weight score sandwich differs from the local-moment diagnostic, which uses \(\hat\Gamma^{-2}\sum_iw_i\hat\psi_i^2/N\).

Further performance, studentization and covariance diagnostics appear in \AffineExternalRef{supplement}{supp:affine_simulation_diagnostics}.

\section{SIPP Illustration: Computation and Localization}\label{app:affine_sipp}

The illustration uses 9,915 complete records from wave 4 of the 1990 SIPP panel, interviewed in 1991, for households with a reference person aged 25--64, an employed member and no self-employed member \citep{benjamin_401k_2003,chernozhukov_hdm_2016}.
Eligibility records employer access to a 401(k), and the statistical estimand averages the difference between the eligible and ineligible outcome regressions over household characteristics conditional on the reported age and income profile.
Under unconfounded eligibility given the adjustment variables and overlap, it identifies the mean causal eligibility effect on net financial assets, an accumulated stock measured in 1991 dollars, at that profile.
The calculation uses no survey weights or sampling-cluster identifiers.

Adjustment includes age and its square, asinh income and its square, education, family size, marriage, sex, two-earner status, defined-benefit pension coverage, home ownership and education-category indicators.
IRA participation is excluded.
Pension and housing variables may reflect sorting or responses to access, making their timing relevant to unconfoundedness.
Age and asinh income use the same standardization for observations and reporting points, with the latter defined using full-sample income percentiles.
The three interior incomes are \$19,413, \$31,476 and \$48,584 at the 25th, 50th and 75th percentiles.
Euclidean distance determines ranks, and observations at exactly equal distance share the average of their assigned rank weights.
The default is \(s=397\), with \(N=(\sum_iw_i^2)^{-1}\approx48.9\) at these profiles.

Each nuisance forest has 2,000 trees, with propensity clipped to \([0.01,0.99]\).
All-tree and OOB predictions use the same propensity forest and two outcome forests.
Five-fold predictions for held-out observations use fifteen additional forests trained on the corresponding retained samples, while all prediction methods share final observations and DNN weights.
The covariance calculation freezes the fitted signals and localization order.

\begin{table}[!htbp]
\centering\small
\begin{tabular}{@{}llrrr@{}}
\hline
Income & Prediction & Estimate & SE & Largest (\%) \\
\hline
25\% & All-tree & \(3.12\) & \(2.26\) & \(21.4\) \\
25\% & Single-fit/OOB & \(2.76\) & \(2.67\) & \(26.2\) \\
25\% & Five-fold & \(2.96\) & \(2.64\) & \(19.7\) \\
50\% & All-tree & \(6.22\) & \(4.73\) & \(60.2\) \\
50\% & Single-fit/OOB & \(5.93\) & \(5.39\) & \(62.4\) \\
50\% & Five-fold & \(6.83\) & \(5.86\) & \(67.9\) \\
75\% & All-tree & \(-2.65\) & \(16.45\) & \(83.2\) \\
75\% & Single-fit/OOB & \(-3.41\) & \(18.43\) & \(83.1\) \\
75\% & Five-fold & \(-3.51\) & \(18.87\) & \(83.8\) \\
\hline
\end{tabular}

\caption{Default eligibility contrasts at age 40 and the three interior
full-sample income percentiles. Estimates and SEs are in thousands of 1991
dollars.
Intervals in the main figure use the pointwise normal reference.
The last column gives the largest covariance share in percent.
All methods use \(s=397\).}
\label{tab:affine_sipp_core}
\end{table}

All nine default intervals in Table~\ref{tab:affine_sipp_core} include zero.
For the frozen CATE pseudo-values \(P_j\) from Section~\ref{sec:affine_results}, the fitted covariance share is
\begin{equation}\label{eq:affine_sipp_share}
 h_j=\frac{(P_j-\bar P)^2}{\sum_{\ell=1}^n(P_\ell-\bar P)^2}.
\end{equation}
Each share is indexed by the deleted household and includes the effect of re-ranking the retained observations.
At the upper profile, the leading control household has age 40, income \$48,402 and assets \$232,600, with fitted propensity between 0.45 and 0.50, and supplies about 83 percent of fitted variance under every nuisance prediction method.
Holding signals and \(s\) fixed, its deletion changes the OOB estimate to \$13,392.

\begin{table}[!htbp]
\centering\small\setlength{\tabcolsep}{4pt}
\begin{tabular}{@{}lrrrrrr@{}}
\hline
Income & $s$ & Estimate & SE & $N$ & Radius & Largest (\%) \\
\hline
25\% & \(397\) & \(2.76\) & \(2.67\) & \(48.9\) & \(0.1501\) & \(26.2\) \\
25\% & \(99\) & \(4.57\) & \(1.77\) & \(198.3\) & \(0.2907\) & \(42.2\) \\
50\% & \(397\) & \(5.93\) & \(5.39\) & \(48.9\) & \(0.1143\) & \(62.4\) \\
50\% & \(99\) & \(7.30\) & \(3.58\) & \(198.3\) & \(0.2247\) & \(51.3\) \\
75\% & \(397\) & \(-3.41\) & \(18.43\) & \(48.9\) & \(0.1031\) & \(83.1\) \\
75\% & \(99\) & \(1.91\) & \(5.72\) & \(198.3\) & \(0.2108\) & \(51.7\) \\
\hline
\end{tabular}

\caption{Selected single-fit/OOB localization comparison with fitted
predictions held fixed. Estimates and SEs are in thousands of 1991 dollars.
Radius contains 90 percent of DNN weight in standardized reporting
coordinates.
The final column is the largest covariance share in
\eqref{eq:affine_sipp_share}. The reference order is \(s=397\).
The order \(s=99\) is exploratory.}
\label{tab:affine_sipp_localization_core}
\end{table}

Broader pooling reduces uncertainty at the upper profile, but its effect on concentration varies across profiles (Table~\ref{tab:affine_sipp_localization_core}).
At the upper profile and order 99, the neighborhood containing 90 percent of the DNN weight spans observed ages 38--42 and incomes \$41,535--\$56,994, compared with ages 39--41 and incomes \$45,231--\$52,350 at order 397.
The largest covariance share falls from about 83 to 52 percent, while at the lower profile it rises from 26 to 42 percent.
Under OOB prediction, the intervals exclude zero at the lower profile for orders 99 and 149 and at the median profile for order 99.
These smaller orders are exploratory, and we do not measure the spatial bias they introduce.

These empirical comparisons fall outside the scope of Theorem~\ref{thm:affine_inference}.
Its learner conditions have not been verified for regression forests, and the bounded coefficient envelope in Condition~\ref{cond:affine_bounded_learner} excludes unbounded dollar outcomes because \(b=-\phi\) contains \(Y\).
The reporting design also differs from the theorem: integer-valued age violates the two-dimensional support and small-ball requirements of Condition~\ref{cond:affine_design}, and the reporting profiles and coordinate scales are estimated from the sample.
We therefore interpret the intervals as illustrative normal-reference comparisons.

Further localization and covariance comparisons, influential observations and transformed-outcome results appear in \AffineExternalRef{supplement}{supp:affine_sipp}.

\FloatBarrier
\section*{Author declarations}

\subsection*{Funding}
This research did not receive any specific grant from funding agencies in the public, commercial, or not-for-profit sectors.

\subsection*{Declaration of competing interests}
The author declares no financial or personal competing interests.

\subsection*{Data and code availability}
The SIPP illustration uses the publicly available \texttt{hdm::pension} dataset.
Simulation replication code and saved numerical evidence are available in the manuscript replication bundle of the \texttt{RdistNN} R package at \url{https://github.com/JakobJuergens/RdistNN}.
The \texttt{RdistNN} R package, its documentation, and the Stata companion are available through \url{https://jakobjuergens.com/code-software/}.

\subsection*{Declaration of generative AI and AI-assisted technologies in the manuscript preparation process}
The author used OpenAI ChatGPT and Codex for language editing and proofreading, mathematical development and proof review, and assistance writing and checking simulation and diagnostic code.
The author reviewed the manuscript in detail and takes full responsibility for its arguments, proofs, computational checks, and other content.

\section*{Acknowledgments}
The author thanks Jack R. Porter, Harold D. Chiang, Bruce E. Hansen, and participants at Camp Econometrics XIX and XX, the 2026 annual meetings of the CEA and MEG, and the 2026 CESG meeting for helpful comments and discussions.

\FloatBarrier
\bibliographystyle{elsarticle-harv}
\bibliography{bibliography,affine_scores}
\end{document}


\pagestyle{plain}
\setcounter{page}{37}
\begin{frontmatter}
  \title{Supplementary material for\\\AffinePaperTitle}
  \author{\CoupledPaperAuthor}
  \expandafter\affiliation\expandafter{\CoupledPaperAffiliation}
  \ead{\CoupledPaperEmail}
  \ead[url]{\CoupledPaperWebsite}
  \begin{abstract}
  The Survey of Income and Program Participation (SIPP) application examines
  how similar 401(k) eligibility estimates can coexist with substantial and
  uneven local uncertainty. We report the full reporting grid and localization
  comparisons, alongside simulations relating pointwise coverage to nuisance
  approximation, spatial bias, studentization and covariance estimation.
  The theoretical results, learner
  verification, simulation design and core numerical evidence appear in the paper.
  \end{abstract}
\end{frontmatter}
\section{SIPP Eligibility Contrasts and Local Information}\label{supp:affine_sipp}

The raw-dollar comparisons in \AffineExternalRef{paper}{sec:affine_sipp} use the same 9,915 households, adjustment variables, profiles and final DNN weights across all-tree, single-fit out-of-bag (OOB) and five-fold predictions.
We extend the comparison to all seven profiles and eight orders, examining local information and localization sensitivity with nuisance predictions fixed.

\subsection{Data, parameter of interest and nuisance estimation}

The data distributed as \texttt{hdm::pension} come from wave 4 of the 1990 SIPP panel, interviewed in February--May 1991, and cover households with a reference person aged 25--64, at least one employed member and no self-employed member \citep[pp.~1261--1262]{benjamin_401k_2003}.
Eligibility records whether a household member works for an employer offering a 401(k).
Participation is a separate choice.
The 9,915 complete records include nonpositive incomes, which the inverse-hyperbolic-sine transformation accommodates, and negative net financial assets.
Estimation and the reported intervals use the unweighted household sample without survey-design or sampling-cluster adjustments.

Net financial assets, measured in 1991 dollars, are an accumulated stock comprising financial retirement assets and other financial assets net of nonmortgage debt.
Under overlap and unconfounded eligibility given the adjustment variables, the parameter of interest is \(\tau_0(\bx)=\Ex*{Y(1)-Y(0)\given X=\bx}\), with age and transformed income as reporting coordinates.
At each profile, the effect averages over the conditional distribution of the remaining household characteristics.
Adjustment uses age and its square, asinh income and its square, years of education, family size, marriage, sex, two-earner status, defined-benefit pension coverage, home ownership and four education-category indicators.
IRA participation is excluded.
Pension and housing variables may reflect sorting or responses to plan access \citep{benjamin_401k_2003}, so their timing matters for unconfoundedness under this adjustment set.

Reporting profiles hold age at 40 and use the full-sample income percentiles 5th, 10th, 25th, 50th, 75th, 90th and 95th.
We standardize observations and profiles identically: age is centered at 41.0602 and divided by 10.3445, while asinh income is centered at 10.9896 and divided by 0.776351.
Euclidean distance in these coordinates determines ranks, and observations at exactly equal distance share the average of their assigned rank weights.
The default uses \(s=397\) and \(N=(\sum_iw_i^2)^{-1}\approx48.9\), with all three prediction methods sharing the final observations and weights.

Each nuisance regression uses 2,000 trees.
Propensity predictions are clipped to \([0.01,0.99]\), and their unclipped values are saved for diagnostics.
All-tree and OOB predictions share three fitted forests, for propensity and the two outcome-arm regressions.
OOB predictions evaluate the propensity regression and the outcome regression for an observation's treatment group at its covariates, using only trees trained without that observation.
The regression for the other treatment group was not trained on that observation.
Five-fold predictions use separate forests trained outside each evaluation fold.

\subsection{Eligibility contrasts and local information}

At the three interior profiles used in the paper, all nine pointwise normal-reference intervals include zero.
Table~\ref{tab:affine_sipp} extends the comparison to the full reporting grid.

\begin{table}[H]
\caption{Default eligibility contrasts in net financial assets at age 40.
Income denotes the full-sample percentile. Estimates, SEs and intervals are
in thousands of 1991 dollars. The last column gives the largest covariance
share across household deletions, in percent. All methods use \(s=397\).}
\label{tab:affine_sipp}
\centering\small
\begin{tabular}{@{}llrrrr@{}}
\hline
Income & Prediction & Estimate & SE & 95\% interval & Largest (\%) \\
\hline
5\% & All-tree & \(19.79\) & \(12.09\) & \([-3.91,\ 43.48]\) & \(58.2\) \\
5\% & Single-fit/OOB & \(24.78\) & \(16.29\) & \([-7.15,\ 56.71]\) & \(52.5\) \\
5\% & Five-fold & \(24.68\) & \(16.07\) & \([-6.82,\ 56.18]\) & \(61.3\) \\
10\% & All-tree & \(12.91\) & \(6.83\) & \([-0.47,\ 26.29]\) & \(77.9\) \\
10\% & Single-fit/OOB & \(14.62\) & \(8.29\) & \([-1.62,\ 30.87]\) & \(70.4\) \\
10\% & Five-fold & \(14.16\) & \(7.61\) & \([-0.75,\ 29.07]\) & \(78.1\) \\
25\% & All-tree & \(3.12\) & \(2.26\) & \([-1.31,\ 7.55]\) & \(21.4\) \\
25\% & Single-fit/OOB & \(2.76\) & \(2.67\) & \([-2.47,\ 7.99]\) & \(26.2\) \\
25\% & Five-fold & \(2.96\) & \(2.64\) & \([-2.22,\ 8.13]\) & \(19.7\) \\
50\% & All-tree & \(6.22\) & \(4.73\) & \([-3.04,\ 15.49]\) & \(60.2\) \\
50\% & Single-fit/OOB & \(5.93\) & \(5.39\) & \([-4.64,\ 16.50]\) & \(62.4\) \\
50\% & Five-fold & \(6.83\) & \(5.86\) & \([-4.65,\ 18.30]\) & \(67.9\) \\
75\% & All-tree & \(-2.65\) & \(16.45\) & \([-34.89,\ 29.58]\) & \(83.2\) \\
75\% & Single-fit/OOB & \(-3.41\) & \(18.43\) & \([-39.53,\ 32.72]\) & \(83.1\) \\
75\% & Five-fold & \(-3.51\) & \(18.87\) & \([-40.49,\ 33.47]\) & \(83.8\) \\
90\% & All-tree & \(2.85\) & \(14.61\) & \([-25.78,\ 31.48]\) & \(50.8\) \\
90\% & Single-fit/OOB & \(2.09\) & \(16.01\) & \([-29.29,\ 33.46]\) & \(48.5\) \\
90\% & Five-fold & \(1.41\) & \(15.62\) & \([-29.21,\ 32.02]\) & \(45.7\) \\
95\% & All-tree & \(-20.82\) & \(56.68\) & \([-131.90,\ 90.26]\) & \(94.8\) \\
95\% & Single-fit/OOB & \(-23.72\) & \(63.10\) & \([-147.38,\ 99.95]\) & \(94.8\) \\
95\% & Five-fold & \(-22.40\) & \(65.09\) & \([-149.97,\ 105.17]\) & \(95.1\) \\
\hline
\end{tabular}

\end{table}

Table~\ref{tab:affine_sipp_information} describes the treatment support and covariance concentration underlying local uncertainty, with eligible mass, \(\sum_iw_iT_i\), measuring the local weight on eligible households.
For inverse-probability weights \(a_{1i}=w_iT_i/\hat\pi_i\) and \(a_{0i}=w_i(1-T_i)/(1-\hat\pi_i)\), the Kish effective sample size \((\sum_i a_{ti})^2/\sum_i a_{ti}^2\) measures weight concentration, excluding outcome-residual variation.
Local weight on clipped propensities is zero at every default profile under all three prediction methods.

\begin{samepage}
The CATE pseudo-values \(P_j\) in \AffineExternalRef{paper}{sec:affine_results} give the fitted covariance share
\[
 h_j=\frac{(P_j-\bar P)^2}{\sum_{\ell=1}^n(P_\ell-\bar P)^2}.
\]
Each share is indexed by the deleted household and includes the effect of its removal and re-ranking the retained observations, as in \AffineExternalRef{paper}{eq:affine_sipp_share}.
\end{samepage}

\begin{table}[H]
\caption{Default single-fit/OOB local information. Radius contains 90 percent
of DNN weight in standardized reporting distance. IPW columns give the
Kish effective sample sizes defined in the text.
The final column sums the five
largest shares in \AffineExternalRef{paper}{eq:affine_sipp_share}. These summaries complement
\(N\) by describing treatment support and covariance concentration.}
\label{tab:affine_sipp_information}
\centering\small
\begin{tabular}{@{}lrrrrr@{}}
\hline
Income & Radius & Eligible mass & IPW eligible & IPW control & Top five (\%) \\
\hline
5\% & \(0.2419\) & \(0.136\) & \(4.6\) & \(42.5\) & \(96.5\) \\
10\% & \(0.1950\) & \(0.159\) & \(5.6\) & \(41.7\) & \(97.9\) \\
25\% & \(0.1501\) & \(0.164\) & \(8.8\) & \(37.2\) & \(72.7\) \\
50\% & \(0.1143\) & \(0.256\) & \(17.4\) & \(31.5\) & \(77.5\) \\
75\% & \(0.1031\) & \(0.445\) & \(24.4\) & \(21.9\) & \(93.4\) \\
90\% & \(0.1346\) & \(0.658\) & \(34.1\) & \(14.4\) & \(82.1\) \\
95\% & \(0.1953\) & \(0.677\) & \(31.1\) & \(17.5\) & \(97.8\) \\
\hline
\end{tabular}

\end{table}

Table~\ref{tab:affine_sipp_cases} identifies households with the largest OOB shares and reports estimates after their deletion, with fitted signals and \(s\) fixed.

\begin{table}[H]
\caption{Households with the largest OOB covariance shares at the seven default profiles.
Assets and the frozen retained-sample estimate are in thousands of 1991
dollars. The reported propensity is fitted, and weight is a percentage.
Deleting a row changes retained ranks while keeping nuisance signals fixed.}
\label{tab:affine_sipp_cases}
\centering\small\setlength{\tabcolsep}{3pt}
\begin{tabular}{@{}lrrrrrr@{}}
\hline
Income & Eligible & Assets & Propensity & Rank & Weight (\%) & Deleted estimate \\
\hline
5\% & \(1\) & \(108.22\) & \(0.068\) & \(41\) & \(0.78\) & \(12.98\) \\
10\% & \(1\) & \(28.27\) & \(0.087\) & \(9\) & \(2.89\) & \(7.67\) \\
25\% & \(1\) & \(-3.80\) & \(0.161\) & \(25\) & \(1.50\) & \(4.13\) \\
50\% & \(1\) & \(112.75\) & \(0.307\) & \(30\) & \(1.23\) & \(1.67\) \\
75\% & \(0\) & \(232.60\) & \(0.494\) & \(1\) & \(4.00\) & \(13.39\) \\
90\% & \(1\) & \(295.90\) & \(0.609\) & \(13\) & \(2.45\) & \(-9.07\) \\
95\% & \(0\) & \(1,006.50\) & \(0.545\) & \(9\) & \(2.89\) & \(37.73\) \\
\hline
\end{tabular}

\end{table}

\subsection{Prediction differences and score components}

The prediction methods use the same observations and final DNN weights, so their estimates share sources of sampling variation.
With nuisance predictions fixed, we use paired pseudo-value calculations to account for this dependence when comparing the estimates.
For two fitted signals, the estimated variance of the difference between their local averages is \(V_1+V_2-2C_{12}\), where \(C_{12}\) is the covariance estimate obtained from their frozen pseudo-values:
\[
C_{12}=\frac{1}{n(n-1)}\sum_{j=1}^n
(P_{1j}-\bar P_1)(P_{2j}-\bar P_2),
\qquad
V_a=\frac{1}{n(n-1)}\sum_{j=1}^n(P_{aj}-\bar P_a)^2.
\]
At the three interior profiles, the centered pseudo-values from OOB and five-fold predictions have correlations of 0.9929, 0.9964 and 0.9991.
Accounting for this shared variation gives paired standard errors for the differences between the estimates of \$318, \$667 and \$907, with nuisance predictions held fixed.
The small differences between local estimates can also reflect cancellation across fitted signals: at the upper profile, moving from OOB to five-fold predictions changes the estimate by \(-\$105\), even though the DNN-weighted absolute changes in the signals sum to \$4,466.
The individual fitted signals therefore move more than the final contrast alone would suggest.
Table~\ref{tab:affine_sipp_pairs} reports all three pairwise comparisons at every profile.

\begin{table}[H]
\caption{Paired comparisons of the default nuisance predictions. Estimate
differences (Gap), their standard errors (Gap SE) and weighted sums of absolute
changes in fitted signals (Absolute movement) are in thousands of 1991 dollars.
Differences follow the order in the second column. Correlations use centered
frozen pseudo-values.
Standard errors include covariance between the estimates, with nuisance
predictions held fixed.}
\label{tab:affine_sipp_pairs}
\centering\footnotesize\setlength{\tabcolsep}{3pt}
\begin{tabular}{@{}llrrrr@{}}
\hline
Income & Prediction difference & Gap & Gap SE & Absolute movement & Correlation \\
\hline
5\% & Single-fit/OOB $-$ All-tree & \(4.99\) & \(4.30\) & \(7.07\) & \(0.9979\) \\
5\% & Five-fold $-$ All-tree & \(4.89\) & \(4.02\) & \(6.93\) & \(0.9992\) \\
5\% & Five-fold $-$ Single-fit/OOB & \(-0.10\) & \(1.59\) & \(3.47\) & \(0.9953\) \\
10\% & Single-fit/OOB $-$ All-tree & \(1.71\) & \(1.62\) & \(3.15\) & \(0.9956\) \\
10\% & Five-fold $-$ All-tree & \(1.25\) & \(0.82\) & \(2.06\) & \(0.9993\) \\
10\% & Five-fold $-$ Single-fit/OOB & \(-0.47\) & \(1.04\) & \(2.41\) & \(0.9951\) \\
25\% & Single-fit/OOB $-$ All-tree & \(-0.36\) & \(0.49\) & \(1.36\) & \(0.9936\) \\
25\% & Five-fold $-$ All-tree & \(-0.16\) & \(0.51\) & \(1.75\) & \(0.9906\) \\
25\% & Five-fold $-$ Single-fit/OOB & \(0.20\) & \(0.32\) & \(1.09\) & \(0.9929\) \\
50\% & Single-fit/OOB $-$ All-tree & \(-0.29\) & \(0.70\) & \(2.48\) & \(0.9991\) \\
50\% & Five-fold $-$ All-tree & \(0.60\) & \(1.24\) & \(3.40\) & \(0.9951\) \\
50\% & Five-fold $-$ Single-fit/OOB & \(0.89\) & \(0.67\) & \(2.23\) & \(0.9964\) \\
75\% & Single-fit/OOB $-$ All-tree & \(-0.75\) & \(2.04\) & \(6.41\) & \(0.9996\) \\
75\% & Five-fold $-$ All-tree & \(-0.86\) & \(2.58\) & \(8.40\) & \(0.9987\) \\
75\% & Five-fold $-$ Single-fit/OOB & \(-0.10\) & \(0.91\) & \(4.47\) & \(0.9991\) \\
90\% & Single-fit/OOB $-$ All-tree & \(-0.77\) & \(1.66\) & \(7.75\) & \(0.9983\) \\
90\% & Five-fold $-$ All-tree & \(-1.44\) & \(1.60\) & \(8.53\) & \(0.9966\) \\
90\% & Five-fold $-$ Single-fit/OOB & \(-0.68\) & \(1.10\) & \(5.46\) & \(0.9979\) \\
95\% & Single-fit/OOB $-$ All-tree & \(-2.90\) & \(6.47\) & \(14.70\) & \(0.9999\) \\
95\% & Five-fold $-$ All-tree & \(-1.58\) & \(8.67\) & \(21.47\) & \(0.9994\) \\
95\% & Five-fold $-$ Single-fit/OOB & \(1.32\) & \(2.87\) & \(12.19\) & \(0.9995\) \\
\hline
\end{tabular}

\end{table}

The signal decomposition in \AffineExternalRef{paper}{eq:affine_cate_signal} relates these differences to regression and residual corrections:
\begin{align}
 r_i&=\hat\mu_1(V_i)-\hat\mu_0(V_i),\notag\\
 u_i&=\frac{T_i\{Y_i-\hat\mu_1(V_i)\}}{\hat\pi(V_i)},\notag\\
 v_i&=-\frac{(1-T_i)\{Y_i-\hat\mu_0(V_i)\}}{1-\hat\pi(V_i)},\notag\\
 \hat\phi_i&=r_i+u_i+v_i.
 \label{eq:affine_sipp_components}
\end{align}
Table~\ref{tab:affine_sipp_components} reports the local component means, showing how the upper-profile OOB regression contrast of \$6,613 is offset by eligible and ineligible residual corrections of \(-\$2,748\) and \(-\$7,270\) to give the adjusted estimate.

\begin{table}[H]
\caption{Default local CATE signal component means, in thousands of 1991 dollars.
The three components in \eqref{eq:affine_sipp_components} sum to the estimate.
Rounding can affect the displayed sum.}
\label{tab:affine_sipp_components}
\centering\small\setlength{\tabcolsep}{3pt}
\begin{tabular}{@{}llrrrr@{}}
\hline
Income & Prediction & Contrast & Eligible correction & Control correction & Total \\
\hline
5\% & All-tree & \(6.37\) & \(13.23\) & \(0.19\) & \(19.79\) \\
5\% & Single-fit/OOB & \(6.32\) & \(18.25\) & \(0.21\) & \(24.78\) \\
5\% & Five-fold & \(6.10\) & \(18.26\) & \(0.32\) & \(24.68\) \\
10\% & All-tree & \(5.94\) & \(7.10\) & \(-0.13\) & \(12.91\) \\
10\% & Single-fit/OOB & \(5.90\) & \(8.89\) & \(-0.16\) & \(14.62\) \\
10\% & Five-fold & \(5.94\) & \(8.34\) & \(-0.13\) & \(14.16\) \\
25\% & All-tree & \(3.90\) & \(-1.31\) & \(0.53\) & \(3.12\) \\
25\% & Single-fit/OOB & \(3.90\) & \(-1.76\) & \(0.62\) & \(2.76\) \\
25\% & Five-fold & \(3.96\) & \(-1.65\) & \(0.64\) & \(2.96\) \\
50\% & All-tree & \(8.12\) & \(-1.93\) & \(0.03\) & \(6.22\) \\
50\% & Single-fit/OOB & \(8.29\) & \(-2.34\) & \(-0.01\) & \(5.93\) \\
50\% & Five-fold & \(7.87\) & \(-1.52\) & \(0.48\) & \(6.83\) \\
75\% & All-tree & \(6.24\) & \(-2.22\) & \(-6.68\) & \(-2.65\) \\
75\% & Single-fit/OOB & \(6.61\) & \(-2.75\) & \(-7.27\) & \(-3.41\) \\
75\% & Five-fold & \(6.61\) & \(-2.15\) & \(-7.96\) & \(-3.51\) \\
90\% & All-tree & \(13.65\) & \(-7.91\) & \(-2.88\) & \(2.85\) \\
90\% & Single-fit/OOB & \(14.16\) & \(-8.76\) & \(-3.31\) & \(2.09\) \\
90\% & Five-fold & \(14.74\) & \(-11.23\) & \(-2.10\) & \(1.41\) \\
95\% & All-tree & \(7.83\) & \(7.71\) & \(-36.36\) & \(-20.82\) \\
95\% & Single-fit/OOB & \(9.10\) & \(8.35\) & \(-41.17\) & \(-23.72\) \\
95\% & Five-fold & \(11.21\) & \(8.64\) & \(-42.24\) & \(-22.40\) \\
\hline
\end{tabular}

\end{table}

The same signal components help explain the estimated uncertainty, because linearity of frozen deletion permits the covariance to be decomposed into their variances and twice each pairwise covariance (Table~\ref{tab:affine_sipp_component_covariance}).
At the upper OOB profile, fitted variance is concentrated in the control residual correction: its diagonal contribution is 309.546, compared with 38.236 for the eligible correction and 0.372 for the regression contrast, in units of \(10^6\) dollars squared.
The component covariances partly offset these separate variances: the off-diagonal entries sum to \(-8.495\), reducing the total to 339.659, the reported SE squared.
This decomposition describes which parts of the signal contribute to fitted variance, while \AffineExternalRef{paper}{eq:affine_sipp_share} describes how that variance is concentrated across deletion indices.

\begin{table}[H]
\caption{Frozen jackknife component covariance, in units of \(10^6\) dollars
squared. Here \(V_r,V_u,V_v\) are variances of the local component averages,
and each \(2C\) column includes both symmetric covariance entries.
All six contributions sum to total variance before rounding.}
\label{tab:affine_sipp_component_covariance}
\centering\ifpreprint\scriptsize\else\footnotesize\fi\setlength{\tabcolsep}{3pt}
\begin{tabular}{@{}llrrrrrrr@{}}
\hline
Income & Prediction & $V_r$ & $V_u$ & $V_v$ & $2C_{ru}$ & $2C_{rv}$ & $2C_{uv}$ & Total \\
\hline
5\% & All-tree & \(0.1\) & \(144.8\) & \(0.2\) & \(1.1\) & \(-0.0\) & \(-0.1\) & \(146.2\) \\
5\% & Single-fit/OOB & \(0.1\) & \(266.6\) & \(0.2\) & \(-1.3\) & \(-0.1\) & \(-0.2\) & \(265.4\) \\
5\% & Five-fold & \(0.1\) & \(258.3\) & \(0.2\) & \(-0.1\) & \(-0.1\) & \(-0.2\) & \(258.3\) \\
10\% & All-tree & \(0.0\) & \(45.2\) & \(0.2\) & \(1.1\) & \(0.0\) & \(0.1\) & \(46.6\) \\
10\% & Single-fit/OOB & \(0.0\) & \(68.0\) & \(0.2\) & \(0.3\) & \(-0.0\) & \(0.1\) & \(68.7\) \\
10\% & Five-fold & \(0.1\) & \(57.1\) & \(0.2\) & \(0.4\) & \(-0.0\) & \(0.1\) & \(57.9\) \\
25\% & All-tree & \(0.0\) & \(3.2\) & \(1.6\) & \(0.1\) & \(0.1\) & \(-0.0\) & \(5.1\) \\
25\% & Single-fit/OOB & \(0.0\) & \(5.3\) & \(2.0\) & \(-0.1\) & \(-0.1\) & \(-0.0\) & \(7.1\) \\
25\% & Five-fold & \(0.0\) & \(5.1\) & \(1.9\) & \(0.0\) & \(-0.1\) & \(0.0\) & \(7.0\) \\
50\% & All-tree & \(0.2\) & \(16.9\) & \(4.6\) & \(0.7\) & \(-0.1\) & \(0.0\) & \(22.3\) \\
50\% & Single-fit/OOB & \(0.2\) & \(24.0\) & \(5.9\) & \(-0.4\) & \(-0.6\) & \(0.0\) & \(29.1\) \\
50\% & Five-fold & \(0.1\) & \(28.7\) & \(6.1\) & \(-0.3\) & \(-0.4\) & \(0.0\) & \(34.3\) \\
75\% & All-tree & \(0.6\) & \(29.0\) & \(224.8\) & \(0.6\) & \(17.4\) & \(-2.0\) & \(270.5\) \\
75\% & Single-fit/OOB & \(0.4\) & \(38.2\) & \(309.5\) & \(-3.1\) & \(-2.6\) & \(-2.8\) & \(339.7\) \\
75\% & Five-fold & \(0.5\) & \(38.6\) & \(328.9\) & \(-4.6\) & \(-5.0\) & \(-2.4\) & \(356.0\) \\
90\% & All-tree & \(1.1\) & \(169.4\) & \(31.5\) & \(13.2\) & \(0.0\) & \(-1.9\) & \(213.4\) \\
90\% & Single-fit/OOB & \(1.0\) & \(214.1\) & \(51.7\) & \(-4.6\) & \(-3.3\) & \(-2.5\) & \(256.3\) \\
90\% & Five-fold & \(1.1\) & \(207.6\) & \(50.5\) & \(-7.9\) & \(-5.8\) & \(-1.7\) & \(243.9\) \\
95\% & All-tree & \(5.2\) & \(121.3\) & \(2,943.6\) & \(3.3\) & \(122.6\) & \(16.1\) & \(3,212.1\) \\
95\% & Single-fit/OOB & \(4.7\) & \(156.6\) & \(3,923.5\) & \(-12.1\) & \(-111.4\) & \(19.9\) & \(3,981.2\) \\
95\% & Five-fold & \(5.5\) & \(150.5\) & \(4,146.6\) & \(-4.0\) & \(-84.0\) & \(21.8\) & \(4,236.5\) \\
\hline
\end{tabular}

\end{table}

The leading household deletion helps interpret the large control-residual contribution.
The household with the largest upper-profile covariance share is ineligible, aged 40, with income \$48,402 and assets \$232,600 (Table~\ref{tab:affine_sipp_upper_case}), so its eligible residual correction is zero.
Relative to OOB prediction, the higher all-tree control mean reduces the magnitude of its control residual, while the propensity change also affects the fitted signal.
These changes in predictions leave its influence substantial: the household accounts for 83.20, 83.05 and 83.84 percent of fitted variance under all-tree, OOB and five-fold prediction.
Deleting it with predictions frozen gives retained-sample estimates of \$12,347, \$13,392 and \$13,767, showing that the sensitivity to this household persists across the tested prediction methods.

\begin{table}[H]
\caption{Predictions for the leading upper-profile ineligible household.
All monetary entries are in thousands of 1991 dollars. The control correction
\(v_i\) is defined in \eqref{eq:affine_sipp_components}.
Propensity is fitted
and all methods use the same observed assets.}
\label{tab:affine_sipp_upper_case}
\centering\ifpreprint\scriptsize\else\footnotesize\fi\setlength{\tabcolsep}{3pt}
\begin{tabular}{@{}lrrrrrr@{}}
\hline
Prediction & Propensity & Control mean & Eligible mean & Contrast & Control correction & Signal \\
\hline
All-tree & \(0.453\) & \(38.73\) & \(30.73\) & \(-8.00\) & \(-354.32\) & \(-362.32\) \\
Single-fit/OOB & \(0.494\) & \(23.34\) & \(30.73\) & \(7.39\) & \(-413.49\) & \(-406.10\) \\
Five-fold & \(0.492\) & \(15.46\) & \(25.37\) & \(9.91\) & \(-427.62\) & \(-417.71\) \\
\hline
\end{tabular}

\end{table}

\subsection{Sensitivity to localization}

\begin{figure}[H]
\centering
\begingroup%
\makeatletter%
\let\Gread@transgrouptrue\Gread@transgroupfalse%
\makeatother%
\includegraphics[width=\textwidth]{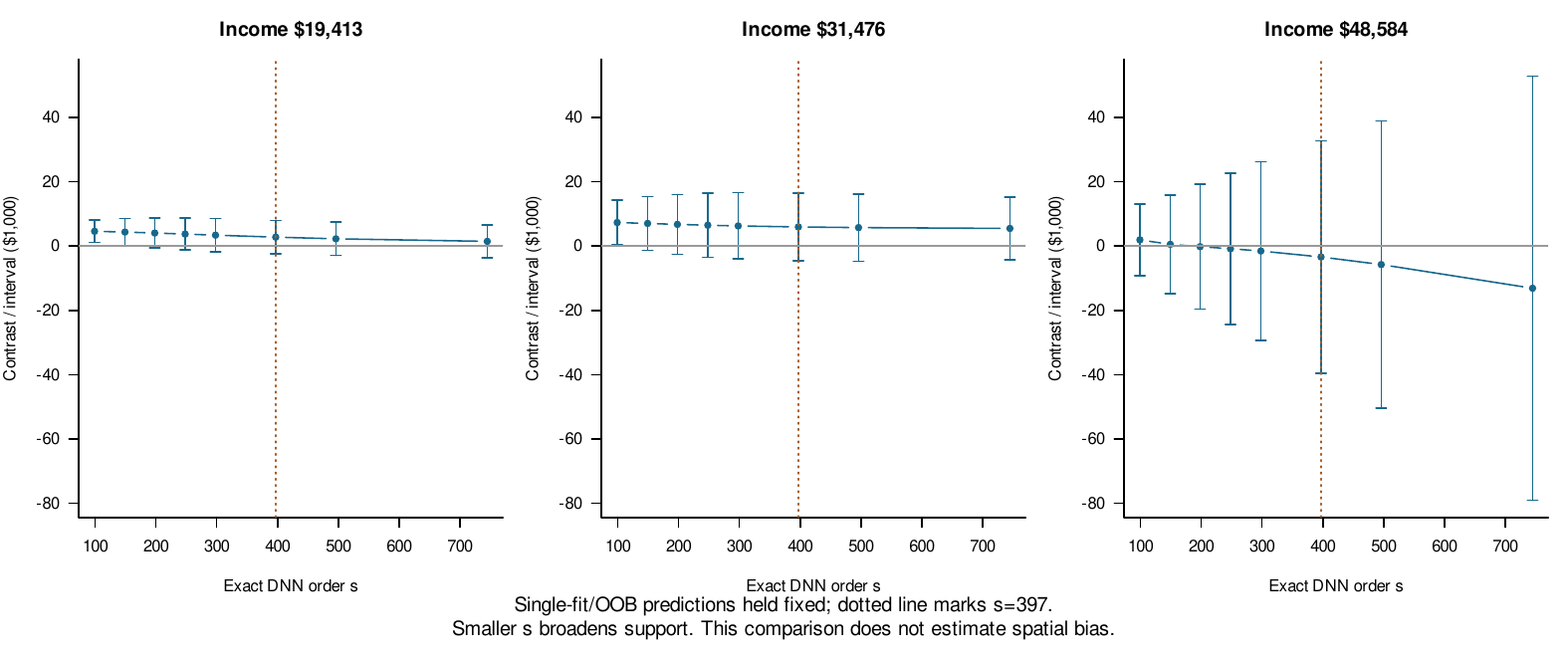}%
\endgroup
\caption{Localization sensitivity under single-fit/OOB prediction at the three
interior profiles, with nuisance predictions held fixed. Bars are pointwise
normal-reference intervals.
The dotted line marks the default \(s=397\).
All profiles use the same dollar scale. Four smaller orders are exploratory.}
\label{fig:affine_sipp_orders}
\end{figure}

The prediction comparisons keep the neighborhood fixed; changing the DNN order instead shows how precision and influence depend on the extent of local averaging.
With nuisance predictions fixed, we compare the nearby orders \(s=298,397,496,745\) with smaller orders 99, 149, 198 and 248, explored in response to the observed variance concentration at the default order.
Figure~\ref{fig:affine_sipp_orders} shows the OOB paths for the three interior profiles, while Tables \ref{tab:affine_sipp_orders_1}--\ref{tab:affine_sipp_orders_3} compare all three prediction methods at each of the seven profiles and eight orders, with \(s=397\) as the reference specification.

At the upper profile, order \(s=99\) assigns weight \(99/9915=0.009985\) to the nearest household, with local effective sample size \(N=198.3\).
The increase in effective size comes from spreading weight over a broader neighborhood.
The radius containing 90 percent of the DNN weight is 0.2108 in standardized reporting coordinates.
It spans observed ages 38--42 and incomes \$41,535--\$56,994, compared with ages 39--41 and incomes \$45,231--\$52,350 at order 397.
The eligible and control IPW Kish effective sample sizes rise from 24.4 and 21.9 to 101.0 and 84.6, respectively, yet the same household still accounts for 51.7 percent of fitted variance.

Broader pooling therefore lowers uncertainty at the upper profile while leaving more than half the fitted variance associated with the same household; its effect on concentration also varies across profiles.
At the lower profile, reducing the order from 397 to 99 lowers SE from \$2,668 to \$1,773 while raising the largest share from 26.2 to 42.2 percent.
At the median profile, the share falls from 62.4 to 51.3 percent, with nonmonotonic changes at intermediate orders.
Under OOB prediction, the intervals exclude zero at the lower profile for orders 99 and 149 and at the median profile for order 99.
The precision gains still come with broader neighborhoods and potentially more spatial bias, which this comparison leaves unmeasured.

\begin{table}[H]
\caption{Localization sensitivity at income percentiles 5--25.
Each prediction column gives estimate (SE) in thousands of 1991 dollars.
\(N\) is the local effective sample size computed from the DNN weights.
Radius contains 90 percent of DNN weight.
The final column gives the largest OOB covariance share. All three
prediction methods share weights at each profile and order.}
\label{tab:affine_sipp_orders_1}
\centering\footnotesize\setlength{\tabcolsep}{3pt}
\begin{tabular}{@{}lrrrrrrr@{}}
\hline
Income & $s$ & $N$ & All-tree & Single-fit/OOB & Five-fold & Radius & Largest (\%) \\
\hline
5\% & \(99\) & \(198.3\) & \(16.06\;(8.31)\) & \(19.51\;(10.75)\) & \(19.61\;(11.25)\) & \(0.5025\) & \(88.4\) \\
5\% & \(149\) & \(131.7\) & \(18.11\;(10.29)\) & \(22.22\;(13.35)\) & \(22.44\;(13.93)\) & \(0.4124\) & \(86.3\) \\
5\% & \(198\) & \(98.9\) & \(19.15\;(11.41)\) & \(23.62\;(14.86)\) & \(23.89\;(15.42)\) & \(0.3767\) & \(82.2\) \\
5\% & \(248\) & \(78.8\) & \(19.68\;(11.97)\) & \(24.37\;(15.69)\) & \(24.62\;(16.14)\) & \(0.3204\) & \(76.5\) \\
5\% & \(298\) & \(65.4\) & \(19.88\;(12.17)\) & \(24.71\;(16.08)\) & \(24.88\;(16.35)\) & \(0.2956\) & \(69.5\) \\
5\% & \(397\) & \(48.9\) & \(19.79\;(12.09)\) & \(24.78\;(16.29)\) & \(24.68\;(16.07)\) & \(0.2419\) & \(52.5\) \\
5\% & \(496\) & \(38.9\) & \(19.44\;(11.96)\) & \(24.54\;(16.49)\) & \(24.12\;(15.71)\) & \(0.2158\) & \(57.1\) \\
5\% & \(745\) & \(25.6\) & \(18.96\;(12.90)\) & \(24.42\;(18.51)\) & \(23.20\;(16.58)\) & \(0.1942\) & \(85.7\) \\
10\% & \(99\) & \(198.3\) & \(10.62\;(3.23)\) & \(12.14\;(4.03)\) & \(11.79\;(3.79)\) & \(0.3804\) & \(34.4\) \\
10\% & \(149\) & \(131.7\) & \(11.10\;(3.81)\) & \(12.66\;(4.73)\) & \(12.23\;(4.37)\) & \(0.3054\) & \(46.1\) \\
10\% & \(198\) & \(98.9\) & \(11.51\;(4.44)\) & \(13.10\;(5.47)\) & \(12.63\;(5.03)\) & \(0.2695\) & \(56.2\) \\
10\% & \(248\) & \(78.8\) & \(11.91\;(5.07)\) & \(13.54\;(6.21)\) & \(13.05\;(5.71)\) & \(0.2319\) & \(62.9\) \\
10\% & \(298\) & \(65.4\) & \(12.29\;(5.69)\) & \(13.95\;(6.94)\) & \(13.46\;(6.37)\) & \(0.2166\) & \(67.0\) \\
10\% & \(397\) & \(48.9\) & \(12.91\;(6.83)\) & \(14.62\;(8.29)\) & \(14.16\;(7.61)\) & \(0.1950\) & \(70.4\) \\
10\% & \(496\) & \(38.9\) & \(13.39\;(7.83)\) & \(15.13\;(9.48)\) & \(14.71\;(8.70)\) & \(0.1780\) & \(70.8\) \\
10\% & \(745\) & \(25.6\) & \(14.03\;(9.67)\) & \(15.72\;(11.69)\) & \(15.48\;(10.71)\) & \(0.1413\) & \(67.6\) \\
25\% & \(99\) & \(198.3\) & \(4.64\;(1.45)\) & \(4.57\;(1.77)\) & \(4.74\;(1.74)\) & \(0.2907\) & \(42.2\) \\
25\% & \(149\) & \(131.7\) & \(4.47\;(1.76)\) & \(4.34\;(2.14)\) & \(4.55\;(2.11)\) & \(0.2228\) & \(42.0\) \\
25\% & \(198\) & \(98.9\) & \(4.22\;(1.96)\) & \(4.04\;(2.39)\) & \(4.27\;(2.35)\) & \(0.2023\) & \(38.7\) \\
25\% & \(248\) & \(78.8\) & \(3.94\;(2.10)\) & \(3.71\;(2.54)\) & \(3.94\;(2.50)\) & \(0.1934\) & \(34.4\) \\
25\% & \(298\) & \(65.4\) & \(3.66\;(2.18)\) & \(3.38\;(2.62)\) & \(3.60\;(2.58)\) & \(0.1747\) & \(29.7\) \\
25\% & \(397\) & \(48.9\) & \(3.12\;(2.26)\) & \(2.76\;(2.67)\) & \(2.96\;(2.64)\) & \(0.1501\) & \(26.2\) \\
25\% & \(496\) & \(38.9\) & \(2.66\;(2.29)\) & \(2.24\;(2.65)\) & \(2.41\;(2.63)\) & \(0.1267\) & \(25.2\) \\
25\% & \(745\) & \(25.6\) & \(1.89\;(2.37)\) & \(1.45\;(2.60)\) & \(1.56\;(2.60)\) & \(0.1007\) & \(28.8\) \\
\hline
\end{tabular}

\end{table}

\begin{table}[H]
\caption{Localization sensitivity at income percentiles 50--75.
Definitions and units follow Table~\ref{tab:affine_sipp_orders_1}.}
\label{tab:affine_sipp_orders_2}
\centering\footnotesize\setlength{\tabcolsep}{3pt}
\begin{tabular}{@{}lrrrrrrr@{}}
\hline
Income & $s$ & $N$ & All-tree & Single-fit/OOB & Five-fold & Radius & Largest (\%) \\
\hline
50\% & \(99\) & \(198.3\) & \(7.29\;(3.14)\) & \(7.30\;(3.58)\) & \(7.53\;(3.79)\) & \(0.2247\) & \(51.3\) \\
50\% & \(149\) & \(131.7\) & \(7.08\;(3.73)\) & \(7.02\;(4.26)\) & \(7.45\;(4.59)\) & \(0.1943\) & \(61.3\) \\
50\% & \(198\) & \(98.9\) & \(6.86\;(4.13)\) & \(6.73\;(4.73)\) & \(7.32\;(5.13)\) & \(0.1659\) & \(66.0\) \\
50\% & \(248\) & \(78.8\) & \(6.64\;(4.42)\) & \(6.46\;(5.06)\) & \(7.18\;(5.50)\) & \(0.1527\) & \(67.5\) \\
50\% & \(298\) & \(65.4\) & \(6.47\;(4.59)\) & \(6.24\;(5.26)\) & \(7.05\;(5.73)\) & \(0.1357\) & \(67.0\) \\
50\% & \(397\) & \(48.9\) & \(6.22\;(4.73)\) & \(5.93\;(5.39)\) & \(6.83\;(5.86)\) & \(0.1143\) & \(62.4\) \\
50\% & \(496\) & \(38.9\) & \(6.05\;(4.68)\) & \(5.72\;(5.32)\) & \(6.63\;(5.74)\) & \(0.1020\) & \(54.9\) \\
50\% & \(745\) & \(25.6\) & \(5.84\;(4.45)\) & \(5.47\;(4.96)\) & \(6.21\;(5.26)\) & \(0.0978\) & \(30.3\) \\
75\% & \(99\) & \(198.3\) & \(2.27\;(5.13)\) & \(1.91\;(5.72)\) & \(1.83\;(5.87)\) & \(0.2108\) & \(51.7\) \\
75\% & \(149\) & \(131.7\) & \(0.93\;(7.00)\) & \(0.52\;(7.83)\) & \(0.47\;(8.02)\) & \(0.1838\) & \(62.8\) \\
75\% & \(198\) & \(98.9\) & \(0.25\;(8.84)\) & \(-0.20\;(9.90)\) & \(-0.22\;(10.14)\) & \(0.1406\) & \(69.9\) \\
75\% & \(248\) & \(78.8\) & \(-0.33\;(10.73)\) & \(-0.83\;(12.02)\) & \(-0.85\;(12.31)\) & \(0.1179\) & \(74.9\) \\
75\% & \(298\) & \(65.4\) & \(-0.99\;(12.63)\) & \(-1.55\;(14.16)\) & \(-1.58\;(14.49)\) & \(0.1081\) & \(78.4\) \\
75\% & \(397\) & \(48.9\) & \(-2.65\;(16.45)\) & \(-3.41\;(18.43)\) & \(-3.51\;(18.87)\) & \(0.1031\) & \(83.1\) \\
75\% & \(496\) & \(38.9\) & \(-4.77\;(20.29)\) & \(-5.76\;(22.74)\) & \(-5.97\;(23.28)\) & \(0.0978\) & \(85.9\) \\
75\% & \(745\) & \(25.6\) & \(-11.40\;(30.02)\) & \(-13.12\;(33.64)\) & \(-13.62\;(34.43)\) & \(0.0921\) & \(90.0\) \\
\hline
\end{tabular}

\end{table}

\begin{table}[H]
\caption{Localization sensitivity at income percentiles 90--95.
Definitions and units are those in Table~\ref{tab:affine_sipp_orders_1}.}
\label{tab:affine_sipp_orders_3}
\centering\footnotesize\setlength{\tabcolsep}{3pt}
\begin{tabular}{@{}lrrrrrrr@{}}
\hline
Income & $s$ & $N$ & All-tree & Single-fit/OOB & Five-fold & Radius & Largest (\%) \\
\hline
90\% & \(99\) & \(198.3\) & \(5.99\;(8.33)\) & \(5.37\;(9.37)\) & \(5.37\;(9.32)\) & \(0.2844\) & \(29.8\) \\
90\% & \(149\) & \(131.7\) & \(4.80\;(9.44)\) & \(4.08\;(10.57)\) & \(3.98\;(10.41)\) & \(0.2220\) & \(28.7\) \\
90\% & \(198\) & \(98.9\) & \(4.34\;(10.53)\) & \(3.58\;(11.73)\) & \(3.37\;(11.49)\) & \(0.2002\) & \(36.6\) \\
90\% & \(248\) & \(78.8\) & \(3.99\;(11.65)\) & \(3.21\;(12.91)\) & \(2.89\;(12.61)\) & \(0.1878\) & \(42.0\) \\
90\% & \(298\) & \(65.4\) & \(3.64\;(12.71)\) & \(2.85\;(14.02)\) & \(2.40\;(13.69)\) & \(0.1687\) & \(45.5\) \\
90\% & \(397\) & \(48.9\) & \(2.85\;(14.61)\) & \(2.09\;(16.01)\) & \(1.41\;(15.62)\) & \(0.1346\) & \(48.5\) \\
90\% & \(496\) & \(38.9\) & \(1.96\;(16.27)\) & \(1.24\;(17.76)\) & \(0.37\;(17.34)\) & \(0.1183\) & \(48.0\) \\
90\% & \(745\) & \(25.6\) & \(-0.82\;(19.78)\) & \(-1.38\;(21.50)\) & \(-2.57\;(21.07)\) & \(0.1005\) & \(39.2\) \\
95\% & \(99\) & \(198.3\) & \(8.75\;(18.96)\) & \(8.86\;(21.16)\) & \(10.27\;(21.71)\) & \(0.3513\) & \(84.4\) \\
95\% & \(149\) & \(131.7\) & \(5.25\;(26.58)\) & \(5.19\;(29.62)\) & \(6.59\;(30.49)\) & \(0.2921\) & \(90.5\) \\
95\% & \(198\) & \(98.9\) & \(0.66\;(33.61)\) & \(0.18\;(37.44)\) & \(1.54\;(38.57)\) & \(0.2581\) & \(92.7\) \\
95\% & \(248\) & \(78.8\) & \(-4.70\;(40.22)\) & \(-5.75\;(44.79)\) & \(-4.41\;(46.18)\) & \(0.2234\) & \(93.8\) \\
95\% & \(298\) & \(65.4\) & \(-10.25\;(46.27)\) & \(-11.93\;(51.52)\) & \(-10.60\;(53.13)\) & \(0.2088\) & \(94.4\) \\
95\% & \(397\) & \(48.9\) & \(-20.82\;(56.68)\) & \(-23.72\;(63.10)\) & \(-22.40\;(65.09)\) & \(0.1953\) & \(94.8\) \\
95\% & \(496\) & \(38.9\) & \(-30.14\;(65.21)\) & \(-34.13\;(72.58)\) & \(-32.78\;(74.90)\) & \(0.1742\) & \(94.7\) \\
95\% & \(745\) & \(25.6\) & \(-47.02\;(79.87)\) & \(-52.96\;(88.89)\) & \(-51.47\;(91.78)\) & \(0.1259\) & \(92.8\) \\
\hline
\end{tabular}

\end{table}

\subsection{Eligibility contrasts on a transformed scale}

We also compare OOB and five-fold eligibility estimates for asinh of net financial assets measured in dollars, taking the mean eligibility contrast on this transformed scale as the estimand (Table~\ref{tab:affine_sipp_asinh}).
All three interior transformed intervals include zero under both prediction methods, while those at the 10th and 95th profiles are positive.

\begin{table}[H]
\caption{Asinh-outcome contrasts at all seven income profiles.
Each prediction column gives estimate (SE).
Intervals use a pointwise normal
reference. The transformation is applied to 1991 dollars. The estimand is
the mean contrast on the transformed scale.}
\label{tab:affine_sipp_asinh}
\centering\small
\begin{tabular}{@{}lrrrr@{}}
\hline
Income & Single-fit/OOB & 95\% interval & Five-fold & 95\% interval \\
\hline
5\% & \(6.46\;(6.40)\) & \([-6.09,\ 19.01]\) & \(6.94\;(5.63)\) & \([-4.09,\ 17.98]\) \\
10\% & \(8.72\;(3.86)\) & \([1.15,\ 16.29]\) & \(8.48\;(3.61)\) & \([1.40,\ 15.55]\) \\
25\% & \(-0.37\;(2.58)\) & \([-5.43,\ 4.70]\) & \(-0.26\;(2.63)\) & \([-5.41,\ 4.89]\) \\
50\% & \(-2.41\;(2.52)\) & \([-7.35,\ 2.53]\) & \(-2.15\;(2.55)\) & \([-7.15,\ 2.85]\) \\
75\% & \(0.20\;(3.29)\) & \([-6.25,\ 6.65]\) & \(0.34\;(3.27)\) & \([-6.06,\ 6.75]\) \\
90\% & \(0.47\;(2.18)\) & \([-3.80,\ 4.74]\) & \(0.73\;(2.11)\) & \([-3.41,\ 4.87]\) \\
95\% & \(4.35\;(2.15)\) & \([0.14,\ 8.56]\) & \(4.22\;(2.13)\) & \([0.04,\ 8.40]\) \\
\hline
\end{tabular}

\end{table}

The paper's pointwise inference theorem does not cover this application.
Its learner conditions have not been verified for regression forests, and the bounded coefficient envelope excludes unbounded dollar outcomes because \(b=-\phi\) contains \(Y\).
Integer-valued age violates the two-dimensional support and small-ball requirements of \AffineExternalRef{paper}{cond:affine_design}.
The reporting profiles and coordinate scales are also estimated from the sample.
We therefore report the intervals as illustrative normal-reference comparisons.

\FloatBarrier
\section{Additional Simulation Diagnostics}\label{supp:affine_simulation_diagnostics}

Agreement between full-sample and cross-fitted estimates leaves open how accurately either estimates the parameter at the reporting point.
Using the main simulation designs and draws (\AffineExternalRef{paper}{sec:affine_numerical}), we examine how nuisance approximation, spatial bias and studentization contribute to coverage shortfalls.
Comparisons at fixed point estimates then assess the effect of covariance estimation.

\subsection{Bias, studentization and interval coverage}
\label{app:affine_production_calibration}

Coverage comparisons within each cell hold DNN weights fixed and vary nuisance predictions.

\begin{table}[H]
\caption{Selected DNN performance, with 10,000 replications per row.
Bias and RMSE are for full-sample fitting.
Coverage compares full-sample fitting,
pooled five-fold fitting and true nuisances under the frozen-score jackknife.
The parameter of interest equals \(1/8\). The C2 multiplier is shown in the case label.}
\label{tab:affine_production_performance}
\centering\small
\begin{tabular}{@{}llrrrrr@{}}
\hline
 &  & \multicolumn{2}{c}{Full-sample fitting} & \multicolumn{3}{c}{Coverage} \\
\cline{3-4}\cline{5-7}
Case $(n,\lambda_s)$ & App. & Bias & RMSE & Full & Five-fold & True \\
\hline
C0 $(512,1)$ & CATE & 0.0001 & 0.1143 & 0.9125 & 0.9122 & 0.8963 \\
C0 $(512,1)$ & Slope & -0.0004 & 0.1136 & 0.9127 & 0.9127 & 0.8893 \\
C1 $(2\,048,1)$ & CATE & 0.0303 & 0.1103 & 0.9186 & 0.9182 & 0.9209 \\
C1 $(2\,048,1)$ & Slope & 0.0301 & 0.1102 & 0.9188 & 0.9190 & 0.9125 \\
C2 $(512,0.4)$ & CATE & 0.0863 & 0.1124 & 0.7535 & 0.7561 & 0.6729 \\
C2 $(512,0.4)$ & Slope & 0.0856 & 0.1116 & 0.7513 & 0.7511 & 0.6713 \\
\hline
\end{tabular}

\end{table}

\begin{samepage}
The C1 results link the observed bias to nuisance approximation: for CATE at \(n=2048\), \(M=1\), the unpenalized population projection has signal bias 0.03125, close in magnitude to the fitted learner's fresh-point score mean of \(-0.03138\) and estimator bias of 0.03031.
The score has the opposite sign because it is written as parameter minus signal.

\end{samepage}

Mean estimated SE can be close to Monte Carlo SD even when the studentized errors have heavier tails than the normal reference.
On the same draws, we compare the ratio of mean SE to Monte Carlo SD, \(\overline{\hat{\mathrm{SE}}}/\mathrm{SD}_{\rm MC}(\hat a)\), with the corresponding ratio using the root mean square SE, \(\sqrt{\overline{\hat V}}/\mathrm{SD}_{\rm MC}(\hat a)\).
Their difference reflects variability in the estimated scale.
To separate the role of centering from that of the estimated scale, Table~\ref{tab:affine_production_calibration} also reports coverage of the empirical Monte Carlo mean and each draw's weighted target, defined in \AffineExternalRef{paper}{eq:affine_realized_weighted_truth}.
The empirical mean is shared across draws, so an ordinary independent-binomial interval is not reported for its diagnostic coverage.

\begin{table}[H]
\caption{Scale and coverage for full-sample DNN jackknife intervals, with
10,000 matched draws per row. Mean and RMS SE ratios use Monte Carlo SD.
The last two columns give coverage of the empirical Monte Carlo mean and
the weighted target.
Point coverage is in
Tables~\ref{tab:affine_production_performance} and \ref{tab:affine_production_endpoint}.}
\label{tab:affine_production_calibration}
\centering\small
\begin{tabular}{@{}llrrrr@{}}
\hline
Case $(n,\lambda_s)$ & App. & \shortstack{Mean\\SE/SD} & \shortstack{RMS\\SE/SD} & \shortstack{Empirical\\center} & \shortstack{Weighted\\truth} \\
\hline
C0 $(512,1)$ & CATE & 1.006 & 1.046 & 0.9125 & 0.9127 \\
C0 $(512,1)$ & Slope & 1.006 & 1.047 & 0.9129 & 0.9120 \\
C2 $(512,0.4)$ & CATE & 0.994 & 1.009 & 0.9345 & 0.9388 \\
C2 $(512,0.4)$ & Slope & 0.995 & 1.009 & 0.9339 & 0.9381 \\
C0 $(531\,441,1)$ & CATE & 1.015 & 1.020 & 0.9456 & 0.9456 \\
C0 $(531\,441,1)$ & Slope & 1.014 & 1.018 & 0.9441 & 0.9441 \\
\hline
\end{tabular}

\end{table}

\begin{figure}[H]
\centering
\begingroup%
\makeatletter%
\let\Gread@transgrouptrue\Gread@transgroupfalse%
\makeatother%
\includegraphics[width=.92\linewidth]{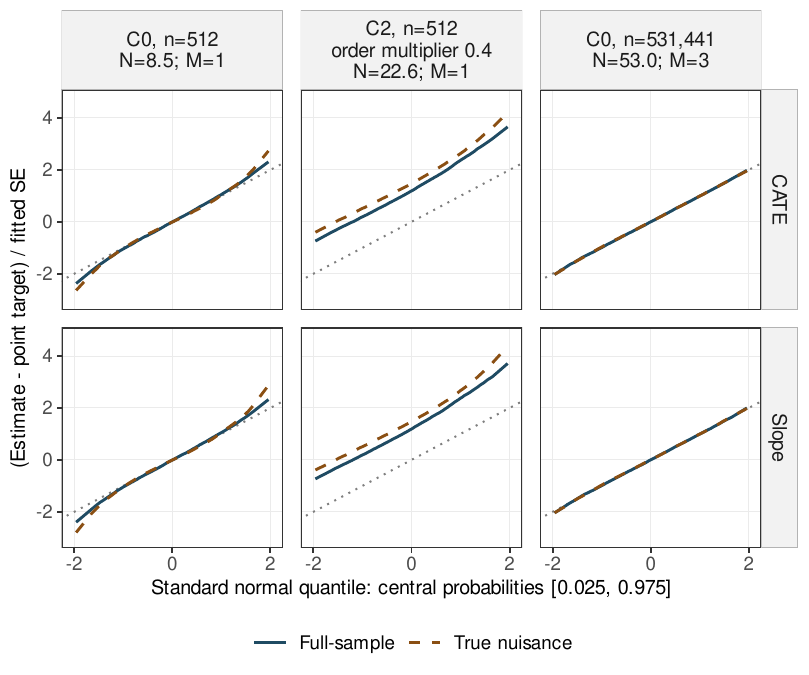}%
\endgroup
\caption{QQ plots centered at the point target for full-sample fitting and
true nuisances, with 10,000 studentized values per curve. Local information
increases from left to right.
The plotted probability range \([0.025,0.975]\) includes both quantile levels
used for the confidence limits. The line is the normal reference.}
\label{fig:affine_production_studentization}
\end{figure}

\begin{samepage}
In C0 at \(n=512\), CATE's studentized 2.5 and 97.5 percent quantiles are \(-2.373\) and 2.308, with point coverage of 0.9125 despite mean-SE/SD of 1.006.
Its studentized excess kurtosis is 1.614, compared with \(-0.102\) after standardization by Monte Carlo SD.
In C2 with multiplier 0.4, lower and upper studentized-tail probabilities of 0.0014 and 0.2451 reflect a positive shift relative to the point target.
Thus C0's shortfall accompanies heavier studentized tails, while C2 adds spatial bias.

\end{samepage}

\begin{table}[H]
\caption{Paired interval and MSE comparisons, with 10,000 full-sample fits
and no exclusions. Kernel changes are jackknife minus conventional covariance
at the same point estimate. DNN-minus-kernel MSE is in units of \(10^{-3}\).
Positive loss favors the kernel. Parentheses give paired Monte Carlo standard
errors (MCSEs) for coverage and loss. The C2 order multiplier is 0.4.}
\label{tab:affine_production_pairs}
\centering\small
\begin{tabular}{@{}llrrr@{}}
\hline
Case $(n,\lambda_s)$ & App. & \shortstack{Kernel JK$-$conv.\\coverage (MCSE)} & \shortstack{Length\\cost} & \shortstack{DNN$-$kernel MSE\\$10^{-3}$ (MCSE)} \\
\hline
C0 $(2\,048,1)$ & CATE & +0.0303 (0.0017) & +0.040 & -0.153 (0.049) \\
C0 $(2\,048,1)$ & Slope & +0.0302 (0.0017) & +0.040 & -0.144 (0.048) \\
C2 $(512,0.4)$ & CATE & +0.0292 (0.0017) & +0.016 & +0.101 (0.028) \\
C2 $(512,0.4)$ & Slope & +0.0290 (0.0017) & +0.016 & +0.096 (0.027) \\
C2S $(2\,048,1)$ & CATE & +0.0342 (0.0018) & +0.041 & +0.003 (0.052) \\
C2S $(2\,048,1)$ & Slope & +0.0349 (0.0018) & +0.041 & +0.011 (0.051) \\
\hline
\end{tabular}

\end{table}

The DNN jackknife and conventional kernel intervals differ in both covariance estimation and localization: in C0 at \(n=2048\), the former gains 0.0345 in coverage (MCSE 0.00197), with mean interval length increasing by 0.04094.
Holding the kernel estimate fixed isolates the covariance change and gives a coverage gain of 0.0303 (MCSE 0.00171), with mean interval length increasing by 0.04043.
The wider intervals therefore accompany a coverage gain even when the local estimate is unchanged.

For point estimation, changing localization affects both bias and variance: the kernel has lower CATE MSE in C2 at \(n=512\), multiplier 0.4, by 0.000101 (paired MCSE 0.000028).

Agreement in aggregate coverage can also conceal differences between training methods on individual draws.
In C0 at \(n=65{,}536\), the full-sample and pooled five-fold CATE coverage indicators differ in 14 of 10,000 replications despite equal coverage rates (paired MCSE 0.000374).

Under heteroskedasticity, C0H has a coverage--length pattern similar to C0 in \AffineExternalRef{paper}{tab:affine_kernel_comparison}.
Reporting density changes the extent of local averaging: moving from C2 to C2S at \(n=2048\) increases the mean DNN weighted radius from 0.02862 to 0.05724 while leaving \(N=12.46\) unchanged.
The wider neighborhood is accompanied by a decline in full-sample CATE coverage from 0.9229 to 0.8705, a paired change of \(-0.0524\) (MCSE 0.00281), and an increase in DNN bias from 0.01826 to 0.05974, with kernel bias of 0.05910 in C2S.
Equal effective sample sizes therefore need not imply comparable spatial averaging or comparable accuracy for the point parameter.

The same 21 replications in C0 and C1 at \(n=4096\) fall outside the Gram event, but ridge regularization and clipping leave every fit defined, so all draws are retained.
Monte Carlo quantities therefore use 10,000 replications, with shared draws for paired comparisons.

\subsection{Covariance estimation and dictionary refinement}

For the slope, the fitted coefficient varies across observations, so changing the parameter value changes the frozen scores by different amounts.
The covariance comparison must therefore account for parameter-estimation error as well as nuisance-estimation error.
We separate these contributions to see how each affects the fitted pseudo-values and their covariance.

\begin{samepage}
For a scalar score, write \(Q_j(a,\veceta)\) for the frozen score pseudo-value in \AffineExternalRef{paper}{eq:affine_score_pseudovalues}, and \(Q_{B,j}(\veceta)\) for the same pseudo-value formula applied to \((B(W_i;\veceta))_{i=1}^n\), as defined before \AffineExternalRef{paper}{lem:affine_frozen_quadratic}.
Parameter values and nuisance predictions are held fixed within each calculation.
The operator \(\operatorname{center}\) subtracts the mean over deletion indices.
Set \(q^{\rm or}=\operatorname{center}(Q(a_0(\bx),\veceta_0))\) and \(q^{\rm fit}=\operatorname{center}(Q(\hat a,\hat{\veceta}))\), and \(f=N/[n(n-1)]\).
Their difference decomposes as
\[
C=\operatorname{center}\{Q(a_0(\bx),\hat{\veceta})
                              -Q(a_0(\bx),\veceta_0)\},\qquad M=(\hat a-a_0(\bx))\operatorname{center}(Q_B(\hat{\veceta})),
 \qquad D=q^{\rm fit}-q^{\rm or}=C+M.
\]
The term \(C\) captures nuisance-estimation error with the parameter held at its true value, while \(M\) records the additional change from estimating the parameter through the fitted coefficient pseudo-values.
\end{samepage}
The scaled squared pseudo-value difference and score-covariance difference satisfy
\[
T=f\sum_jD_j^2=f\sum_jC_j^2+f\sum_jM_j^2+2f\sum_jC_jM_j,\qquad F-O=T+2f\sum_jq_j^{\rm or}D_j.
\]
Here \(F\) and \(O\) are the feasible and oracle jackknife score covariance estimates, and \(T\) measures the scaled squared discrepancy between fitted and oracle pseudo-values.
The discrepancy between pseudo-values does not by itself determine the covariance difference: the oracle cross term \(F-O-T\) can increase or offset the contribution of \(T\).
The cross term between the nuisance and parameter components is also signed, so their separate squared contributions need not add up to the total discrepancy.
Centering removes parameter-estimation error for CATE's constant coefficient, while the slope retains this contribution.
Across the reported sizes, the nuisance component exceeds the parameter component.

\begin{table}[H]
\caption{C0 slope score covariance under full-sample fitting, with 10,000 draws per row.
Entries are means of draw-specific ratios to oracle jackknife score
covariance \(O\): nuisance and parameter components, signed nuisance--parameter
cross term, total scaled squared difference, and oracle cross term \((F-O-T)/O\).
The signed cross column includes its factor two.}
\label{tab:affine_production_covariance}
\centering\small
\begin{tabular}{@{}rrrrrr@{}}
\hline
$n$ & Component & Parameter & Cross & $T$ & \shortstack{Oracle cross\\$F-O-T$} \\
\hline
2\,048 & 0.8703 & 0.00073 & -0.00143 & 0.8696 & +0.01245 \\
65\,536 & 0.1505 & 0.00432 & -0.00129 & 0.1535 & -0.00654 \\
531\,441 & 0.2523 & 0.00213 & -0.00115 & 0.2533 & -0.00276 \\
\hline
\end{tabular}

\end{table}

Table~\ref{tab:affine_production_endpoint} assesses variance calibration using the asymptotic point variance as reference.
For CATE at the three-bin partition (M3), the ratio of mean estimated score covariance to mean oracle jackknife covariance is 1.227; averaging the ratios from individual draws gives 1.229.

\begin{table}[H]
\caption{Full-sample C0 results: two bins per adjustment coordinate (M2)
at \(n=65{,}536\) and three bins (M3) at
\(n=531{,}441\), with 10,000 draws each.
Fresh-point risk sums propensity and both outcome-mean MSEs for CATE, and
propensity and marginal-outcome MSEs for slope, conditional on the
reporting point. Projection risk uses the unpenalized population projections.
Coverage bounds are exact binomial 95 percent intervals. Gap quantiles are
95th percentiles of absolute paired differences.
Normalized gaps use
\(\sqrt N/\sqrt{\Omega_{\bx}}\), and raw gaps use estimate units. Normalized variance ratios
use the asymptotic conditional point constant.}
\label{tab:affine_production_endpoint}
\centering\small
\begin{tabular}{@{}lrrrr@{}}
\hline
Metric & \shortstack{M2\\CATE} & \shortstack{M2\\Slope} & \shortstack{M3\\CATE} & \shortstack{M3\\Slope} \\
\hline
Bias & 0.0002 & 0.0002 & 0.0000 & 0.0000 \\
RMSE & 0.0488 & 0.0487 & 0.0383 & 0.0383 \\
Point coverage & 0.9369 & 0.9352 & 0.9455 & 0.9441 \\
95\% coverage lower & 0.9320 & 0.9302 & 0.9409 & 0.9394 \\
95\% coverage upper & 0.9416 & 0.9399 & 0.9499 & 0.9485 \\
Mean interval length & 0.193 & 0.192 & 0.152 & 0.152 \\
Fresh-point risk sum & 0.00600 & 0.00379 & 0.00993 & 0.00625 \\
Analytic projection risk & 0.00592 & 0.00372 & 0.00988 & 0.00621 \\
\shortstack[l]{Absolute full/five-fold q95\\(asymptotic SD units)} & 0.024 & 0.023 & 0.020 & 0.019 \\
\shortstack[l]{Absolute full/true-nuisance q95\\(asymptotic SD units)} & 0.751 & 0.744 & 0.967 & 0.960 \\
\shortstack[l]{Absolute full/true-nuisance q95\\(estimate units)} & 0.0343 & 0.0339 & 0.0338 & 0.0335 \\
$\Ex{\widehat\Omega}/\Omega_x$ & 1.172 & 1.168 & 1.249 & 1.247 \\
$N\Varbi{\rm MC}{\widehat a}/\Omega_x$ & 1.143 & 1.141 & 1.202 & 1.205 \\
RMS SE/MC SD & 1.012 & 1.012 & 1.020 & 1.018 \\
\hline
\end{tabular}

\end{table}

Moving from M2 to M3 changes cell boundaries as well as dictionary dimension: within-cell sine variance rises from 0.09472 to 0.15804, alongside higher projection and fresh-prediction risk (Table~\ref{tab:affine_production_endpoint}).
Despite this increase in approximation and prediction error, paired CATE coverage improves by 0.0086 (MCSE 0.00327), with mean interval-length change \(-0.04027\).
These outcomes occur along a joint schedule for \(n,N,M,K,s\), so the endpoints combine increasing sample size and local information with changes in localization and dictionary approximation.

Holding the DNN point estimate fixed, Table~\ref{tab:affine_dnn_covariance_routes} compares jackknife, direct-moment and squared-weight covariance using the paper's simulation-appendix formulas.
Both comparisons show larger coverage gains and interval-length increases at \(n=512\), when local information is more limited.
The coverage difference narrows at \(n=65{,}536\), where CATE's jackknife/direct-moment difference is 0.0003 (paired MCSE 0.0009), with longer jackknife intervals.

\begin{table}[H]
\caption{C0 CATE intervals with alternative DNN covariance estimators under full-sample fitting, with 10,000 shared draws per
cell. Upper panel: point coverage (mean interval length). Lower panel:
jackknife-minus-alternative coverage and mean length differences, with
paired coverage MCSEs in parentheses. All three estimators use the same point
estimate and normal 95 percent reference.}
\label{tab:affine_dnn_covariance_routes}
\centering\small
\begin{tabular}{@{}llrrr@{}}
\hline
$n$ & App. & Jackknife & Direct moment & Direct squared \\
\hline
512 & CATE & 0.9125 (0.450) & 0.9036 (0.409) & 0.8620 (0.383) \\
65\,536 & CATE & 0.9369 (0.193) & 0.9366 (0.187) & 0.9256 (0.184) \\
\hline
$n$ & App. & Alternative & \shortstack{Coverage gain\\(paired MCSE)} & Length cost \\
\hline
512 & CATE & Moment & +0.0089 (0.0016) & +0.041 \\
512 & CATE & Squared & +0.0505 (0.0022) & +0.068 \\
65\,536 & CATE & Moment & +0.0003 (0.0009) & +0.006 \\
65\,536 & CATE & Squared & +0.0113 (0.0011) & +0.008 \\
\hline
\end{tabular}

\end{table}


\bibliographystyle{elsarticle-harv}
\bibliography{bibliography,affine_scores}